\documentclass[fleqn,usenatbib]{mnras}

\usepackage{newtxtext,newtxmath}

\usepackage[T1]{fontenc}

\DeclareRobustCommand{\VAN}[3]{#2}
\let\VANthebibliography\thebibliography
\def\thebibliography{\DeclareRobustCommand{\VAN}[3]{##3}\VANthebibliography}

\usepackage{url}
\usepackage[mathscr]{euscript}
\usepackage[cal=boondoxo]{mathalfa}
\usepackage{booktabs}
\usepackage{listings}
\usepackage{ulem}
\usepackage[dvipsnames]{xcolor}

\def\aj{AJ}
\def\araa{ARA\&A}
\def\apj{ApJ}
\def\apjl{ApJ}
\def\apjs{ApJS}

\def\aap{A\&A}

\def\aaps{A\&AS}
\def\mnras{MNRAS}
\def\pasp{PASP}

\usepackage{graphicx}	
\usepackage{amsmath}	
\usepackage{soul}
\usepackage{acronym}

\acrodef{SED}{spectral energy distribution}
\defcitealias{Coelho14}{Coelho14}
\defcitealias{Castelli+Kurucz2003}{C\&K03}
\defcitealias{LopesSanjuan+2019}{LS+19}

\definecolor{darkgreen}{rgb}{0.09, 0.45, 0.27}

\newcommand{\ionnote}[2]{{\textrm{#1}}{\textrm{\sc #2}}}

\title[Data Release 2 of S-PLUS]{Data Release 2 of S-PLUS: accurate template-fitting based photometry covering $\sim$1000 square degrees in 12 optical filters}

\author[F. Almeida-Fernandes et al.]{
F.~Almeida-Fernandes$^{1}$\thanks{E-mail: felipe.almeida.fernandes@usp.br, felipefer42@gmail.com},
L.~Sampedro$^{1}$,
F.R. Herpich$^{1}$,
A. ~Molino$^{1}$,
C. E. ~Barbosa$^{1}$,
M.L. ~Buzzo$^{1}$,
\newauthor{
R. A. ~Overzier$^{2}$,
E. V.R. de ~Lima$^{1}$,
L.M.I. ~Nakazono$^{1}$,
G.B. ~Oliveira Schwarz$^{3}$,
H. D. ~Perottoni$^{1}$,
}
\newauthor{
G. F. ~Bolutavicius$^{1}$,
L.A. ~Guti{\'e}rrez-Soto$^{1}$,
T. ~Santos-Silva$^{1}$,
A. Z. ~Vitorelli$^{1}$,
A. ~Werle$^{1, 4, 11}$,
}
\newauthor{
D. D. Whitten$^{5}$,
M.V. ~Costa Duarte$^{1}$,
C.R. ~Bom$^{6,7}$,
P. ~Coelho$^{1}$,
L. ~Sodr{\'e} Jr.$^{1}$,
V. M. ~Placco$^{8}$,
}
\newauthor{
G. S. M. Teixeira$^{6}$,
J. Alonso-Garc\'{i}a$^{9,10}$,
T.C. Beers$^{5}$,
A. ~Kanaan$^{11}$,
T. ~Ribeiro$^{12}$,
W. ~Schoenell$^{13}$, 
}
\newauthor{
C. ~Mendes de Oliveira$^{1}$
}
\\
$^{1}$Departamento de Astronomia, IAG, Universidade de S\~ao Paulo, Rua do Mat\~ao, 1226, 05509-900, S\~ao Paulo, Brazil\\
$^{2}$Observat\'orio Nacional/MCTIC, R. Gen. Jos\'e Cristino, 77,  20921-400, Rio de Janeiro, Brazil\\
$^{3}$Universidade Presbiteriana Mackenzie, R. da Consola\c{c}\~ao, 930 - Consola\c{c}\~ao, S\~ao Paulo, Brazil\\
$^{4}$INAF- Osservatorio Astronomico di Padova, Vicolo Osservatorio 5, 35122 Padova, Italy \\
$^{5}$Department of Physics and JINA Center for the Evolution of the Elements, University of Notre Dame, Notre Dame, IN 46556, USA\\
The remaining institutions are at the end of the paper.
}

\date{Accepted XXX. Received YYY; in original form ZZZ}

\pubyear{2021}

\begin{document}
\label{firstpage}
\pagerange{\pageref{firstpage}--\pageref{lastpage}}
\maketitle

\begin{abstract}
The Southern Photometric Local Universe Survey (S-PLUS) is an ongoing survey of $\sim$9300 deg$^2$ in the southern sky in a 12-band photometric system. This paper presents the second data release (DR2) of S-PLUS, consisting of 514 tiles covering an area of 950 deg$^2$. The data has been fully calibrated using a new photometric calibration technique suitable for the new generation of wide-field multi-filter surveys. This technique consists of a $\chi^2$ minimisation to fit synthetic stellar templates to already calibrated data from other surveys, eliminating the need for standard stars and reducing the survey duration by $\sim$15\%. We compare the template-predicted and S-PLUS instrumental magnitudes to derive the photometric zero-points (ZPs). We show that these ZPs can be further refined by fitting the stellar templates to the 12 S-PLUS magnitudes, which better constrain the models by adding the narrow-band information. We use the STRIPE82 region to estimate ZP errors, which are $\lesssim10$ mmags for filters J0410, J0430, $g$, J0515, $r$, J0660, $i$, J0861 and $z$; $\lesssim 15$ mmags for filter J0378; and $\lesssim 25$ mmags for filters $u$ and J0395. We describe the complete data flow of the S-PLUS/DR2 from observations to the final catalogues and present a brief characterisation of the data. We show that, for a minimum signal-to-noise threshold of 3, the photometric depths of the DR2 range from 19.9 mag to 21.3 mag (measured in Petrosian apertures), depending on the filter. The S-PLUS DR2 can be accessed from the website: \href{https://splus.cloud}{https://splus.cloud}.

\end{abstract}

\begin{keywords}
surveys -- techniques: photometric  -- catalogues -- astronomical data bases: miscellaneous -- stars: general -- galaxies: general 
\end{keywords}



\section{Introduction}

Wide-field photometric surveys are essential for research in astronomy, especially because of the large volume of data they are able to provide in a reasonable amount of time and with more extensive sky coverage compared to spectroscopic surveys. Surveys such as the Sloan Digital Sky Survey \citep[SDSS,][]{York+2000}, 2MASS \citep{Skrutskie+2006}, ATLAS \citep{Shanks+2015}, and PanSTARRS, \citep{Chambers+2016}, to mention a few, have contributed to the development of countless areas in astronomy: from the study of asteroids to the large scale structure of the Universe.

Following the success of these past surveys, several ongoing and planned projects are being executed to complement and supplement the available data in terms of increasing i) the sky-coverage: mainly by including the southern hemisphere (e.g. DES, \citealp{Abbott+2018}; Skymapper \citealp{Wolf+2018}); ii) the photometric-depth: reaching fainter magnitudes (e.g. LSST; \citealp{Ivezic+2019}); or iii) the wavelength range and resolution: extending or increasing the number of pass-bands in previous filter systems and even replacing broadband with narrow-band filters to widen the spectral feature sensitivity (e.g. the Pristine Survey, \citealp{Starkenburg+2017}).

Regarding the topic of expanding the wavelength resolution, three surveys clearly stand out: the Javalambre Physics of the Accelerating Universe Astrophysical Survey (JPAS, \citealp{Benitez+2014}; and miniJPAS, \citealp{Bonoli+2020}), the Javalambre Photometric Local Universe Survey \citep[J-PLUS,][]{Cenarro+2019} and the Southern Photometric Local Universe Survey \citep[S-PLUS,][]{MendesDeOliveira+2019}. Of these, J-PAS is the most ambitious and plans to cover an area of 8500 deg$^2$ of the northern sky, observing in 54 equally-spaced narrow-band filters. J-PLUS has a crucial role in the calibration of J-PAS and stands out as a photometric survey of its own. J-PLUS is observing the northern sky in 12 filters (5 broad bands and 7 narrow bands) with an 83 cm telescope. S-PLUS, which is the subject of this paper, is the J-PLUS counterpart in the southern hemisphere, with many similarities to J-PLUS. It uses an identical 83 cm telescope and the same 12 filter system. S-PLUS plans to cover an area of $\sim9300$ deg$^2$ in the southern sky. It is important to mention that there is currently no counterpart for J-PAS under development in the southern hemisphere. Therefore, with its 12-filters system, S-PLUS will remain the highest resolution large photometric survey in this area in the near future. 

The extended resolution of these surveys, covering key spectral features, enables their application in many different fields. In particular, S-PLUS data has been used to study clusters of galaxies, considering accurate photometric redshifts in Stripe-82 using template fitting \citep{Molino+2020} and machine learning (Vinicius-Lima et al., submitted), ultra-diffuse galaxies \citep{Barbosa+2020}, lenticular galaxies \citep{Cortesi+21}, the Hydra cluster galaxies  \citep{Lima-Dias+2021}, conduct searches for quasars (Nakazono et al. submitted), determine galaxy morphology (Bom et al. submitted), perform star/galaxy separation \citep{Costa-Duarte+2019}, analyse stellar populations in and around the Milky Way, including stellar groups in the CMa OB1 association (Santos-Silva et al. submitted), determine and study the photometric metallicity and carbon distributions of stars in the Milky Way’s Halo (Whitten et al. accepted), find and characterize compact planetary nebulae \citep{GutierrezSoto+2020} and ultra metal-poor stars (Placco et al., in prep) and study active low-mass stars in CMa R1 star-forming region \citep{Gregorio-Hetem+2021}, as well as several ongoing projects. Given this wide range of applications, it is of utmost importance to provide precise and accurate photometry that is reliable for both point and extended sources.

The photometric calibration is the process of translating photon counts measured at the detector into physical fluxes at the top of the atmosphere or, equivalently, into magnitudes calibrated in relation to a given reference. The calibration is usually represented by the estimation of the zero-point (ZP), which is measured in units of magnitudes and corresponds to the quantity that needs to be added to the instrumental magnitudes to obtain the calibrated magnitudes. The ZP is affected not only by the sensitivity of the instruments at the time of the observation but also by the atmospheric conditions and the airmass in front of the target. This process is a part of every astronomical observation since the use of photometric plates. However, it still presents challenges that require the development of new techniques, especially regarding large surveys such as S-PLUS.

The most traditional calibration technique is the observation of spectro-photometric standard stars \citep{Gregg+2006, LeBorgne+2003} at different airmasses throughout the night. By convolving the stellar spectra with a filter system, it is possible to compare expected and observed magnitudes and use the difference between them to estimate the zero-points. In particular, this traditional approach is not suitable for observational programs that image large sky regions under many pass-bands (such as S-PLUS, J-PLUS or J-PAS). This technique requires multiple observations in each filter throughout the night. However, the time dedicated for calibration observations scales linearly with the number of filters being used, leaving less time to observe science targets. In S-PLUS, the observation of standard stars can require as much as 1h15m of observations for each night.

One way to avoid the need to observe standard stars is to use spectral surveys (e.g. SDSS) to convolve the spectra of stars that are already included in the observations. This is not a viable option for surveys in the southern hemisphere because there are not enough spectroscopic observations available to ensure that every field will have enough reference stars to be calibrated. Even for the northern hemisphere, the spectral coverage of these surveys is usually not enough for the calibration of the bluer and redder filters. Another possibility for the calibration is the direct comparison with another photometric survey with similar transmission curves (although colour-terms need to be included in the comparison to account for the small discrepancies between different filter systems). This technique is also not suitable for the new generation of multi-filter surveys when these are the first to observe a given region of the sky in a specific filter system, which is usually the case for the narrow bands.

To overcome these challenges, J-PLUS employs a technique known as stellar locus regression \citep{Covey+2007, High+2019, Kelly+2014}. This approach relies on the fact that stars with different stellar parameters populate a specific and well defined region (the stellar locus) in colour-colour space. Overall, the process consists of an iterative relative calibration between several J-PLUS observations achieved by matching the stellar locus in all 2145 possible colour-colour combinations, followed by an absolute calibration that relies on anchoring the $i$ band to the similar broad band photometry of either SDSS or PS1 \citep[][hereafter, LS+19]{LopesSanjuan+2019}. Recently, \citep{LopezSanjuan+2021} present a revised version of the J-PLUS calibration, taking into account the systematic impacts of metallicity.

Concurrently, we have independently developed a novel technique for the calibration of S-PLUS. Like LS+19, we also use \texttt{SExtractor} \citep{Bertin+Arnouts1996} to measure instrumental magnitudes and take advantage of other previously calibrated large-area surveys for the photometric calibration. Still, there are several differences in our implementation. Our technique consists of fitting a library of stellar models (which are a set of convolved synthetic spectral energy distributions) to previously calibrated reference catalogues (such as SDSS, Skymapper, PanSTARRS, etc.), and using the best-assigned models to predict the calibrated S-PLUS magnitudes for these reference stars. Instead of correcting the catalogues for interstellar medium (ISM) extinction, we leave it as another free parameter to be fit by the models. Therefore, we do not rely on photometric transformations and extinction maps, which normally contribute to increase the final uncertainties. We also avoid the use of stellar locus for filters like J0378 and J0395, which have a naturally large spread in colour-colour diagrams, making it difficult to properly characterise the stellar locus even if the photometric uncertainties are negligible. Another great advantage of our approach is that it allows each S-PLUS field to be individually calibrated.

Our calibration approach also allows us to find offsets that correlate to the source position in the CCD, an effect also noted by LS+19 for the J-PLUS observations. We discuss the correction of this effect in Section \ref{sec:CCD_correction}. Finally, we show that we can use a stellar locus calibration in combination with the model fitting technique when the wavelength range of the available reference catalogue is more limited than the one of S-PLUS -- which is the case for a few S-PLUS Main Survey observations and for a whole S-PLUS sub-survey (the `Shorts survey', which consists of the observation of the same footprint but with total exposure times of 1/36th of the Main Survey). We discuss this stellar locus calibration in Section \ref{sec:stellar_locus_calibration}.

The technique developed in this work was used to calibrate the S-PLUS Data Release 2 (DR2), which includes observations taken between 2016 and 2020. We use the DR2 calibration to validate our method both in terms of external checks against reference catalogues as well as by analysing the internal consistency of our data. Then we provide a full characterisation of the DR2 data, as well as a description of the complete data-flow from observations to the access of the final catalogues.

S-PLUS DR2 covers a few regions that may be of interest for particular studies. It includes the STRIPE82 region, which is fundamental in terms of calibration and comparison to other catalogues. We also provide S-PLUS observations in the Hydra cluster region, which can provide information about the morphology, environment, and characteristics of the galaxies within and nearby this cluster \citep{Lima-Dias+2021}. DR2 covers the full G12 region of the Galaxy and Mass Assembly survey \citep[GAMA,][]{Driver+2009}, which supplements S-PLUS with precise spectral information for galaxies. Finally, DR2 also includes multiple fields of galactic latitude between 30 and 60 degrees, which, in synergy with Gaia DR3 and other surveys like 2MASS, WISE \citep{Wright+2010} and GALEX \citep{Morrissey+2005} enables the study of the stellar properties of nearby populations and of the Galactic halo.  

In Section \ref{sec::SPLUS} we present an overview of the S-PLUS project, including a description of the data-flow from observations to final catalogues, where we also highlight the differences between S-PLUS DR2 and DR1. In Section \ref{sec::Calibration_description} we describe the novel calibration technique developed for S-PLUS and in Section \ref{sec::Calibration_validation} we validate this technique through the analysis of the STRIPE82 calibration. We characterise the whole DR2 data in Section \ref{sec::DR2}, and we describe the DR2 data in Section \ref{sec::Catalogs}. Finally, we summarise this work in Section \ref{sec::Summary}. 

Unless specified otherwise, all magnitudes here are presented in the AB system.

\section{The S-PLUS DR2}
\label{sec::SPLUS}

\subsection{S-PLUS Overview}

S-PLUS is a 12-band optical photometric survey of two non-contiguous fields, covering a total area of $\sim$8000 deg$^{2}$, at high Galactic latitudes ($b~>~30~$deg) using a dedicated 0.83m robotic telescope, the T80-South (T80S), located at Cerro Tololo, Chile. S-PLUS will cover an additional $\sim$1300 deg$^{2}$ of the Galactic plane and bulge to enable Galactic studies. In this work, we focus on the aspects that are of particular interest to the second data release of the S-PLUS main survey. Additional information about S-PLUS can be found in \citet{MendesDeOliveira+2019}.

\begin{figure}
\begin{center}
\includegraphics[width=.48\textwidth]{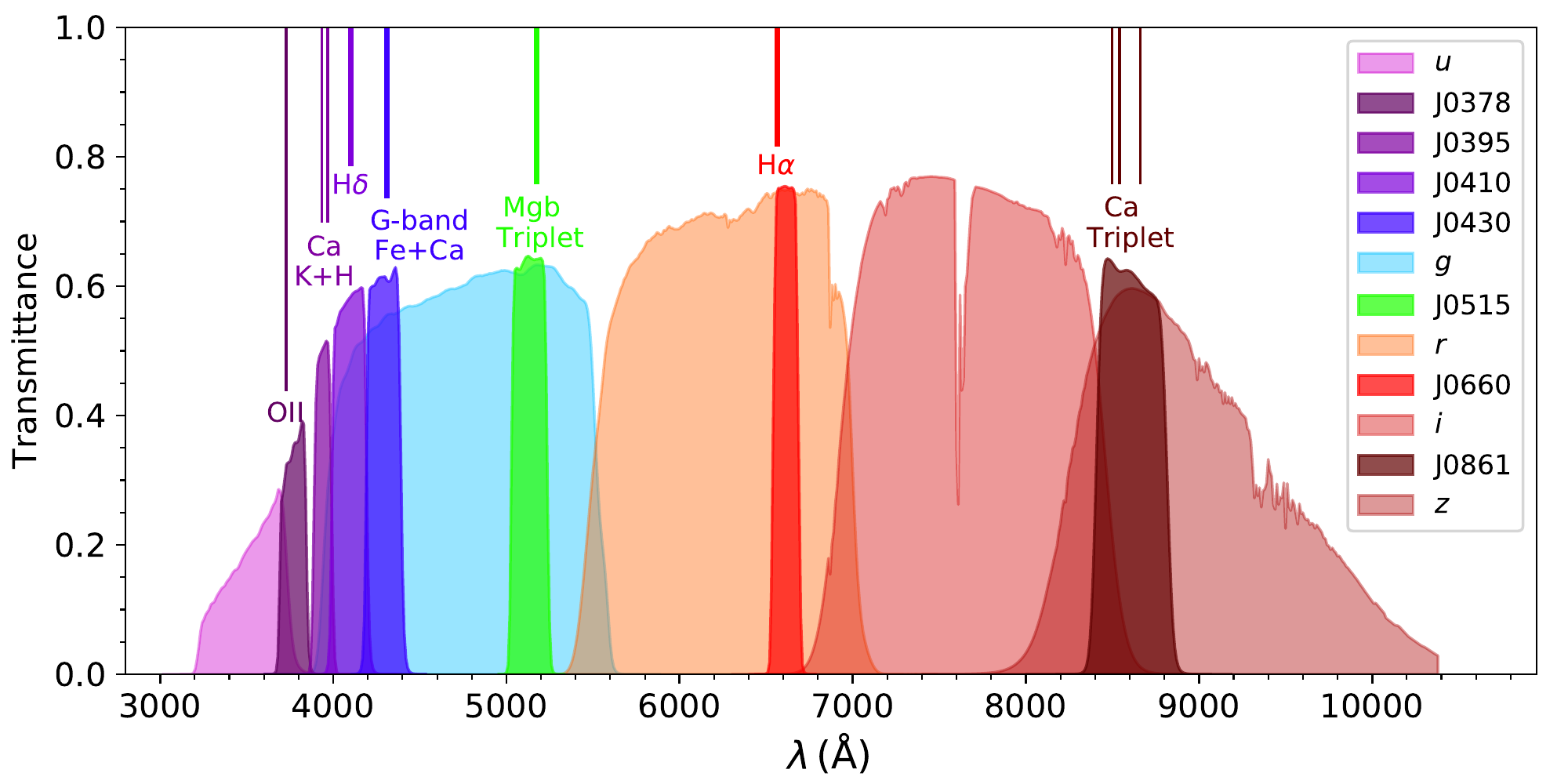}
\caption{\label{fig:SPLUS_filters}The S-PLUS filter system. The 7 narrow-band filters are centred on prominent spectral features (\ionnote{O}{ii}, J0378;  Ca H+K, J0395; H$\delta$, J0410; G-band, J0430; Mgb triplet, J0515; H$\alpha$, J0660 and Ca Triplet, J0861). }
\end{center}
\end{figure}

The telescope is equipped with a $9232 \times 9216$ 10 $\mu$m pixel e2v\footnote{\url{https://imaging.teledyne-e2v.com/}} detector with a plate scale of 0.55 arcsec pixel$^{-1}$ with an effective field-of-view (FoV) of 2 deg$^2$. S-PLUS uses the Javalambre filter system \citep{MarinFranch+2012}, shown in Figure \ref{fig:SPLUS_filters}, consisting of 5 SDSS-like broad-band filters ($u$, $g$, $r$, $i$, $z$) and 7 narrow-band filters centred on prominent spectral features (i.e., \ionnote{O}{ii}, Ca H+K, Dn4000, H$\delta$, Mg$b$, H$\alpha$ and CaT). S-PLUS is designed primarily for Milky Way and nearby Universe sciences like searches for low-metallicity stars, planetary nebula, star-forming regions, open and globular clusters, mapping the large-scale structure of the nearby universe using accurate photo-zs, studies of star formation in nearby galaxies, quasar searches and studies of transients and variable sources, to name a few.

The S-PLUS project will simultaneously conduct several sub-surveys, in most cases using different observational strategies aiming at tackling diverse scientific cases. In this work, we focus on the Main Survey (hereafter MS), and we refer the interested reader to \citet{MendesDeOliveira+2019} for an in-depth description of these observational programs. The MS is motivated to conduct various Galactic and extragalactic science projects. The footprint has been designed to have large overlapping areas with other existing or incoming deep extragalactic surveys such as DES \citep{Abbott+2018}, KiDS (de Jong et al. 2015), ATLAS \citep{Shanks+2015} or LSST \citep{Ivezic+2008, Ivezic+2019}. These common regions will serve as much for calibration purposes of the S-PLUS observations as to provide improved photometric redshifts for objects in these fields down to $i=21$ AB.

\subsection{DR2 Data Flow - From observations to final catalogues}
\label{sec:DR2_dataflow}

 Here we discuss the data flow of the second S-PLUS data release, DR2, detailing every step from observations to the final catalogues. Overall, the process is very similar to the one applied for the S-PLUS DR1, but as we continue to improve our pipelines, there are slight differences between the current and former steps. When this is the case, the differences are highlighted and discussed in detail.

\subsubsection{Observations}
\label{sec:observations}

\begin{figure}
\begin{center}
\includegraphics[width=0.47\textwidth]{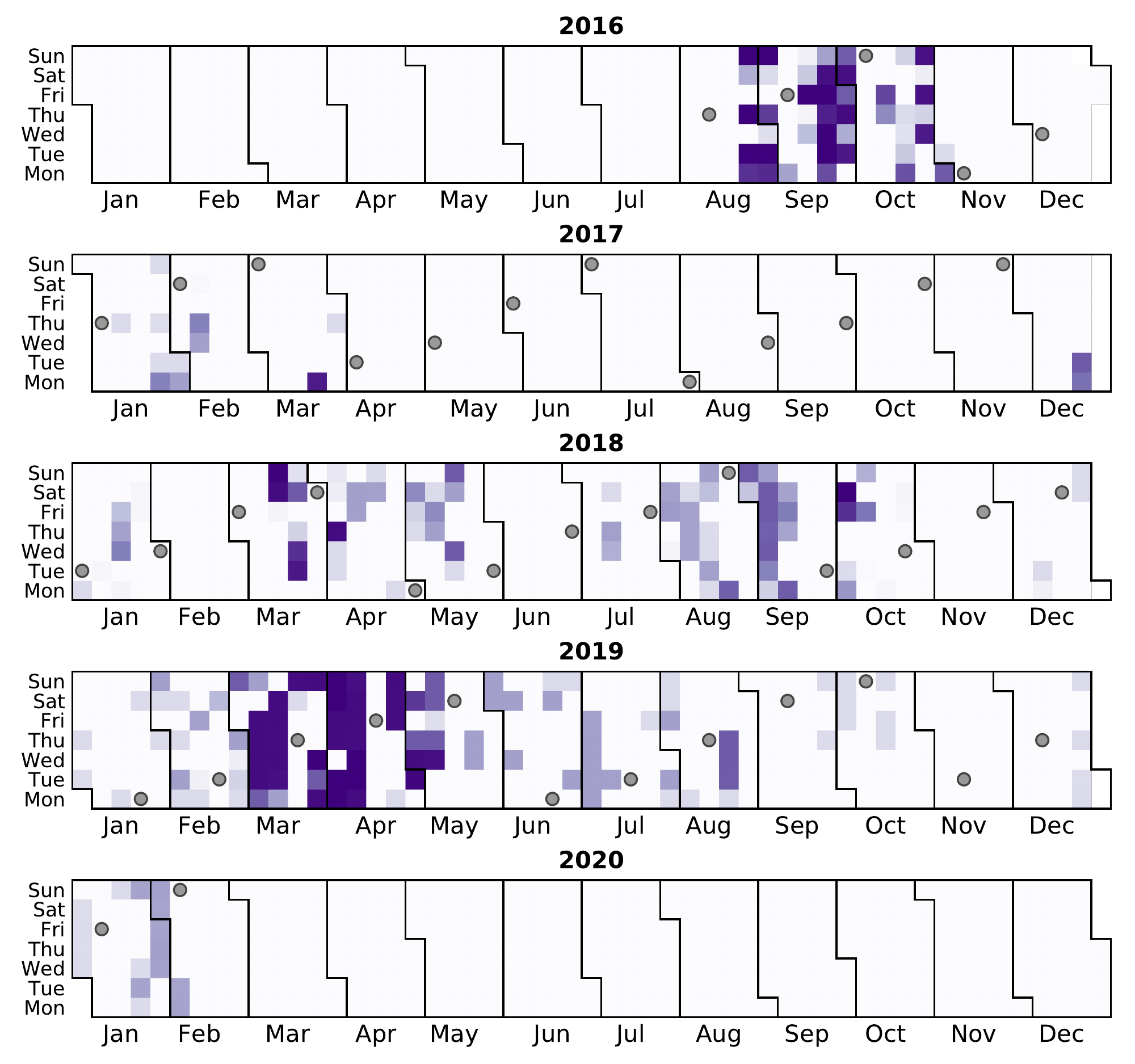}
\caption{\label{fig:DR2-ObsDate-heatmap}Heat-map calendar of the observations included in the S-PLUS DR2. The grey circles indicate the dates of full-moon. The lack of observations near the full moon is a design choice to have the depth as homogeneous as possible for the Main Survey.}
\end{center}
\end{figure}

The DR2 observations span a broad time window, from August 2016 to February 2020. During this time, the only significant interruption happened between April and November 2017, when a series of technical issues were identified and fixed during the science verification phase. In total, the DR2 contains observations taken during 273 nights, corresponding to a total exposure time of 630~hours (considering all 12 filters and excluding overheads). Figure \ref{fig:DR2-ObsDate-heatmap} shows the nights where at least one DR2 observation took place. The colour represents the number of observations in each night, where the dark purple dates are the ones with the highest number of observations.

\begin{figure}
\begin{center}
\includegraphics[width=.45\textwidth]{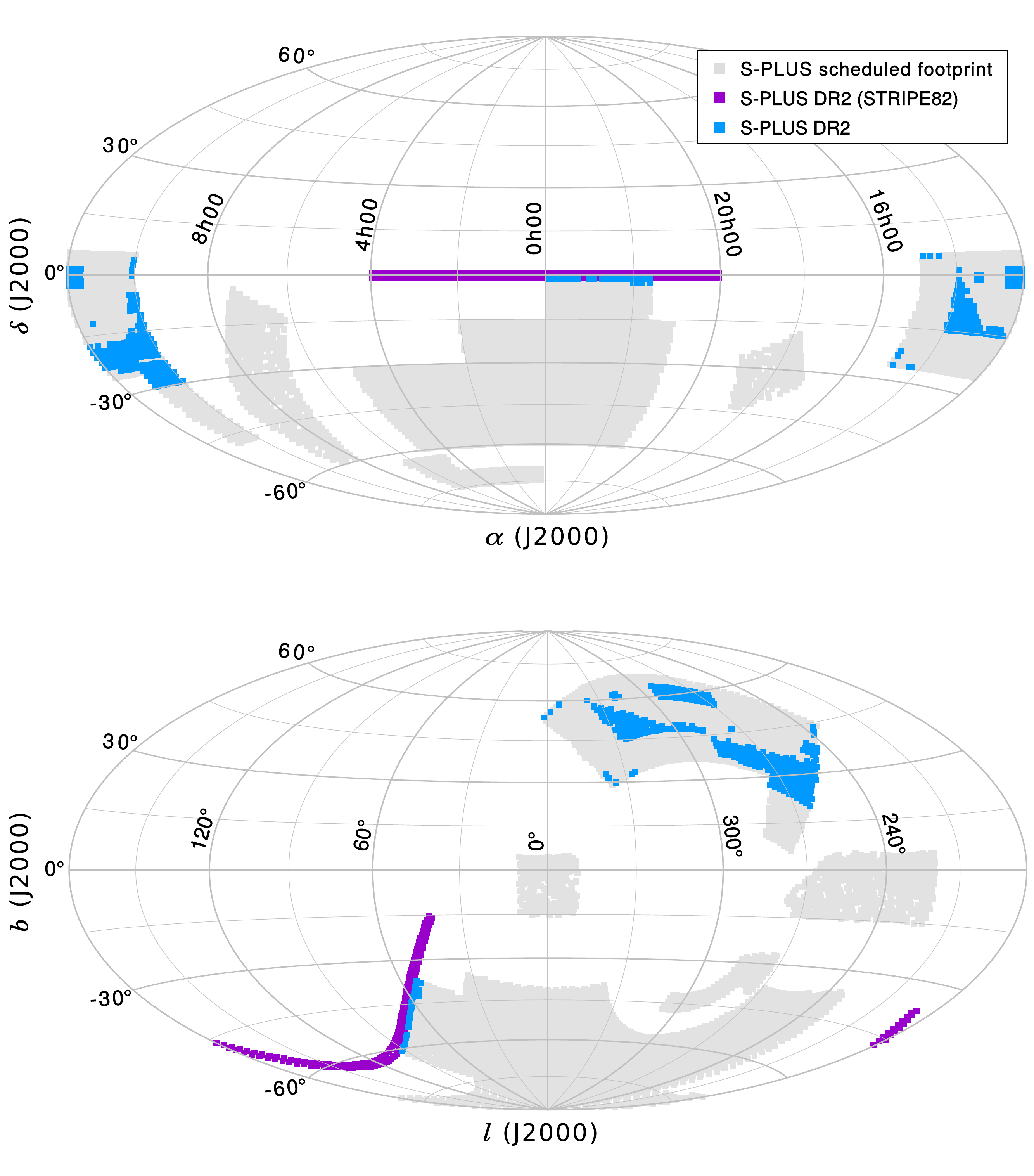}
\caption{\label{fig:DR2_footprint}Equatorial (upper panel) and galactic (lower panel) footprint of the S-PLUS DR2. The grey area represents the total S-PLUS planned footprint, while the blue and purple regions are the fields that are included in DR2. The DR2 area corresponds to $\sim$11.5\% of the total programmed footprint. The STRIPE82 region is highlighted in purple.}

\end{center}
\end{figure}

The DR2 footprint is shown in Figure \ref{fig:DR2_footprint}, both in equatorial and galactic coordinates. The S-PLUS DR2 includes the DR1 observations of STRIPE82 (but now with improved astrometry and data quality checks), as well as novel fields observed in the northern galactic hemisphere. In total, it comprises the observations of 514 MS fields, which account for an area of 950.5 deg$^2$. The purple region in Figure \ref{fig:DR2_footprint} corresponds to the 170 S-PLUS fields in the SDSS Stripe-82 region. This region, which we refer to as "STRIPE82", was observed multiple times by SDSS and many other surveys and is ideal for testing the S-PLUS calibration described in Section \ref{sec::Calibration_description}.

The T80S operation is automatically performed by the \texttt{chimera} Observatory Control System \citep{HenriqueSilva+2017}. Each night, observations are automatically planned to consider the date, the moon's position and brightness, and a set of predefined fields and sub-surveys priorities. During the observations, an automatic supervisor algorithm is present to take the weather conditions into account and execute the necessary changes to the observation schedule. Human intervention is also occasionally necessary to ensure the best usage of the telescope time. 

During standard conditions, the MS, which comprises the DR2 data, is the one that receives the highest priority. It is removed from the schedule during the nights close to the full moon ($\mathrm{moon\ brightness} > 84$ per cent), as well as when the seeing conditions are over $\sim2.0$ arcsec, ensuring that the data have the best quality T80S can provide.

\subsubsection{Image reduction and astrometry}
\label{sec:astrometry}

The reduction process of each individual image includes overscan subtraction, trimming, bias subtraction, master flat-field correction, cosmetic corrections (e.g. satellite and aeroplane trails, cosmic rays and bad pixels masking),  and fringing pattern subtraction (which is usually only necessary for the $z$ filter). This process is performed using version 0.9.9 of the data processing pipeline \texttt{jype}, which was designed for the data reduction of J-PLUS and J-PAS \citep{2014SPIE.9152E..0OC}. Since this process remained the same as for DR1 (only with refined astrometry and improved image quality control), we refer the reader to \citet{MendesDeOliveira+2019} for details.

\begin{figure}
\begin{center}
\includegraphics[width=.5\textwidth]{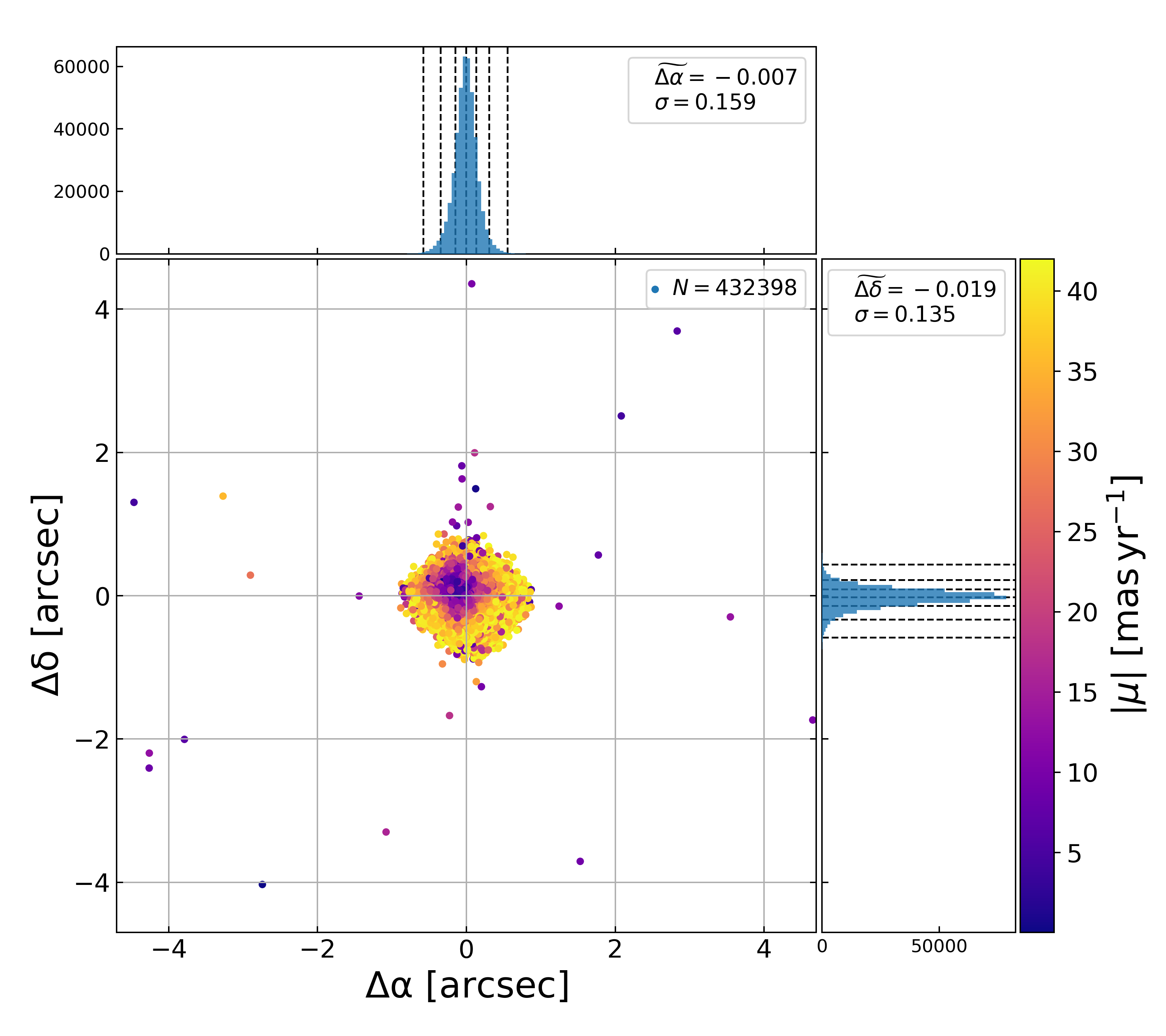}
\caption{Difference between the S-PLUS positions relative to Gaia's, $\Delta\alpha = \mathrm{RA(}$S-PLUS$) - \mathrm{RA(Gaia)} \times \cos \delta$ versus $\Delta\delta = \mathrm{DEC(}$S-PLUS$) - \mathrm{DEC(Gaia)}$, coloured by the absolute proper motion $|\mu|$. The histograms show the distribution for $\Delta\alpha$ and $\Delta\delta$. The dashed lines correspond to the percentiles 0.15, 2.5, 16, 50, 84, 97.5 and 99.85. The median $\widetilde{\Delta\alpha}$ and $\widetilde{\Delta\delta}$ (which correspond to the percentile 50) is listed in the legend of the histograms with the corresponding standard deviation ($\sigma$).}
\label{fig:astrometry}
\end{center}
\end{figure}

Astrometry is calculated by the software \texttt{SCAMP} \citep{SCAMP2006ASPC..351..112B} using the 2MASS catalogue as the reference with all positions being set for the epoch J2000. In order to quantify the S-PLUS astrometric precision relative to a more modern and precise survey, we did the comparison with Gaia's Data Release 2 positions. To calculate the relative spatial offsets, we selected all objects for $g$ band with good photometry (no deblending, no saturation, or any other possible source of errors originated by photometric issues; \texttt{SExtractor}'s $\texttt{PhotoFlag} = 0$, see Section \ref{sec::Catalogs}), as well as only stars via \texttt{SExtractor}'s $\texttt{CLASS\_STAR} > 0.95$. We also added a constraint to get only detections with $\mathrm{FWHM}(g) < 2.5$ arcsec and $13 < g < 18$ to select only well-sampled stars far from the detectability limits of S-PLUS. With this selection, we used the \texttt{Astropy} \texttt{SkyCoord.match\_to\_catalog\_3d} method \citep{2013A&A...558A..33A,2018AJ....156..123A} for the pair matching with Gaia's catalogue using a broad 5-arcsec radius to avoid introducing biases to the sample. After this process, we identified a considerable number of stars with high proper motion $\mu > 100~\mathrm{mas~yr^{-1}}$ for both right ascension and declination (RA and DEC). To account for these objects without actually correcting the positions for the proper motion, we calculated its absolute value defined as $|\mu| = |\mu_\alpha| + |\mu_\delta|$, which gives us the modular distance to the origin. In order to ``decontaminate'' our sample for high proper motion stars, we selected only those with $|\mu| < 41.988~\mathrm{mas~yr^{-1}}$ which corresponds to the 95th percentile of $|\mu|$. Figure \ref{fig:astrometry} shows the difference between the S-PLUS positions relative to Gaia's, $\Delta\alpha = (\mathrm{RA(S-PLUS) - RA(Gaia)}) \times \cos \delta$ versus $\Delta\delta = \mathrm{DEC(S-PLUS) - DEC(Gaia)}$, coloured by $|\mu|$ along with the histograms for $\Delta\alpha$ and $\Delta\delta$. The histograms also show the lines corresponding to the percentiles 0.15, 2.5, 16, 50, 84, 97.5 and 99.85. We obtained a median value of $\widetilde{\Delta\alpha} = -0.007$ arcsec and $\widetilde{\Delta\delta} = -0.019$ arcsec, both listed in the the histograms legend along with their corresponding standard deviation ($\sigma$). The mean values for the differences are $\overline{\Delta\alpha} = -0.010$ arcsec and $\overline{\Delta\delta} = -0.027$ arcsec.

In Figure \ref{fig:astrometry} we can see a few objects that have large discrepancies between S-PLUS and Gaia positions despite having a reasonably low proper motion. To understand the reason why these objects have such big differences in positions relative to Gaia's, we selected all those with $|\Delta\alpha| > 1$ or $|\Delta\delta| > 1$ arcsec, which corresponds to 28 out of 432\,398 (0.006 per cent). Analysing these objects individually, we found that 21 of them are located in the edges of their respective S-PLUS fields, where we know there are more artefacts and other sources of contamination (such as bad pixels, light contamination by bright stars outside of the field-of-view, etc.). The other 7 are located in central parts of the CCD and are mismatched with nearby stars, given that the magnitudes of the stars truly located in those coordinates are higher than 19 mag in the $g$ band (hence excluded by construction from our sample). Still, the number of outliers is negligible and does not affect the statistical relevance of the astrometric precision estimation.










As previously mentioned, for the MS, each field is observed three times for each filter, with a few exceptions when an image for a given filter had to be discarded due to artefacts or for a few fields that were observed multiple times. Therefore, the last step of the reduction pipeline is combining the multiple images for each filter, which is accomplished with the software \texttt{SWarp} \citep{SWarp2002ASPC..281..228B}. The final images are normalised to 1-second observations. All the aforementioned reduction steps are performed on-site, and only the final images are downloaded for the application of the next steps.

\subsubsection{Source Detection}

\begin{figure}
\begin{center}
\includegraphics[width=.47\textwidth]{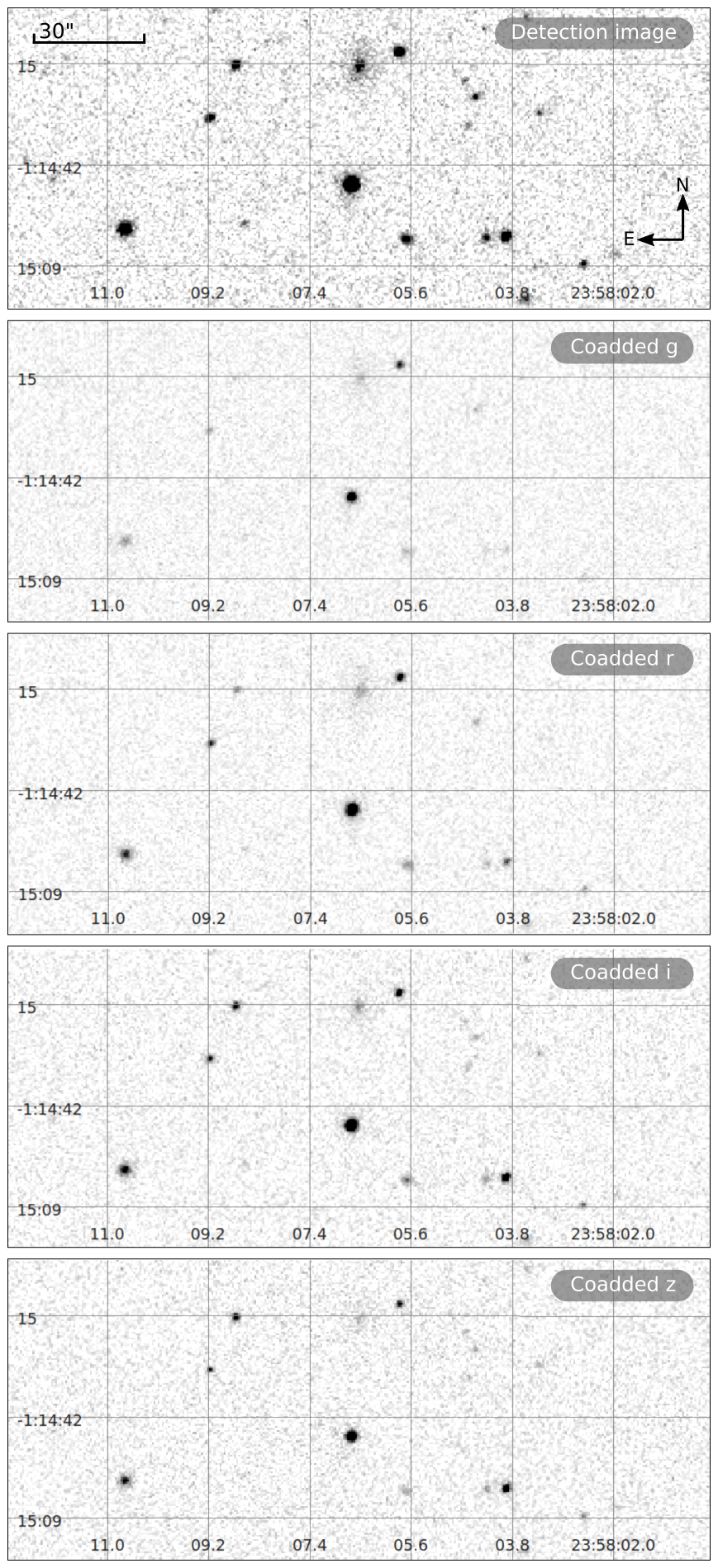}
\caption{\label{fig:detection_image}Cutout of a typically crowded region in the S-PLUS field STRIPE82-0001. The detection image, showed on top, is obtained from a weighted average of the also shown $g$, $r$, $i$ and $z$ images. The darkness represents counts in each pixel, which are displayed in the same range for all images. It can be seen how the detection image improves the S/N, allowing for the detection and proper characterisation of fainter sources.}
\end{center}
\end{figure}

From the source detection to the creation of the final catalogue, a new version of the S-PLUS pipeline is being used for DR2, although many algorithms remain exactly the same. This is the case for the source detection, which, just like for DR1 is performed by applying \texttt{SExtractor} to a combined reduced image consisting of the weighted-sum of the reddest broad bands ($griz$), which we denominate detection image\footnote{The only difference between DR1 and DR2 in this regard is that, both for the detection and individual images, due to schedule and technical issues, DR2 does not take into account the weight images produced by SWarp when measuring the photometry.}. In Figure \ref{fig:detection_image} we show an example of this process to highlight the benefits of using the detection image for source detection.

We note that the blue bands are not used in the construction of the detection image, creating a possible bias towards redder sources. This is a necessary trade-off because the bluer filters are much less transparent and hence have a poorer signal-to-noise ratio.

For DR2, in terms of detection threshold, we set a limit of 1.1~sigma in the detection image (as can be seen in the \texttt{SExtractor}'s input configuration file in Appendix \ref{ap:SExtractor}). This very conservative limit ensures that we identify sources as deep as possible, but it also means that some spurious detections may be included.

\subsubsection{Photometry and aperture definitions}

Forced aperture photometry is obtained from the individual images for each source's position and apertures derived from the detection image. We obtain the photometry for each individual image by running \texttt{SExtractor} in dual-mode, using the detection image to detect the sources and the individual science images for the measurements. An example of a \texttt{SExtractor} configuration file used for this step can be found in Appendix \ref{ap:SExtractor}. When the flux of a particular source found in the detection image is below the detection threshold in the individual image, we attribute the value 99 for the magnitude and substitute the magnitude error for a $2\sigma$ upper limit for the magnitude. 

We estimate the fluxes in several different apertures, as they are optimised for different applications: (i) The AUTO aperture is defined in terms of the Kron elliptical aperture \citep{Kron1980}, and is designed to integrate the total flux of extended sources, at the cost of a lower signal-to-noise ratio (S/N) . (ii) The Petrosian  magnitude, named PETRO, is similar to the former, but defined in terms of the Petrosian radius \citep{Petrosian1976}, and is best suitable for deriving the physical properties of extended sources. (iii) Fixed circular aperture photometry is obtained for 32 different diameters ranging from 1 to 50 arcsec around each detection. This set of apertures is used to estimate aperture corrections for the 3-arcsec aperture, which are important to correct the total magnitude of point sources used for the calibration. (iv) Finally, we also include Isophotal photometry, named ISO, which is derived from all the object pixels above the defined threshold and is the one that best preserves its shape. 

It is important to notice that the \texttt{SExtractor} parameters used to define the AUTO and PETRO apertures were modified for DR2 in relation to DR1. The changes were taken in order to better represent the total magnitude of the extended sources. The SExtractor parameters \texttt{PHOT\_AUTOPARAMS} were changed from 1.00, 1.00 to 3.00, 1.82 and \texttt{PHOTO\_PETROPARAMS} from 1.00, 3.00 to 2.00, 2.73. The changes result in a larger aperture and the minimum aperture in both cases to be set at a diameter of 3 arcsec.

\subsubsection{Zero-point variation across the field}

As will be further discussed in Section \ref{sec:CCD_correction}, we still find a residual systematic offset in the photometric zero-points (ZPs) that correlates with the position in the detector. These offsets do not appear to correlate with the filter, time of observation, or airmass. The range of the offsets is typically between $-20$ and $20$ mmags in the final calibrated catalogues. \citet{LopesSanjuan+2019} found similar $xy$ correlated offsets in the J-PLUS images, which are reduced by the same software, but in their case the offsets correlate differently for different filters.

A ZP variation across the field is a common issue for telescopes with a large FoV. Given its systematic nature, correction maps (see Figure \ref{fig:XY_offsets}) were derived and applied to the photometric catalogues before proceeding with the calibration. We note that this issue was not known at the time of DR1.

\subsubsection{Aperture correction}
\label{sec:aper_corr}

In what concerns aperture photometry, the fixed circular apertures are the ones that provide the best results for point sources. We point out that, due to the point spread function (PSF) of the observation, the fraction of the total flux measured depends on the chosen aperture diameter. A large enough diameter to contain the total flux of the source would also include too much background noise and is more likely to be contaminated by nearby sources. Therefore, the best approach is to use a smaller aperture diameter that maximises the  S/N and then correct for the amount of flux expected to be lost in this aperture.

We calculated that the best aperture for point sources in S-PLUS is the 3-arcsec diameter. This aperture has a sufficiently high S/N, while still being large enough to avoid undersampling. The aperture correction is the amount of magnitude that needs to be added to this 3-arcsec measurement for the measured instrumental magnitude to correspond to the total instrumental flux of the star\footnote{Even though we apply the aperture correction for all the detections, it only has a physical meaning for those that behave as point sources.}. The corrections are obtained for the coadded image of each filter while assuming that the PSF variation across the field can be neglected. 

\begin{figure}
\begin{center}
\includegraphics[width=.48\textwidth]{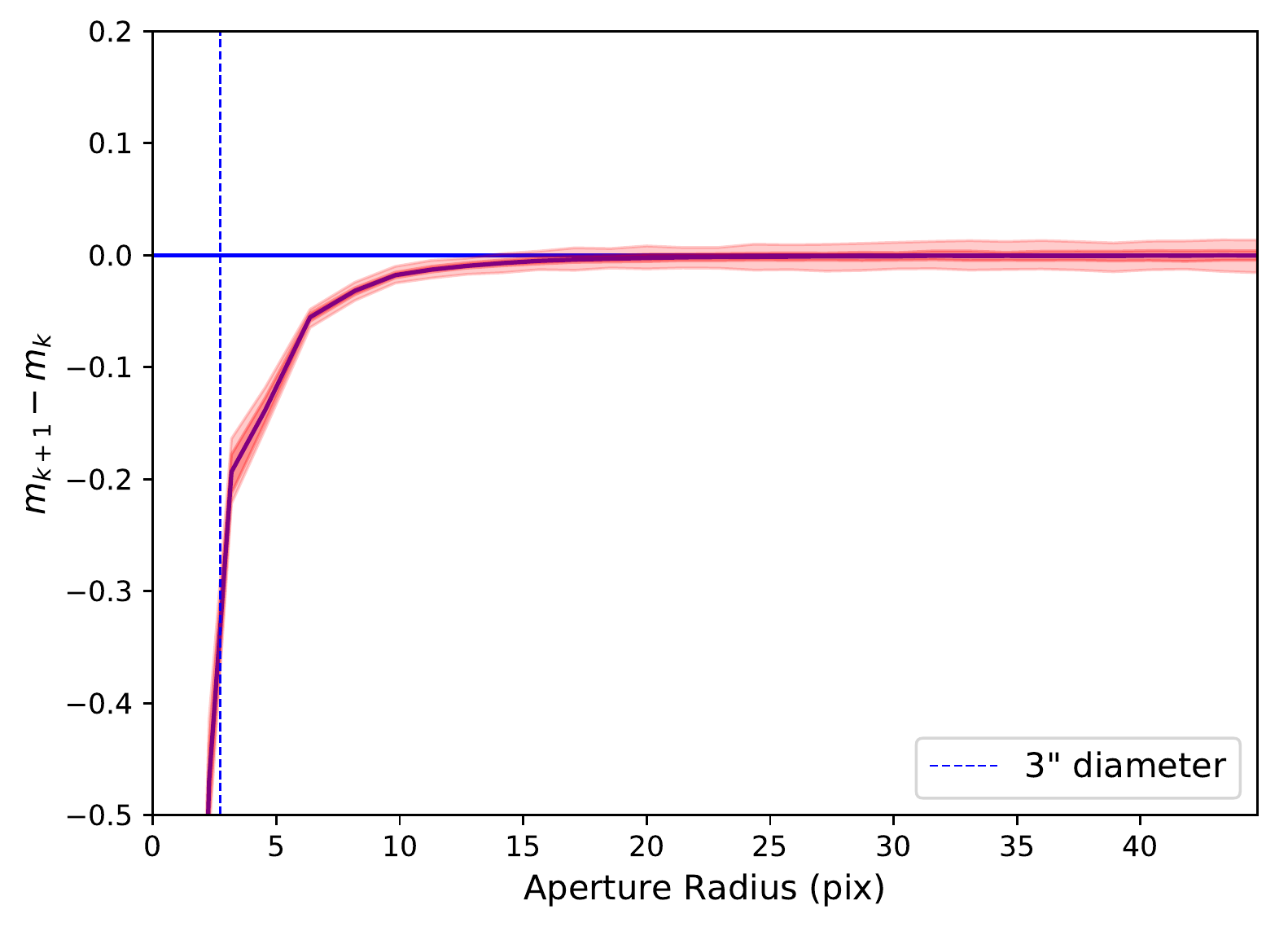}
\caption{\label{fig:example_growth_curve}Example of the Growth Curve obtained for the filter $r$ of field STRIPE82-0001. The average difference between consecutive apertures (purple line) is computed for 32 fixed apertures from 1 to 50 arcsec. The red shaded regions represent the 68 and 95 per cent confidence intervals. The pipeline checks for stability at the 40-arcsec apertures. The cumulative differences from the 3-arcsec diameter to the point of stability are added to the instrumental 3-arcsec magnitudes to correct for total magnitudes. Only stars with good photometry are used for the computation. In this case, 523 stars were selected.} 
\end{center}
\end{figure}

We measure the aperture corrections from the observation's growth curves, which represent the change in magnitudes as a function of aperture. We built this curve by measuring the instrumental magnitudes in 32 different fixed circular apertures, ranging from 1 to 50 arcsec in diameter. Each growth curve corresponds to the average magnitude difference obtained for a sample of a hundred stars in each observation that satisfies the following selection criteria: (i) \texttt{SExtractor}'s photometric flag equal to 0, to remove sources that might have contamination of nearby objects; (ii) \texttt{SExtractor}'s \texttt{CLASS\_STAR} parameter greater than 0.9 (to ensure the selected sources behave like point sources); and (iii) S/N between 30 and 1000 (to ensure the source is neither saturated nor too faint to be dominated by photometric errors). An example of the growth curve obtained for the filter $r$ of the field STRIPE82-0001 is given in Figure \ref{fig:example_growth_curve}.

Aperture correction was not applied in The DR1, where photometric ZPs were obtained in relation to the 3-arcsec diameter apertures. Considering the usual variation in the seeing between the observation of different filters, the aperture correction is filter dependent, affecting not only the calibrated magnitudes but also the colours. This is one of the reasons why we expect an improvement in the photometric ZPs precision for DR2.

\subsubsection{Reference catalogue used in the calibration}

The instrumental magnitudes are calibrated to the AB system using the photometric calibration technique described in Section \ref{sec::Calibration_description}. This technique takes advantage of many wide-area surveys that already have reliable photometric calibrations and overlap with the S-PLUS footprint. Using these catalogues as a reference for the S-PLUS calibration, we avoid the need to observe spectrophotometric standard stars.

The large FoV of S-PLUS ensures that there are always at least a few hundred stars (that satisfies the selection criteria for the calibration) in common with a combination of reference catalogues spanning the whole wavelength range of S-PLUS. The DR1 calibration used the Ivezi{\'c}'s SDSS \citep{Ivezic+2007} catalogue as a reference to calibrate the STRIPE82 region. In DR2, we keep the Ivezi{\'c} catalogue as the fundamental reference for the calibration, but due to the expanded footprint, we also make use of the ATLAS All-Sky Stellar Reference Catalogue (ATLAS Refcat2; \citealp{Tonry+2018}) and the GALEX, in particular DR6/7 \citep{Bianchi+2014}, to calibrate the areas other than the STRIPE82. We also take advantage of the highly uniform data provided by the Gaia DR2 \citep{GaiaCollaboration+2018} across the whole sky to ensure that all fields are scaled to the same magnitude.

\subsubsection{Fitting of stellar templates}

The translation between the reference catalogue filter system and that of S-PLUS is done by fitting spectral stellar templates to the reference catalogues and using the best fit model to predict the S-PLUS magnitudes for the star.

This technique follows the one applied for the S-PLUS DR1, yet with significant improvements for DR2. The most important changes are (i) the use of theoretical templates instead of empirical; (ii) the inclusion of the ISM extinction as a free parameter (absent from DR1); and (iii) the implementation of an optimised fitting algorithm directly included in the pipeline. In the case of DR1, the model templates were fit using the software LePhare \citep{Arnouts+1999, Ilbert+2006}, which involved the computation of several steps not relevant for calibration purposes and was dropped in favour of an internally implemented fitting algorithm.

For DR2, we make use of the stellar templates of \citet{Coelho14}, hereafter denominated as Coelho14, which provide theoretical data covering a wide range of stellar parameters. We also tested the ATLAS9 theoretical stellar templates of \citep[][hereafter C\&K03]{Castelli+Kurucz2003}, and the second version of the empirical Next Generation Spectral Library \citep[NGSL][]{Gregg+2006, Heap2007}. We present the results comparing the different libraries in Section \ref{sec:spectral_library_compaison}.

\subsubsection{Photometric calibration}

Overall, the calibration consists of three main steps: (i) the external calibration, in which the models are fitted to the reference catalogue magnitudes; (ii) the internal calibration, in which we fit the models to the S-PLUS externally calibrated magnitudes, refining the ZPs; and (iii) the Gaia scale calibration, which brings the S-PLUS magnitudes to the AB magnitude scale performed using Gaia as a reference (see Section \ref{sec::Calibration_description}).

The external and internal calibrations were also applied to the DR1 photometry, but the latter consisted of an iterative process that kept re-calibrating the magnitudes until convergence was achieved. Using simulated data, we found that this iterative process could lead to the convergence towards systematic offsets and that a single application of the internal calibration step is enough to improve the external calibration. Therefore, we use a single internal calibration step in DR2. The Gaia scale calibration is a new addition to DR2 and corresponds to an absolute shift in the magnitudes that does not affect the colours.

\begin{figure*}
\begin{center}
\includegraphics[width=\textwidth]{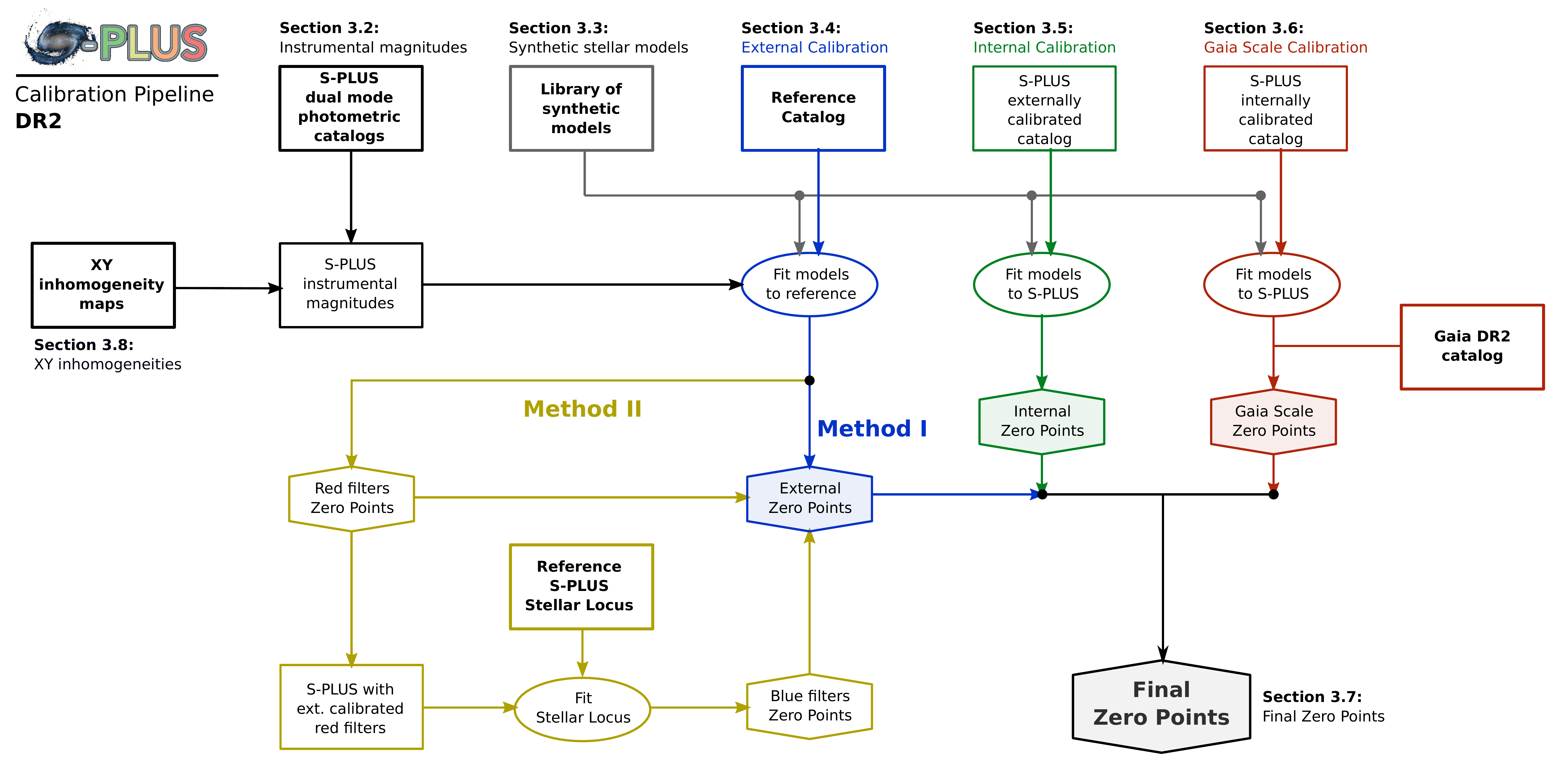}
\caption{\label{fig:diagram}Diagram detailing the steps of the S-PLUS calibration pipeline used for the DR2 and indicating the Section of the paper that covers each step. Both Methods I and II are represented by the diagram and only differ by a few additional steps for Method II during the external calibration. Boxes with text in boldface represent the inputs of the pipeline. The represented end-product in this diagram is the Final Zero-Points, which are subsequently used to generate the final calibrated catalogues.}
\end{center}
\end{figure*}

\subsubsection{Master Catalogue}

The final calibrated master catalogue includes data that is concatenated from the \texttt{SExtractor} detection catalogue, as well as each filter photometric catalogue measured in the dual-mode photometry. It includes the identifiers, position and morphological properties for each source in the detection image, as well as magnitudes, errors and S/N measured in several different apertures for each filter. The apertures include the previously described auto, petro and iso measurements and the fixed circular apertures for 3 and 6 arcsec diameters. The 3-arcsec aperture corrected magnitudes are included as "PStotal", as they represent the total magnitudes of the point sources. The photometric ZPs, obtained in relation to the PStotal magnitudes, are added to all apertures to provide magnitudes calibrated to the AB system.

The measured FWHM and the \texttt{SExtractor} photometric flags and estimated \texttt{CLASS\_STAR} parameter are included for all filters and for the detection image. The positions (both in the sky and in the CCD) and all morphological parameters (e.g. elongation, ellipticity,  Kron radius, Petrosian radius, etc.) are given only for the detection image.

The fields are individually calibrated and separately added to the database. Therefore, we warn the users that multiple detections are to be expected for sources located in regions where fields overlap. Only the current data release photometry identifiers are included in the main catalogues. These identifiers are not related in any way to those included in DR1 and, since each field is calibrated separately, duplicated sources in DR2 will have different photometry IDs. For those interested in combining DR1 and DR2 data, we suggest a sky coordinate crossmatch considering a maximum radius of 1 arcsec. In addition to the photometric IDs of the main catalogues, unique individual IDs that will carry over through future data releases are provided as a value-added catalogue.

\subsubsection{Value Added Catalogues}

In addition to the data mentioned above, we also provide value-added catalogues (VACs) that come from subsequent analyses of the generated master catalogue. At the time of DR2, the included VACs are (i) star/galaxy/quasar classification, (ii) photometric redshifts, (iii) flags to allow masking the detections around very bright stars, and (iv) individual object identifier to flag and simplify the removal of duplicated entries and enable direct comparison between different data releases. These VACs are individually discussed in Section \ref{sec::Catalogs}.

The possibility of further including VACs provides flexibility that was not present in DR1. Nevertheless, DR1 also included \mbox{star/galaxy} classification and photo-zs directly in the master catalogues. It is also important to emphasise that the techniques employed to derive both the star/galaxy classification and the photo-zs for the DR2 VACs are not the same as the ones applied for DR1.

\subsubsection{Database}

The new S-PLUS database is a server built on multiple programming languages, from \texttt{Python}, \texttt{Java}, \texttt{Javascript} to \texttt{C/C++}. The database allows the user to download catalogue data, obtain fits images of specific areas for a given position, make colour images, and have access to a \texttt{Python} package that may be useful to integrate database queries into a research workflow.

As for the fits images or png cutouts, there is an API (Application Programming Interface) that allows users to make requests that the server processes in real-time. The advantage of an API is that it enables interoperability between the data in the database and other applications, as it works through simple network requests. 

The S-PLUS database follows multiple International Virtual Observatory Alliance (IVOA) standards on how to connect to the server and how to visualise or download the desired data. It is based on a Table Access Protocol (TAP) service that is capable of handling astronomical data query language (ADQL) to interact with all the catalogues, allowing access using known programs in Astronomy such as \texttt{TOPCAT} \citep[Tool for Operations on Catalogues And Tables;][]{2005ASPC..347...29T} or astroquery (\texttt{Python}).

\section{The novel model fitting calibration pipeline}
\label{sec::Calibration_description}

In this section, we describe in detail the methodology developed to perform the photometric calibration of S-PLUS and provide a few examples when necessary. The validation of the calibration, an estimation of its precision and accuracy, and more details about DR2 data are provided in the next sections.

\subsection{The calibration workflow overview}

The full calibration process for any given reduced image is summarised in Figure \ref{fig:diagram}, including three main steps: (i) the external calibration, where we perform the \ac{SED} fitting of stars in an external reference catalogue to determine a model for their spectra and the expected S-PLUS magnitudes; (ii) the internal calibration, where we do the SED fitting to the previously calibrated S-PLUS magnitudes to take advantage of the information in the 12 filters to better constrain the stellar models and refine the ZPs; finally, (iii) the Gaia scale calibration, which consists in finding the necessary flux offset to bring the S-PLUS magnitudes to the AB reference set with Gaia standard stars.

The methodology relies on catalogues, or combinations of catalogues, of other wide-field surveys, which already provide accurate photometry for millions of stars, including SDSS, Skymapper, PanSTARRS, Gaia, and DES. Naturally, it is desirable that the photometric bands of the reference catalogue cover the entire spectral range of the S-PLUS filters, such that the expected stellar magnitudes for S-PLUS can be well constrained (but that is limited by the footprint overlap between S-PLUS and the reference surveys). Whenever that is the case, we may simply use the reference catalogue to perform the photometric calibration, which we denominate as ``Method I''. However, if that is not the case, we can overcome this issue once we have enough S-PLUS stars already calibrated and use them to calculate the expected stellar locus in S-PLUS colour-colour diagrams for calibration of stars for which the reference catalogues do not have all information needed, which we denominate ``Method II''.

Our technique is designed to avoid the use of complex transformation equations between different photometric systems, which can be an additional source of uncertainties and are sometimes inaccurate. In particular, the S-PLUS narrow bands are designed to measure specific stellar spectral features. They are, therefore sensible to stellar parameters that the reference broad-bands are not usually able to measure. This is the case, for example, for the J0515 filter, which is very sensitive to $\log g$, while the SDSS-like $ugriz$ filters are not. In cases like this, the transformation equations would not be able to provide reliable S-PLUS narrow-band magnitudes from the comparison with the reference catalogues. By fitting stellar templates to the reference stars, we can treat them as if they were spectrophotometric standard stars and avoid the transformation equations. We also avoid the need to use extinction maps to correct the data from ISM extinction by leaving it as a free parameter of the models.

As mentioned before, another great advantage of this methodology is that it optimises the use of the telescope for science observations. In particular, most traditional calibration strategies require the observation of standard spectrophotometric stars under good photometric conditions at different airmasses throughout the night, so the positional, temporal and wavelength extinction dependency can be predicted and corrected. For large observational programs aiming at imaging large sky-regions under many passbands (such as S-PLUS, J-PLUS or J-PAS), the repetitive observation of these “standard fields” represents a multiplicative factor to the total observational time requested for the project to complete the observations. In the case of S-PLUS, the observation of standard stars in all 12 filters requires more than an hour every night. Therefore, following the development of this pipeline, we can now allocate this time to further observation of scientific targets, increasing the time-efficiency of the telescope by 15-20 per cent during the observations of the main S-PLUS programs.

All the reference catalogues used for the DR2 calibration provide AB magnitudes in 4 to 6 different passbands. It can be argued that this number of filters is not enough to reliably fit the spectrum of a single star. However, the power of this technique comes from the number of reference stars in each S-PLUS pointing, which ranges from 500 to 10000 stars, depending on the density of the field and the coverage of the reference catalogue. Therefore, even if the fitted spectrum is not perfect for a given star, we are still using hundreds (or thousands) of them to estimate a single ZP, minimising the random biases from the fitting process. 

\subsection{Instrumental Magnitudes}

The calibration pipeline begins after the image reduction, and source detection, photometry and aperture corrections are carried out as described in Section \ref{sec:DR2_dataflow}. The aperture correction ($AC$) is calculated for each filter from the estimated growth curve. It is defined as the total flux converted to magnitudes that need to be added to the 3-arcsec aperture measurements in the way that the resulting magnitude represents the total flux of the source (which is assumed to be point-like). Aperture corrections are individually calculated for each S-PLUS field. In terms of the CCD ADU counts, for a given filter $\mathcal{m}$\footnote{Hereafter we use the notation $\mathcal{m}$ to represent the magnitude in any given S-PLUS filter or of the considered reference catalogues. When referring to a particular filter, we simplify the notation by using the filter name to represent the magnitude in this filter (e.g. $F378_\mathrm{inst}$ represents the instrumental magnitude of filter J0378).}, over the 3-arcsec circular diameter ($\left. \mathrm{ADU}_\mathcal{m}\right|_{3"}$), the instrumental magnitudes ($ \mathcal{m}_{\mathrm{instr}}$) are given by:
%
\begin{equation}
    \mathcal{m}_{\mathrm{instr}} = -2.5 \log_{10}(\left. \mathrm{ADU}_\mathcal{m} \right|_{\mathrm{3"}}) + 20 + AC_\mathcal{m} + \xi_\mathcal{m}(X,Y)\,\mathrm{,}
\label{eq:instrumental_mags}
\end{equation}
where the number 20 is the initial guess for the ZP, and $\xi_m$ represents the offset correction that is obtained from the inhomogeneities maps (Section \ref{sec:CCD_correction}).

For each S-PLUS field, we obtain the instrumental magnitudes for the 12 S-PLUS filters and compile a master instrumental catalogue, which also includes the positions from the detection image. We use the \texttt{Java} package STILTS \citep{Taylor2006} to crossmatch this compiled catalogue to the chosen external catalogue that will be used as a reference for the calibration using a maximum radius of 1 arcsec.

By the end of this step, we produced a catalogue with S-PLUS instrumental magnitudes (corrected for the total flux of the source) and the reference catalogue AB magnitudes, which are assumed to be calibrated. 

\subsection{Stellar Models}
\label{sec:models}

The use of synthetic models for the spectral libraries also provides a few advantages. It ensures that the S-PLUS magnitudes are properly calibrated to the AB system so that the relative differences between filters perfectly represent the physical discrepancies in flux emissions at each passband. Synthetic models are also initially free from ISM dust extinction, which allows us to simulate its effects for different dust laws without including additional correction-related uncertainties. However, it is important to point out that any systematic errors in the synthetic models will also be present in the calibrated magnitudes. In Section \ref{sec:spectral_library_compaison} we compare the calibration results using different spectral libraries to quantify the influence of the chosen spectral library on both systematic and random ZP deviations.

\begin{figure}
\begin{center}
\includegraphics[width=.48\textwidth]{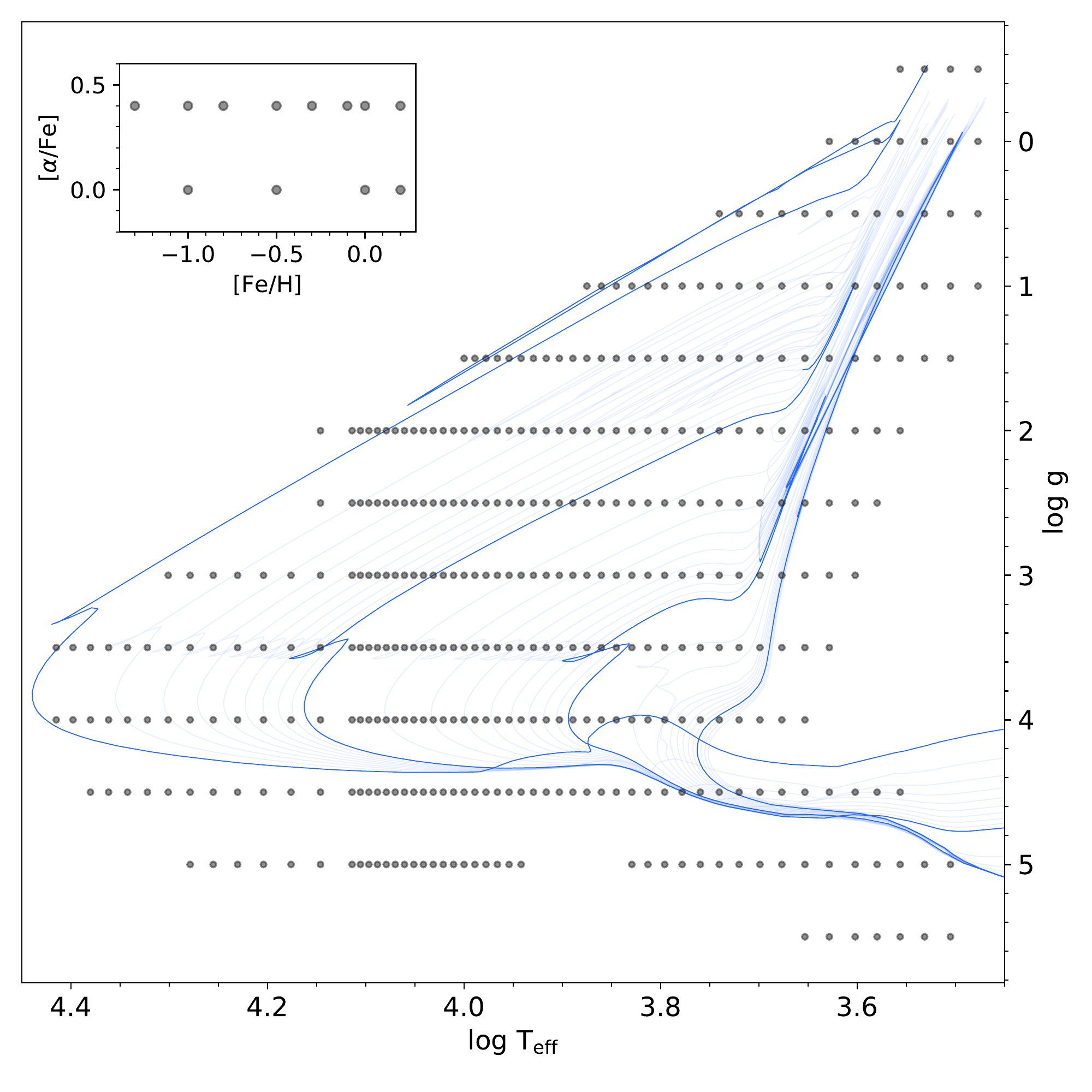}
\caption{\label{fig:modelsHR}The grid of \citep{Coelho14} theoretical spectral models in the space parameter of $T_\mathrm{eff}$ and $\log g$, and also [Fe/H] and [$\alpha$/Fe] (inner panel). A set of texttt{PPARSEC} isochrones \citep{Bressan+2012, Marigo+2017} with solar metallicity is shown in blue. The highlighted isochrones correspond to the ages of 0.01, 0.1, 1, and 10 Gyr.}
\end{center}
\end{figure}

For the DR2 calibration, we have chosen to work with the spectral library of \citetalias{Coelho14} because it provides a complete grid of synthetic models over a large range of stellar parameters and chemical abundances, which is shown in Figure \ref{fig:modelsHR} on top of \texttt{PARSEC} isochrones \citep{Bressan+2012, Marigo+2017}. The main panel corresponds to the $T_\mathrm{eff}$ versus $\log g$ space, while the inner panel indicates the metallicities and $\alpha$-enhancements also covered by the models. The highlighted isochrones correspond to ages of 0.01, 0.1, 1, 10 Gyr from left to right. As we can see, except for a few pre-main-sequence stages (that are not relevant for calibration purposes), the synthetic library covers the whole parameter space for stars of all different ages and evolutionary phases. 

For each spectral model, we also simulate ISM extinction for 20 values of $E_\mathrm{B-V}$ in a variable spaced grid between 0.025 and 1 mag. We employed the extinction curve of \citet{Cardelli+1989} and adopted $R_V = 3.1$. When fitting the models to the magnitudes, the extinction is treated as any other free parameter. In the case of \citetalias{Coelho14}, there is a total of 3727 unique spectra before simulating the ISM extinction. Adding the 20 spectra with simulated extinction per each model brings the total size of the library to 78267 spectra.

After simulating the ISM extinction, we pre-computed the convolved magnitudes for each spectrum in several different photometric systems (see Figure \ref{fig:filter_system}), obtaining a library of synthetic stellar photometry. We find the best match between the entries in the photometric library and a given observation by minimising the $\chi^2$ taking into account the magnitudes in all observed filters and assuming Gaussian errors for the measurements. Since the photometric library contains theoretical magnitudes in several different photometric systems, this best match allows us to convert from one photometric system to another. We refer to this process of finding the best match between the observation (of either S-PLUS or the reference catalogue) and the theoretical photometry as fitting the models to the observed magnitudes.

\begin{figure*}
\begin{center}
\includegraphics[width=\textwidth]{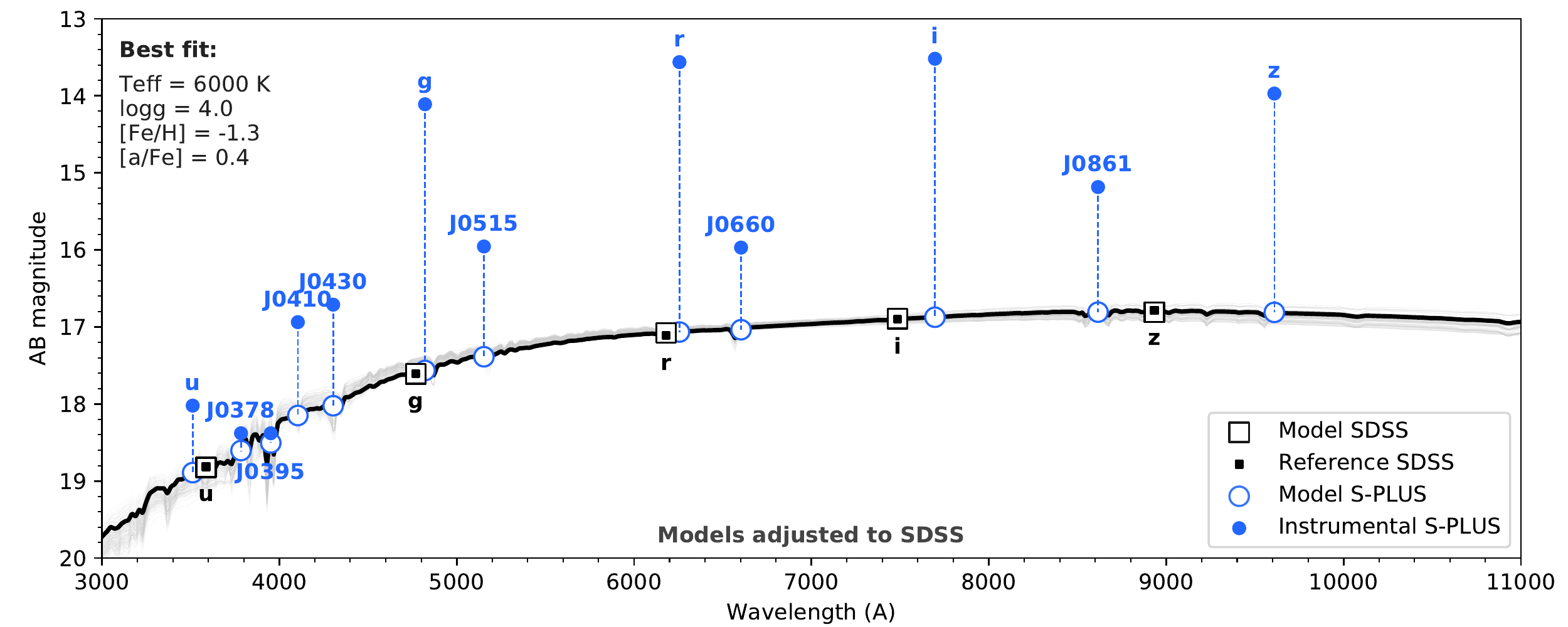}
\caption{An example of the external calibration model fitting process for star SDSS J000027.92-004122.4. The reference SDSS magnitudes \citep{Ivezic+2007} of this star is shown as filled black squares. The black line represents the spectral model that is found as the best fit for these magnitudes and which spectral parameters are shown in the upper left corner. The subsequent 200 best fits are also shown as grey lines for comparison. The large black open squares represent the SDSS magnitudes convolved from the best-fit model. The blue filled circles represent the instrumental S-PLUS magnitudes for this star that we seek to calibrate, while the blue open circles correspond to the S-PLUS magnitudes convolved from the best-fit spectra. The dashed lines represent the difference between the instrumental and the model predicted S-PLUS magnitudes. Together with thousands of other stars, they are used to characterise the calibration ZPs for each filter. }
\label{fig::model_fit_sdss_example}
\end{center}
\end{figure*}

\begin{figure}
\begin{center}
\includegraphics[width=.48\textwidth]{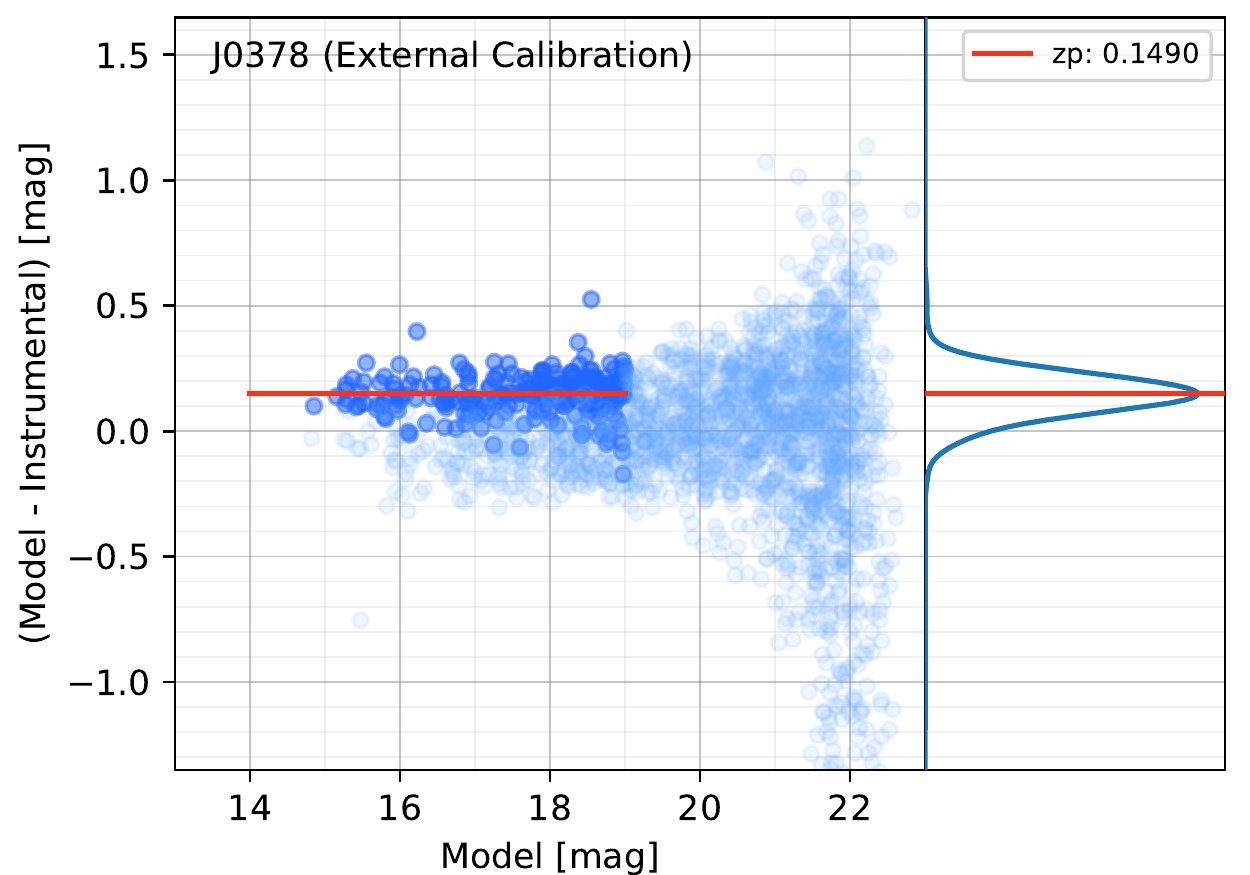}
\caption{The external ZP characterisation for filter J0378 in the field STRIPE82-0001. The x-axis shows the model J0378 magnitudes, while the y-axis shows the difference between the model and instrumental magnitudes for this filter. Only the stars with magnitudes between 14 and 19, and whose model $\log g$ is higher than 3, are taken into account and are highlighted. The red line represents the calculated ZP, which corresponds to the mode of the density distribution that is obtained from a kernel density estimation of the differences between model and instrumental magnitudes.}
\label{fig::external_zp_fitting_example}
\end{center}
\end{figure}

\subsection{External Calibration}

The external calibration step begins with the fitting of the photometric library to the reference catalogue magnitudes $\mathcal{m}_\mathrm{ref}$, along with its reported uncertainty $\delta\mathcal{m}_\mathrm{ref}$. Considering the set of reference passbands $\mathscr{R}$, the $\chi^2$ for the i'th model is given by Equation~\ref{Eq::chi2_external}\footnote{We adopted symmetrical magnitude errors for the calculation of the $\chi^2$. We note that our results in Section \ref{sec::Calibration_validation} confirm that this simplification is valid for calibration purposes.}:

\begin{equation}
 \chi^2_{\mathrm{ext},i} = \sum_{\mathcal{m} \in \mathscr{R}} \left( \frac{ \mathcal{m}_{\mathrm{mod},i} - \mathcal{m}_{\mathrm{ref}} } {\delta \mathcal{m}_\mathrm{ref}} \right)^2 \,.
 \label{Eq::chi2_external}
\end{equation}

The fitted model, which corresponds to that with the minimum $\chi^2$ for a given observation, provides the \textit{model predicted magnitudes} $\mathcal{m}_\mathrm{mod}$ for both the reference magnitudes (used to compute the $\chi^2$) and also for the S-PLUS filter system magnitudes (as well as for any other filter system present in the synthetic photometry library). Figure \ref{fig::model_fit_sdss_example} shows an example of the external model fitting process for star SDSS J000027.92-004122.4. In this particular case, the reference magnitudes (solid black squares) come from the Ivezi{\'c} catalogue: $\mathscr{R} = \{u_\mathrm{SDSS}, g_\mathrm{SDSS}, r_\mathrm{SDSS}, i_\mathrm{SDSS}, z_\mathrm{SDSS}  \}$. The model magnitudes of the best fit are represented by the big open black squares. The solid black line corresponds to the spectrum of the model that resulted in the convolved model magnitudes of the best fit. For comparison, the next 200 best models (out of 78267 in the whole photometric library) are also represented as shaded grey lines. The S-PLUS predicted magnitudes for this fit are shown as open blue circles. These magnitudes are compared to the S-PLUS instrumental magnitudes (solid blue circles). The difference seen between these two S-PLUS magnitudes can be attributed to the field's photometric calibration ZPs and, to a lesser extent, also to noise, arising mostly from photometric errors and model fitting inaccuracies.

\begin{figure}
\begin{center}
\includegraphics[width=.48\textwidth]{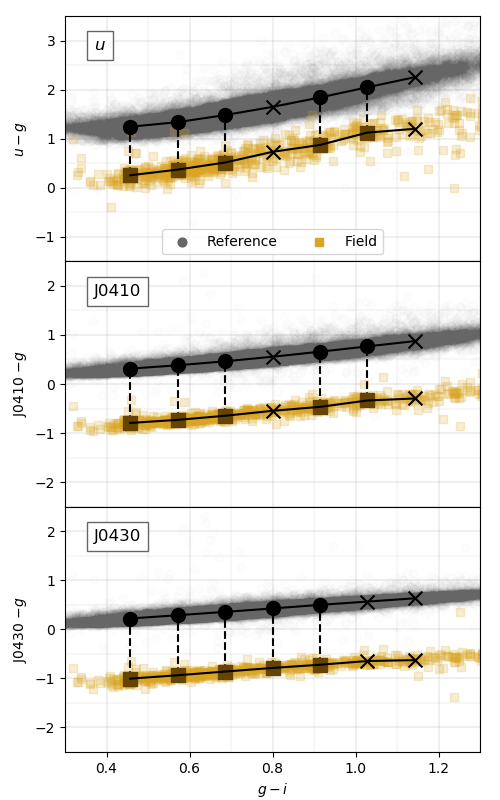}
\caption{\label{fig:stellar_locus_ex} Application of the stellar locus calibration step to the field STRIPE82-0001 for filters $u$, J0410, J0430. The grey circles correspond to the fundamental reference calibration of S-PLUS and act as a reference for the stellar locus calibration. The golden squares are the measured magnitudes for the field that is being calibrated. For the golden squares, filters $g$ and $i$ have been previously calibrated using the external calibration step, while $u$, J0410 and J0430 correspond to the instrumental magnitudes. The data was divided into 7 bins of $g-i$ between 0.4 and 1.2 mag. The average $X - g$ (where $X$ can be $u$, J0410 and J0430) was calculated for each bin and is represented by either circles, squares or crosses. The crosses indicate the bins with the extreme differences between the reference and the field data and are removed from the analysis. The zero-points are estimated from the differences between the averages of the reference and the field (dashed lines).}
\end{center}
\end{figure}

In order to minimise the contribution of the errors, the ZPs are estimated taking into account the instrumental and model predicted magnitudes of hundreds or even thousands of stars in each field. We apply three additional cuts in the previously selected reference stars: (i) a magnitude upper limit to remove the faint stars whose photometric uncertainties are too high (for DR2, the cutting magnitude is 19 mag in all S-PLUS filters); (ii) a magnitude lower limit of 14 mag, to remove possibly saturated stars; and (iii) a $\log g$ cut (using the parameters obtained from the best fit) to remove possibly misclassified giant stars. This last selection cut is necessary because some of the S-PLUS narrow bands are sensitive to $\log g$, while the reference magnitudes used to fit the models usually are not. Given that dwarf stars outnumber giants by a factor of 10 to 100, we expect to have much more dwarf stars mistakenly fitted with a giant star model than the opposite. Also, model atmospheres computed under the assumption of plane-parallel geometry, such as in Coelho14, are known to reproduce the colours of dwarfs better than those of giants \citep{Martins+2007}. Therefore, in the external calibration, we remove from the ZP fitting all the stars whose best fitting model has $\log g < 3$.

After the stars have been selected, the distribution of differences between the predicted and the instrumental S-PLUS magnitudes is used to estimate the ZPs for each respective filter. To avoid the influence of outliers, we use a kernel density estimation to obtain the density profile and characterise the ZPs from the mode of the distribution. We use a Gaussian kernel with a bandwidth of 0.05 mag, which was found to produce the best results for simulated data (see Appendix \ref{ap:zp_characterization}). Figure \ref{fig::external_zp_fitting_example} shows an example of this process for the ZP estimation of the filter J0378 of the S-PLUS field STRIPE82-0001. The stars selected for the characterisation of the ZP are shown as solid blue circles, while the shaded blue circles represent the total crossmatch between S-PLUS and SDSS. We chose to use this filter for the example since it is one of the most challenging to calibrate and highlights the importance of applying the stellar classification cut. As we can see, there is a large number of non-selected stars with magnitudes between 14 and 19 mag deviating from the estimated ZP, represented by the red line. This is because J0378 is highly sensitive to $\log g$, and these are mostly dwarf stars that were misattributed to a giant star template during the model fitting (since SDSS filters are not as sensitive to $\log g$).

After obtaining the external calibration zero-points for each filter ($\mathrm{ZP}_\mathcal{m}^\mathrm{ext}$), we produce an externally calibrated S-PLUS catalogue. We defined the externally calibrated magnitudes in terms of the instrumental magnitudes as:
\begin{equation}
    \mathcal{m}_{\mathrm{ext}} = \mathcal{m}_{\mathrm{instr}} + \mathrm{ZP}_\mathcal{m}^{\mathrm{ext}} \,.
\end{equation}

In the case shown in the example, all 12 S-PLUS filters can be calibrated from the 5 reference SDSS filters. This is also the case for any other reference catalogue (or combination of reference catalogues) whose filters span the whole wavelength range of S-PLUS. We refer to this calibration of the 12 filters from the model fitting of the reference catalogue as ``Method I''. 

\subsubsection{Stellar locus calibration} 
\label{sec:stellar_locus_calibration}

Only a limited fraction of S-PLUS wavelength range is available in the reference catalogues for some particular fields. This situation is not common given the large number of wide-area surveys currently available, but it is frequent enough not to be ignored: for instance, it affects 24 out of the 514 S-PLUS fields in DR2. For these fields, only the $griz$ reference measurements are available to use for the calibration, lacking the necessary magnitude(s) to properly constrain the models in the blue region of the spectrum. We find that this issue can be solved by following the Method I approach to calibrate the filters in the wavelength ranges where the models are well constrained (in this case, $g$, J0515, $r$, J0660, $i$, J0861 and $z$), and then including an additional step that makes use of the expected stellar locus in colour-colour diagrams to estimate the ZPs of the remaining blue filters (in this case $u$, J0410 and J0430). We denominate this approach, ``Method II''.

After the external calibration of the 7 redder filters, we produce 3 colour-colour diagrams of $g_\mathrm{ext} - i_\mathrm{ext}$ versus $\mathcal{m}_\mathrm{inst} - g_\mathrm{ext}$, where $\mathcal{m}$ can be the filters $u$, J0410 and J0430. Then, we compare the stellar locus in these colour-colour diagrams to the stellar locus obtained from a previous S-PLUS calibration performed through Method I. Since the red filters are already calibrated, the difference between the stellar locus of the field and the reference S-PLUS calibration can be attributed to the ZP of the blue filter. We estimate this difference by dividing the stellar locus in 7 bins of colours in the interval $0.4 \le g - i (mag) \le 1.2$ and calculating the average $\mathcal{m}_\mathrm{inst} - g_\mathrm{ext}$ in each bin for the field and for the reference stellar locus. An example of this process, for field STRIPE82-0001, is shown in Figure \ref{fig:stellar_locus_ex}. We remove the lowest and highest difference to avoid the contribution of outliers and estimate the ZP from the average of the differences of the remaining bins.

Due to the natural spread of the stellar locus for filters J0378 and J0395, this technique does not produce acceptable results for their calibration, which, in the case of Method II, is obtained only during the next internal calibration step.

\begin{figure*}
\begin{center}
\includegraphics[width=\textwidth]{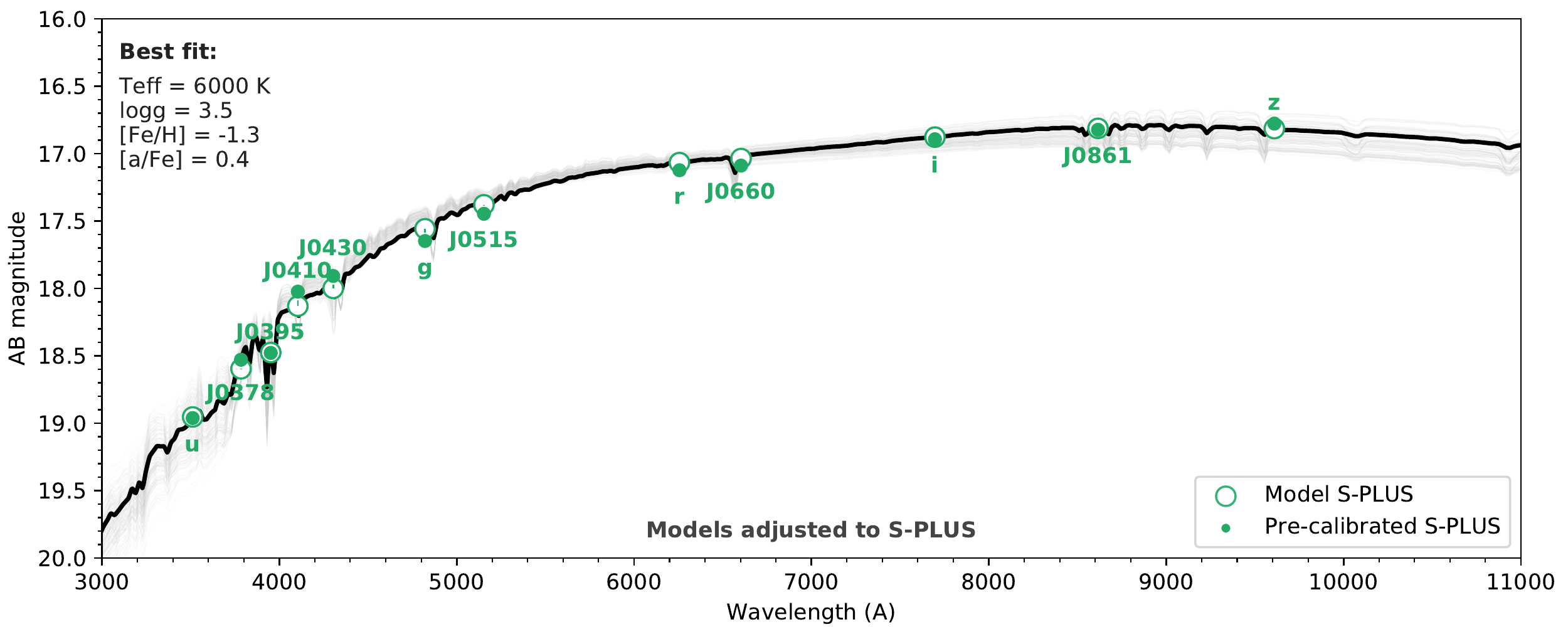}
\caption{\label{fig:model_fit_splus_example} Analogous to Figure \ref{fig::model_fit_sdss_example} for the internal calibration step. In this case, the best model (black line) is fit to the externally calibrated S-PLUS magnitudes, represented by the filled green circles. The open green circles correspond to the model predicted internally calibrated magnitudes.}
\end{center}
\end{figure*}

\begin{figure}
\begin{center}
\includegraphics[width=.48\textwidth]{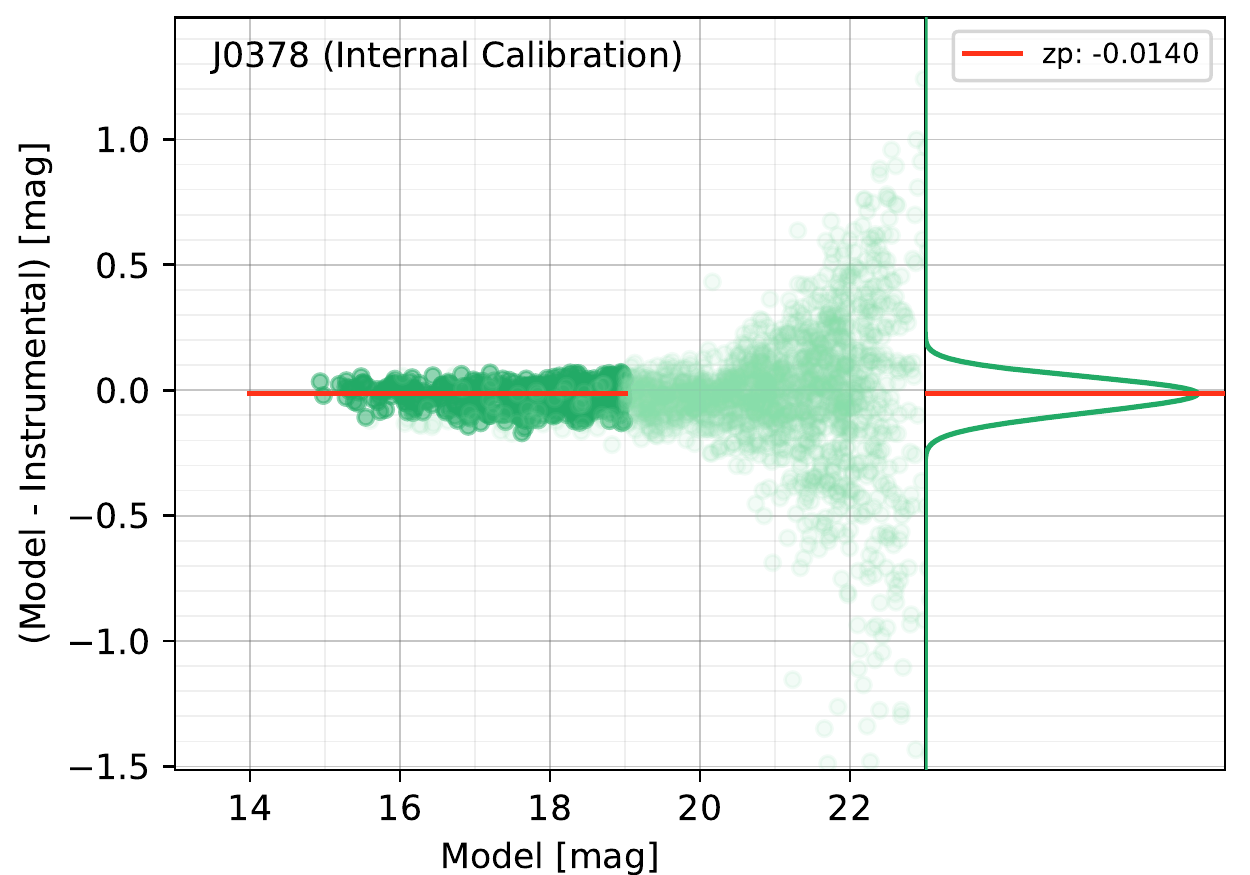}
\caption{\label{fig:model_fit_splus_example2} Analogous to Figure \ref{fig::external_zp_fitting_example} for the ZP characterisation of the internal calibration step for filter J0378 of field STRIPE82-0001. The ZP is characterised as the mode of the distribution of the differences between the externally calibrated and the model predicted S-PLUS magnitudes.}
\end{center}
\end{figure}

\subsection{Internal Calibration}

The S-PLUS narrow bands are strategically centred around important spectral features: OII for J0378;  Ca H+K for J0395; H$\delta$ for J0410; CH G-band for J0430; Mg$b$ triplet in J0515; H$\alpha$ at J0660; and Ca Triplet for J0861. To mention a few examples, the Ca line, measured by the J0395 filter, is particularly sensitive to metallicity, In this wavelength, narrow-band photometry around this feature is usually employed by surveys that search for metal-poor stars \citep{Starkenburg+2017, Keller+2007}. Also, the Mg$b$ triplet, measured by filter J0515, is well known for its surface gravity sensitivity \citep{Geisler1984, Majewski+2000}, and narrow-band photometry has been used by surveys, such as APOGEE, to search for giant stars \citep{Zasowski+2013, Majewski+2016}.

This added information allows the calibrated S-PLUS magnitudes to better constrain the synthetic models in comparison to the reference catalogue used in the previous step. For this reason, a second step of the model fitting calibration is performed in the S-PLUS externally calibrated magnitudes. The $\chi^2$ estimation is analogous to the one used in the former section. The only differences are that the reference magnitudes and errors being used are now taken from the S-PLUS externally calibrated catalogue, and the set of filters are the 12 S-PLUS passbands ($\mathscr{S}$). In the case of Method II, since only 10 filters are calibrated during the first step, the filters J0378 and J0395 are not considered for the $\chi^2$ estimation.

\begin{equation}
 \chi^2_{\mathrm{int},i} = \sum_{\mathcal{m} \in \mathscr{S}} \left( \frac{ \mathcal{m}_{\mathrm{mod}, i} - \mathcal{m}_{\mathrm{ext}} } {\delta \mathcal{m}_\mathrm{ext} } \right)^2 \,.
\end{equation}

The \textit{internal calibration zero-point} ($\mathrm{ZP}_\mathcal{m}^\mathrm{int}$) is once more estimated from the mode of the differences between the catalogue and the model predicted magnitudes. In this case, it is not necessary to apply a surface gravity cut since the models can now be expected to be correctly constrained regardless of the stellar luminosity class. An example of the internal ZP fitting is shown in Figures \ref{fig:model_fit_splus_example} and \ref{fig:model_fit_splus_example2}. Figure \ref{fig:model_fit_splus_example} is analogous to Figure \ref{fig::model_fit_sdss_example} and shows the best fit for star SDSS J000027.92-004122.4, but this time fitting the 12 S-PLUS measurements. The difference between the S-PLUS and the model magnitudes corresponds to a refinement that needs to be added to the ZPs. As we did in the first step, the internal calibration ZP is estimated taking into account all the selected stars in the field, which is shown in Figure \ref{fig:model_fit_splus_example2} for filter J0378. It is important to mention that this image shows a significant reduction in outliers when compared to Figure \ref{fig::external_zp_fitting_example}, which is a result of the improved constraints provided by the S-PLUS narrow bands.

The internal ZPs, which are usually smaller than 0.05 mag, are added to the externally calibrated magnitudes to produce what we call an internal calibrated catalogue for the S-PLUS reference stars:

\begin{equation}
    \mathcal{m}_{\mathrm{int}} = \mathcal{m}_{\mathrm{ext}} + \mathrm{ZP}_\mathcal{m}^{\mathrm{int}} \,.
\end{equation}

\begin{figure*}
\centering
\includegraphics[width=\textwidth]{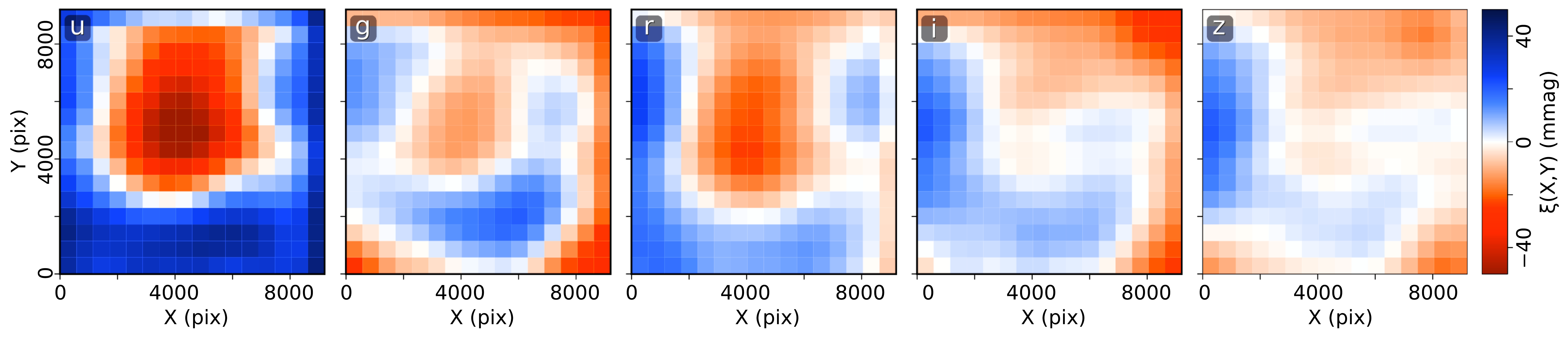}
\caption{16 $\times$ 16 ZP variation across the CCD obtained from the stacked STRIPE82 data by comparing the calibrated S-PLUS magnitudes to the reference SDSS magnitudes. The maps were constructed for filters $u$, $g$, $r$, $i$ and $z$ (presented from left to right). The mean value was subtracted from each map and we applied a smoothing Gaussian kernel with a sigma equivalent to the size of 1 bin.}
\label{fig:XY_offsets}
\end{figure*}

\subsection{Gaia Scale Calibration}
\label{sec:gaia_scale_calibration}

After the external and internal calibrations, there are no more changes to the shape of the SED. Nevertheless, an additional step is performed to bring S-PLUS to the Gaia DR2 \citep{GaiaCollaboration+2018} magnitude scale, which we call the `Gaia scale calibration'. This step is crucial for S-PLUS DR2 (and future data releases) because different reference catalogues are being used for the photometric calibration in different parts of the sky and is subject to magnitude scale offsets between these catalogues. Due to its all-sky nature as well as simultaneously observing fields separated by 106.5 deg, Gaia provides an ideal uniform reference frame for the photometric scale. 

The Gaia scale calibration also makes use of the same model fitting technique employed in the previous steps; However, in this case, the models are fit to the S-PLUS internally calibrated magnitudes to predict the three Gaia magnitudes ($\mathrm{G}$, $\mathrm{G}_\mathrm{BP}$ and $\mathrm{G}_\mathrm{RP}$). The predicted magnitudes for the reference stars are then compared to the Gaia DR2 magnitudes. The offsets are calculated in the same way as the ZP estimation for the previous steps. The Gaia scaled calibration zero-point, $\mathrm{ZP}^\mathrm{GS}$, is characterised as the average between the offsets found for the three Gaia filters and is the same for the 12 S-PLUS filters.

Since this step comes after the internal calibration, when the 12 S-PLUS magnitudes are already calibrated, there is no difference between Methods I and II. It is also important to note that, in this case, we fit the difference of the S-PLUS prediction minus the reference catalogue magnitude instead of the opposite. Therefore, the estimated ZPs (or offsets) need to be subtracted from the S-PLUS magnitudes.

\subsection{Final ZP and calibrated magnitudes}

The final calibration ZP is a combination of the ZP obtained in the external, internal and Gaia scale steps. Given that the S-PLUS catalogue is updated at the end of each step, the Final ZP is simply a sum of the ZPs obtained in each step. As mentioned before, when running \texttt{SExtractor} to obtain the magnitudes from the images, we have used a guess ZP of 20 mag. This initial guess also needs to be added to the Final ZP, which is then given by:

\begin{equation}
    \mathrm{ZP}_\mathcal{m} = 20 + \mathrm{ZP}_\mathcal{m}^\mathrm{ext} + \mathrm{ZP}_\mathcal{m}^\mathrm{int} - \mathrm{ZP}^\mathrm{GS} \,.
\end{equation}

Finally, we apply these ZPs to the photometric catalogues to obtain the final calibrated magnitudes for all the sources detected in the field in all the different apertures considered (in this case, we need to subtract the initial guess of 20 mag, otherwise, it would be included twice given our definition of $\mathcal{m}_\mathrm{instr}$). The S-PLUS calibrated magnitudes are then given by:

\begin{equation}
    \mathcal{m} = \mathcal{m}_\mathrm{instr} + \mathrm{ZP}_\mathcal{m} - 20  \,.
\end{equation}

\begin{table*}

\caption{\label{tab:methods} Denomination of the 4 different calibration strategies applied for DR2. Methods Ia, Ib and Ic only differ by the reference catalogue used in the external calibration. The predicted ZPs, either from model fitting or stellar locus technique, are used to calibrate the S-PLUS instrumental magnitudes. In the internal calibration, the models are fitted to the S-PLUS previously externally calibrated magnitudes to again predict the S-PLUS magnitudes and refine the ZPs. The only exceptions are filters J0378 and J0395 for Method II, where the instrumental magnitudes are only calibrated during the Internal calibration. This table omits the Gaia Scale calibration step, which is the same in all methods: the 12 internally calibrated S-PLUS magnitudes are used to predict Gaia magnitudes, which are compared to Gaia's DR2 to fit and correct for systematic offsets. The last column shows the number of fields in DR2 that ended up being calibrated by each specific method. Method Ic is only used for verification and is not present in the DR2 data.}

\footnotesize
\begin{tabular}{cccccccc}
\hline \hline

& &
\multicolumn{3}{c}{External Calibration} & 
\multicolumn{2}{c}{Internal Calibration}  &  
\\ \cmidrule(lr){3-5}\cmidrule(lr){6-7}

Method & 
Library & 
\begin{tabular}[c]{@{}c@{}}Models are\\ fitted from\end{tabular} & 
\begin{tabular}[c]{@{}c@{}}ZPs from \\ model prediction\end{tabular} &
\begin{tabular}[c]{@{}c@{}}ZPs from \\ stellar locus\end{tabular} &
\begin{tabular}[c]{@{}c@{}}Models are\\ fitted from\end{tabular} &
\begin{tabular}[c]{@{}c@{}}ZPs from \\ model prediction\end{tabular} &
\# fields \\ \hline \hline

Method Ia & 
Coelho14 &
\begin{tabular}[c]{@{}c@{}}SDSS\\ (u, g, r, i, z)\end{tabular} & 
\begin{tabular}[c]{@{}c@{}}S-PLUS \\ (all 12 filters)\end{tabular} & 
\begin{tabular}[c]{@{}c@{}}None\end{tabular} &
\begin{tabular}[c]{@{}c@{}}S-PLUS \\ (all 12 filters)\end{tabular} &
\begin{tabular}[c]{@{}c@{}}S-PLUS\\ (all 12 filters)\end{tabular}  & 
\begin{tabular}[c]{@{}c@{}}170\end{tabular} \\[+1em]

Method Ib & 
Coelho14 &
\begin{tabular}[c]{@{}c@{}}RefCat2 (g, r, i, z) \\ \& GALEX (NUV)\end{tabular} & 
\begin{tabular}[c]{@{}c@{}}S-PLUS \\ (all 12 filters)\end{tabular} &
\begin{tabular}[c]{@{}c@{}}None\end{tabular} &
\begin{tabular}[c]{@{}c@{}}S-PLUS \\ (all 12 filters)\end{tabular} &
\begin{tabular}[c]{@{}c@{}}S-PLUS\\ (all 12 filters)\end{tabular} & 
\begin{tabular}[c]{@{}c@{}}320\end{tabular} \\[+1em]

Method Ic & 
Coelho14 &
\begin{tabular}[c]{@{}c@{}}Skymapper \\ (u, v, g, r, i, z)\end{tabular} & 
\begin{tabular}[c]{@{}c@{}}S-PLUS \\ (all 12 filters)\end{tabular} &
\begin{tabular}[c]{@{}c@{}}None\end{tabular} &
\begin{tabular}[c]{@{}c@{}}S-PLUS \\ (all 12 filters)\end{tabular} &
\begin{tabular}[c]{@{}c@{}}S-PLUS\\ (all 12 filters)\end{tabular} & 
\begin{tabular}[c]{@{}c@{}}0\end{tabular} \\[+1em]

Method II & 
Coelho14 &
\begin{tabular}[c]{@{}c@{}}RefCat2\\ (g, r, i, z)\end{tabular} & 
\begin{tabular}[c]{@{}c@{}}S-PLUS\\ ($g$, J0515, $r$, J0660, \\ $i$, J0861, $z$)\end{tabular} & 
\begin{tabular}[c]{@{}c@{}}S-PLUS \\ ($u$, J0410, J0430)\end{tabular} & 
\begin{tabular}[c]{@{}c@{}}S-PLUS \\ (all, except \\ J0378, J0395)\end{tabular} & 
\begin{tabular}[c]{@{}c@{}}S-PLUS\\ (all 12 filters)\end{tabular} & 
\begin{tabular}[c]{@{}c@{}}24\end{tabular} \\ \hline \hline       

\end{tabular}
\end{table*}

\subsection{XY Inhomogeneities}
\label{sec:CCD_correction}
A first run of the calibration for the STRIPE82 region revealed a correlation between ZP offsets and the $X,Y$ position of the source on the CCD. The offsets were computed in relation to the SDSS reference magnitudes. A similar issue has also been observed in the calibration of J-PLUS. Given the similarities between both surveys, it is no surprise that these offsets are also present in S-PLUS. Some of the reasons that can cause this effect are airmass and/or PSF variations across the field, non-uniform transmission curves across the filter's surface, and the presence of scattered light in the focal plane \citep{Regnault+2009, Starkenburg+2017, LopesSanjuan+2019}.

In DR2, this issue is solved by applying an ad-hoc correction to the instrumental magnitudes. We obtained the correction maps following the steps: (i) we compile a calibrated catalogue for a large number of fields (the whole STRIPE82 region); (ii) compute the average difference between S-PLUS and SDSS for each bin in a 16 $\times$ 16 map across the CCD; (iii) subtract, from each bin, the average value between all bins (which ensures that no systematic corrections are introduced in this step); and (iv) smooth the maps by applying a Gaussian filter with a kernel size of 1 bin (to minimise the noise contribution).

In Figure \ref{fig:XY_offsets} we show the correction maps obtained for the $u$, $g$, $r$, $i$, $z$ filters. In general, the offsets are minimal, ranging from -20~to~20~mmags for the GRIZ bands. Only the $u$ band show offsets as extreme as -40~to~40~mmags. The maps shown in the figure were obtained for the STRIPE82 region, using SDSS as a reference. We have also checked that they do not change significantly when obtained for different sets of S-PLUS fields. In particular, they show no correlation with the observation's airmass and do not change when measured in relation to a different reference catalogue. Therefore, we are convinced that the maps obtained for the STRIPE82 can be used for the correction of all the fields in DR2. We also note that there is very little variation between the correction maps of two adjacent filters, which allows us to apply the corrections for narrow-band filters by using the broad-band map with the closest effective wavelength. Filters J0378 and J0395 are assigned the $u$-band correction map, filters J0410, J0430 and J0515 are assigned the $g$-band map, while filters J0660 and J0861 are assigned the $r$-band and $z$-band maps, respectively.

In terms of the instrumental magnitudes, these corrections are represented by the term $\xi_\mathcal{m}(X,Y)$ in Equation \ref{eq:instrumental_mags} and correspond to the value of the offset in the correction map for the bin that contains the source. Given that the correction maps have already been smoothed out and that the variation between adjacent bins is already smaller than the photometric errors, we do not implement any interpolations for the correction maps.

 
\section{Validation of the photometric calibration}
\label{sec::Calibration_validation}

In this Section we validate the results and characterise the errors of the pipeline by calibrating the STRIPE82 region using four different methods and comparing them with literature data. This region was chosen due to the large number of well-calibrated surveys available to be used as a reference both for the calibration and for the comparisons. We also evaluate the calibration dependency on the chosen reference catalogues and spectral libraries. As an additional check, the internal consistency of the pipeline is tested by comparing the magnitudes of an internal crossmatch of S-PLUS fields with large overlaps. 

\subsection{Choice of the reference catalogue}

The desired properties of an ideal reference catalogue for the S-PLUS calibration are: (i) being able to provide uniform and well-calibrated data across a large FoV (at least for the 2 deg$^2$ of the S-PLUS observations); (ii) a sky coverage that contains the whole area observed by S-PLUS, (iii) a photometric system that covers the whole range of wavelengths of the S-PLUS system, while still having enough filters in the middle to correctly constrain the model spectra in this region; (iv) containing enough sources in each S-PLUS observation to allow for the model-fitting technique to be properly applied (it usually requires at least 100 stars with photometric errors smaller than 0.01 mag).

Even though the S-PLUS footprint has significant overlaps with other surveys, no single catalogue satisfies all the aforementioned criteria. We circumvent this obstacle by employing different strategies according to the availability of data for each observed field: the lack of wavelength coverage can be solved by combining more than one reference catalogue or by employing the stellar locus calibration for this filter. After extensively testing different catalogues, we find that the best strategies for the S-PLUS calibration are three Method I calibrations, which we divide in  `Method Ia', when using SDSS as a reference; `Method Ib' when the reference is a combination of ATLAS RefCat2 and GALEX (only NUV band); and Method Ic, where the reference is Skymapper (although this calibration is only used for verification and is not present in the DR2 data); and also the Method II calibration using ATLAS RefCat2 and the stellar locus technique when the field is in a region without reference photometry for the blue filters. We summarise these methods in Table \ref{tab:methods} and list the magnitudes used as a reference for the model fitting and the predicted magnitudes used for the estimation of the ZPs in each step.

\begin{figure}
\begin{center}
\includegraphics[width=.47\textwidth]{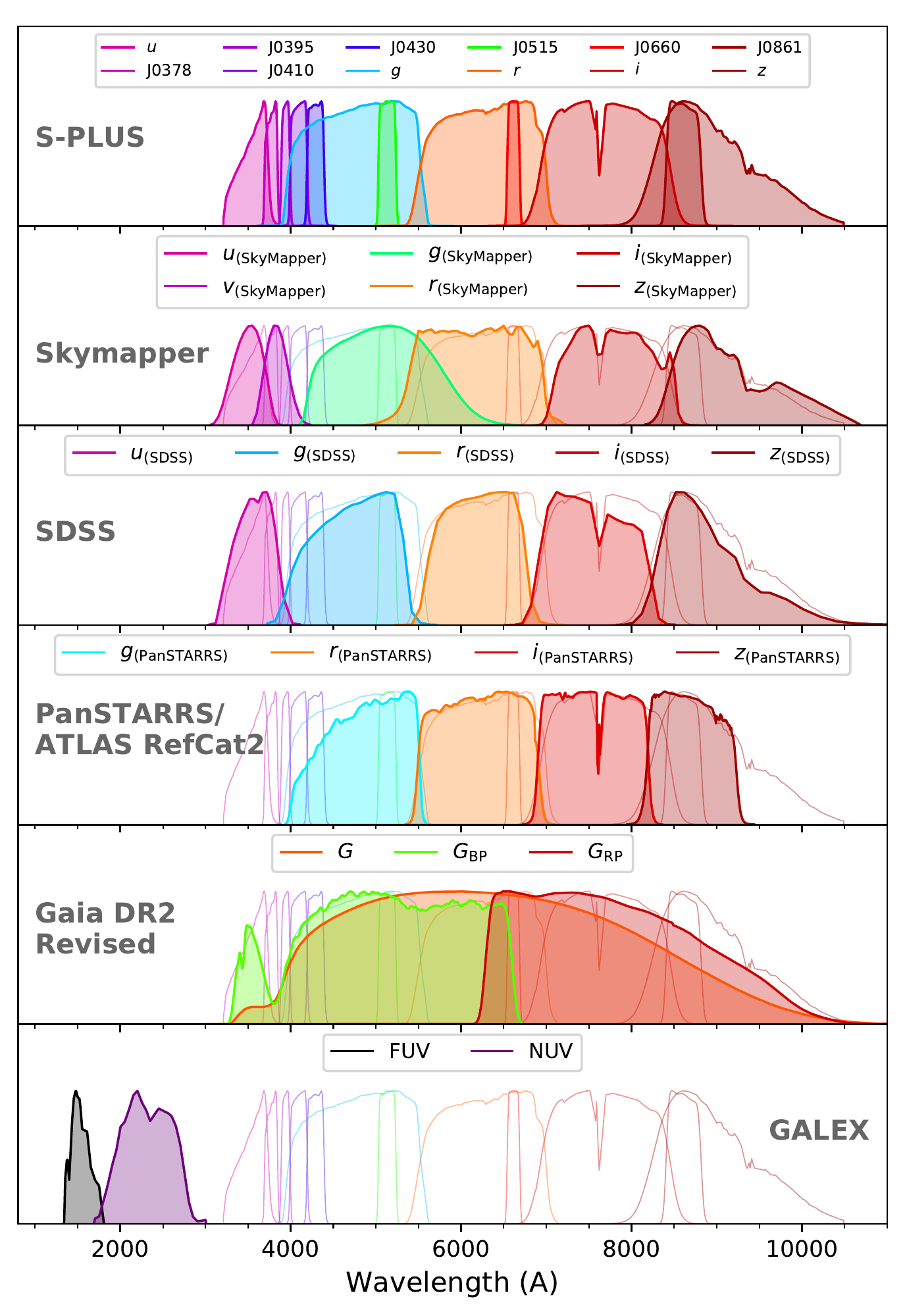}
\caption{\label{fig:filter_system}Passbands of the S-PLUS filter system and a few overlapping surveys. The S-PLUS filters are part of the Javalambre filter system \citep{MarinFranch+2012}. The S-PLUS filter system is shown in each panel for comparison. The included filter systems are the Skymapper DR2 \citep{Bessell+2011}, SDSS \citep{Doi+2010}, ATLAS-REFCAT2/PanSTARRS filter system \citep{Chambers+2016, Tonry+2018}, Gaia DR2 \citep{Evans+2018} and finally GALEX DR6/7 \citep{Morrissey+2005}).}
\end{center}
\end{figure}

As we described in Section \ref{sec:models}, the \citetalias{Coelho14} spectral library is pre-convolved in the passbands of S-PLUS and the reference catalogues to generate the library of synthetic photometry that is used in the pipeline. In Figure \ref{fig:filter_system} we present the photometric system of each survey that was used in the calibration. In each panel, the contours of the S-PLUS filter system are shown for comparison. Filters $g$, $r$, $i$ and $z$ of S-PLUS, SDSS, Skymapper, and PanSTARRS (which is the system adopted in the ATLASRefCat2) are very similar, but not exactly the same, which supports the idea that direct comparisons are not adequate, justifying the use of the model fitting step for the transformation between the different systems. The Gaia system, used in the Gaia Scale calibration is also represented.

\begin{figure*}
\begin{center}
\includegraphics[width=\textwidth]{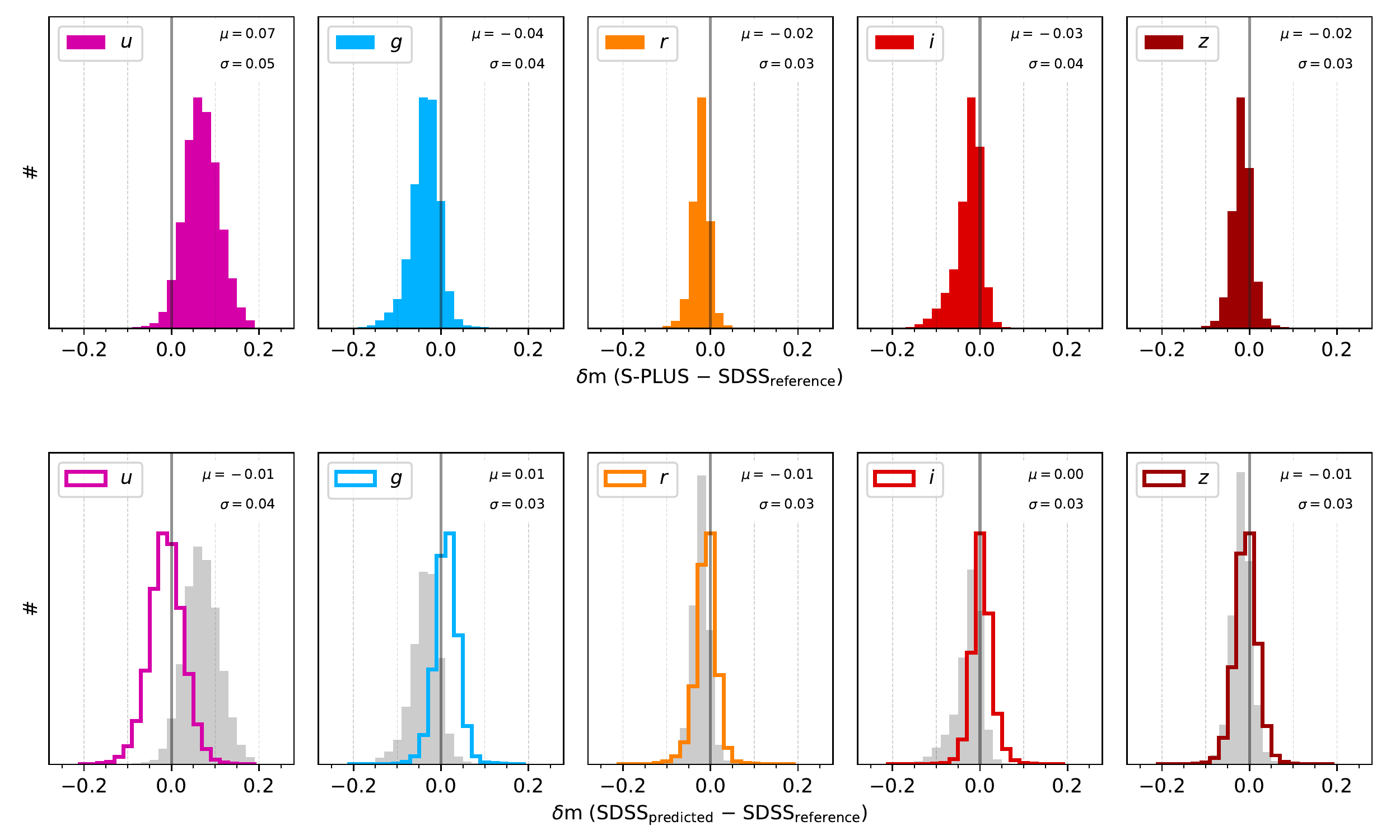}
\caption{\label{fig:SPLUS_SDSS_comparison} Top: Direct comparison between S-PLUS magnitudes (after external calibration step) and the reference SDSS magnitudes for the STRIPE82 objects with magnitudes below 17 mag. The root mean square (rms) of the distribution for the 5 ugriz filters are $0.05$, $0.04$, $0.03$, $0.04$, $0.03$ mag, and the mean differences are $0.07$, $-0.04$, $-0.02$, $-0.03$ and $-0.02$ mag respectively. Bottom: Comparison between the model predicted SDSS magnitudes (obtained from the S-PLUS magnitudes) and the reference SDSS magnitudes. In this case, rms for the ugriz filters are $0.04$, $0.03$, $0.03$, $0.03$ and $0.03$ mag, and mean differences are $-0.01$, $0.01$, $-0.01$, $0.00$, $-0.01$ mag, respectively. The distributions of the upper panels are shown in grey for comparison.}
\end{center}
\end{figure*}

\subsubsection{Method Ia and comparison to the SDSS}

The SDSS catalogue \citep{York+2000}, in particular the Ivezi{\'c} calibration of the STRIPE82 region \citep{Ivezic+2007}, is the closest to fulfil all the desired requirements of an ideal reference catalogue. Its 5-band photometric system \citep{Doi+2010} not only shares the same wavelength coverage as S-PLUS but also has very similar transmission curves to the S-PLUS broad bands, which ensures that the models that properly fit the SED of one system will also properly fit the other. It is also very precise and uniformly calibrated across the entire footprint, with the reported errors of photometry and ZPs being below $\sim0.01$ mag. Twelve of the 170 STRIPE82 fields are not covered by the Ivezi{\'c} calibration, in which case the SDSS reference data comes from DR12 \citep{Alam+15}.

The only significant drawback is the lack of coverage of the southern hemisphere, which prohibits its use for the calibration of almost all of the S-PLUS observations. Nevertheless, given all of its qualities, we have used SDSS as the reference to calibrate the STRIPE82 region and chosen it as a fundamental reference calibration for S-PLUS. This allows us first to employ all other strategies to the STRIPE82 region, compare the results with the Method Ia calibration, correct for any arising systematic offsets, and estimate the ZP uncertainty associated with each different strategy. The stellar locus reference used in Method II is also drawn from this STRIPE82 calibration.

\begin{table}
\caption{Mean offsets in the ZPs between the external calibration obtained from Method Ia and each of the other three methods. In the case of Method II the offsets correlate with the mean ISM extinction of the field. During the calibration, these offsets are corrected after the external calibration step (and before the internal calibration).}
\label{tab:offsets}
\centering
\begin{tabular}{lccr}
\hline \hline
Filter & Method Ib   & Method Ic   & Method II \\ \hline \hline
$u$      & $\;\;0.049$ & $\;\;0.080$ & $ 0.028 - 1.446\,E_\mathrm{B-V}$  \\
J0378   & $   -0.029$ & $   -0.057$ & $ 0.056 - 1.548\,E_\mathrm{B-V}$  \\
J0395   & $   -0.114$ & $   -0.164$ & $-0.027 - 1.539\,E_\mathrm{B-V}$  \\
J0410   & $   -0.034$ & $   -0.048$ & $ 0.011 - 0.713\,E_\mathrm{B-V}$  \\
J0430   & $   -0.065$ & $   -0.101$ & $ 0.001 - 0.518\,E_\mathrm{B-V}$  \\
$g$      & $   -0.014$ & $   -0.026$ & $-0.012 - 0.077\,E_\mathrm{B-V}$  \\
J0515   & $   -0.024$ & $   -0.042$ & $-0.019 - 0.121\,E_\mathrm{B-V}$  \\
$r$      & $   -0.016$ & $   -0.009$ & $-0.010 + 0.029\,E_\mathrm{B-V}$  \\
J0660   & $\;\;0.012$ & $\;\;0.023$ & $ 0.015 - 0.080\,E_\mathrm{B-V}$  \\
$i$      & $   -0.024$ & $   -0.021$ & $-0.016 - 0.053\,E_\mathrm{B-V}$  \\
J0861   & $   -0.047$ & $   -0.048$ & $-0.029 - 0.059\,E_\mathrm{B-V}$  \\
$z$      & $   -0.049$ & $   -0.051$ & $-0.035 - 0.050\,E_\mathrm{B-V}$  \\
\hline \hline
\end{tabular}
\end{table}

We calibrated the 170 STRIPE82 fields and performed two tests to validate this calibration: the first is a direct comparison of the S-PLUS calibrated magnitudes with the reference SDSS magnitudes. Since the goal is to assert the quality of the ZP estimation specifically from this reference catalogue, the comparison is made using the catalogue produced in the external calibration step. We also selected only stars with magnitudes smaller than 17 mag to ensure that only the stars with good photometry are considered. The direct comparison between the S-PLUS magnitudes and SDSS counterparts is shown in the upper panels of Figure \ref{fig:SPLUS_SDSS_comparison}. The root mean square (rms) of the distributions are 0.05, 0.04, 0.03, 0.04 and 0.03 mag for filters $u$, $g$, $r$, $i$ and $z$, respectively. The rms is systematically $\sim0.01$ mag lower than the one calculated for S-PLUS DR1 \citep{MendesDeOliveira+2019}, reflecting the changes made to the calibration pipeline. Nevertheless, offsets as high as 0.07~mag can be observed in the case of the $u$ band. These offsets are not a result of ZP inaccuracies but reflect the slightly different transmission curves between both surveys.

Our second test consists of comparing the magnitudes in the same photometric system. This is achieved by using the same model-fitting algorithm to fit the models to the S-PLUS externally calibrated magnitudes and predict the SDSS magnitudes for the stars. The difference between the predicted and the reference SDSS magnitudes is shown in the lower panels of Figure \ref{fig:SPLUS_SDSS_comparison}. In this case, the calculated rms are even lower: 0.04 mag for the $u$ band and 0.03 mag for the others. As expected, the comparison of the magnitudes in the same photometric system removes the previously observed offsets, which now does not exceed 0.01 mag.

These results provide an upper limit for the uncertainty of the ZP of the broad bands. However, the observed rms cannot be attributed to the errors in ZPs alone, since it also depends on the photometric errors of S-PLUS, the photometric and ZP errors of SDSS as well as the noise introduced in the process of converting again from S-PLUS to SDSS magnitudes. The fact that, even when coupled with all other uncertainties, the rms is smaller than 0.04 mag for all filters already attests to the quality of our calibration technique. Nevertheless, further tests were employed taking into account the calibration using different reference catalogues to better characterise the ZP uncertainties, also including the narrow bands.

\begin{table*}
\caption{Final offsets and scatter between the STRIPE82 calibration of different strategies against the strategy using the Coelho14 models and the SDSS catalogue as reference. These offsets indicate the systematic and random errors that might be present when employing the different strategies.}
\label{tab:ref_comp_final}
\begin{tabular}{lrccrccrcccrccrc}
\hline \hline
& \multicolumn{8}{c}{ZP comparison against Method Ia} & \,\, & \,\, & \multicolumn{5}{c}{ZP comparison against Coelho14} \\ \cmidrule(lr){2-9} \cmidrule(lr){12-16}
& \multicolumn{2}{c}{Method Ib} && \multicolumn{2}{c}{Method Ic}  && \multicolumn{2}{c}{Method II} &&& \multicolumn{2}{c}{C\&K03}  && \multicolumn{2}{c}{NGSL} \\ \cmidrule(lr){2-3}\cmidrule(lr){5-6}\cmidrule(lr){8-9}\cmidrule(lr){12-13}\cmidrule(lr){15-16}
Filter$\quad\quad$ & offset & scatter && offset & scatter && offset & scatter &&& offset & scatter && offset & scatter \\ \hline \hline
$u$    &  0.004 & 0.021 &&  0.001 & 0.023 && -0.001 & 0.023 &&&  0.000 & 0.004 && -0.014 & 0.005 \\
J0378 &  0.009 & 0.013 && -0.002 & 0.021 && -0.004 & 0.020 &&&  0.019 & 0.007 &&  0.006 & 0.008 \\
J0395 & -0.006 & 0.023 &&  0.000 & 0.031 &&  0.000 & 0.032 &&& -0.053 & 0.013 && -0.033 & 0.014 \\
J0410 & -0.002 & 0.008 &&  0.000 & 0.011 &&  0.001 & 0.010 &&&  0.008 & 0.003 &&  0.011 & 0.003 \\
J0430 & -0.009 & 0.011 && -0.004 & 0.011 &&  0.001 & 0.010 &&&  0.002 & 0.004 && -0.001 & 0.005 \\
$g$    &  0.009 & 0.005 && -0.003 & 0.005 &&  0.000 & 0.003 &&& -0.001 & 0.002 && -0.002 & 0.002 \\
J0515 &  0.006 & 0.004 &&  0.002 & 0.004 &&  0.001 & 0.004 &&& -0.009 & 0.002 && -0.013 & 0.003 \\
$r$    &  0.008 & 0.002 &&  0.006 & 0.002 &&  0.000 & 0.002 &&&  0.002 & 0.002 &&  0.002 & 0.001 \\
J0660 & -0.001 & 0.004 && -0.002 & 0.004 &&  0.000 & 0.004 &&&  0.008 & 0.001 && -0.009 & 0.002 \\
$i$    & -0.002 & 0.002 && -0.004 & 0.003 &&  0.000 & 0.002 &&& -0.001 & 0.001 && -0.003 & 0.002 \\
J0861 & -0.001 & 0.004 &&  0.000 & 0.004 && -0.001 & 0.003 &&& -0.005 & 0.001 && -0.014 & 0.002 \\
$z$    &  0.000 & 0.004 &&  0.000 & 0.005 &&  0.000 & 0.002 &&& -0.002 & 0.001 &&  0.005 & 0.002 \\
\hline \hline
\end{tabular}
\end{table*}

\subsubsection{Method Ib - ATLAS Refcat2 and GALEX}

The ATLAS All-Sky Stellar Reference Catalog \citep{Tonry+2018} (hereafter, ATLAS Refcat2) is the most suitable option for the calibration of most S-PLUS fields, as it is available for the whole footprint. It is an all-sky compendium of several catalogues that is expected to be at least 99 per cent complete down to a magnitude of $19$ mag. It includes data from Pan-STARRS DR1 \citep{Chambers+2016, Flewelling+2020}, Gaia DR2 (\citealp{GaiaCollaboration+2018}, which is the source of astrometry), SkyMapper DR1 \citep{Wolf+2018}, APASS DR9 \citep{Henden+2014, Henden+2016},  2MASS \citep{Skrutskie+2006} and Tycho-2 \citep{Hog+2000, Pickles+Depagne2010}, and the Yale Bright Star Catalog as well as two new sources of photometry, the ATLAS pathfinder observations and a re-reduction of APASS. \citet{Tonry+2018} use all the magnitudes available to calculate PanSTARRS  magnitudes for all the stars ATLAS Refcat2 stars, providing a uniform all-sky catalogue in a single photometric system with errors expected to be no larger than 5 mmags.  

The photometric system in the ATLAS Refcat2 does not cover the whole wavelength range of the S-PLUS filters and is incapable of constraining the models to a level that allows us to properly calibrate the blue bands. For this reason, we include near-ultraviolet (NUV) observations from the Galaxy Evolution Explorer (GALEX, \citealp{Morrissey+2005}). Therefore, our reference catalogue in Method Ib consists of a crossmatch between ATLAS RefCat2 and GALEX DR6/7.

Similar to Method Ia, we applied the calibration using Method Ib to calibrate all the 170 fields in the STRIPE82 region. When comparing the results (after the external calibration step), we find systematic offsets in all filters. As shown in Table \ref{tab:offsets}, these offsets are usually smaller than 0.03 mag but can be as large as 0.1 mag (for the J0395 filter). These offsets can be caused by differences in the magnitude scales of the reference catalogues and possibly indicate that the NUV magnitude is not as good for fitting the blue regions of the spectrum as the $u$ band of SDSS. Nevertheless, we adopt the Method Ia calibration as a reference and correct the offsets from the Method Ib externally calibrated magnitudes before proceeding with the internal and the Gaia Scale calibrations. This correction is applied for all Method Ib calibrations, including fields other than the ones in the STRIPE82.

The comparison of the final calibrated magnitudes between Methods Ib and Ia is shown in Figure \ref{fig:ref_comp_final}. The purple circles represent the average of the differences between both ZPs and the errors bars represent the rms of this distribution, which are also presented in Table \ref{tab:ref_comp_final}. We see an excellent agreement between the ZPs obtained from the two methods. The final systematic offsets are always smaller than 10 mmag, and the rms is smaller than 5 mmag for filters $g$, J0515, $r$, J0660, $i$, J0861 and $z$; smaller than 15 mmag for filters J0378, J0410 and J0430; and smaller than 25 mmag for filters $u$ and J0395.

\subsubsection{Method Ic - Skymapper}

We also compare the SDSS calibration to a calibration using Skymapper DR2 \citep{Onken+2019}, which we name Method Ic. Skymapper is a southern hemisphere survey that observed in 6 bands with a 1.3-m telescope located in Siding Spring Observatory, Australia \citep{Keller+2007}. It has 5 SDSS-like filters (u, g, r, i and z) and an additional narrow band, 'v' filter, capable of constraining our models in the blue region of the spectrum (see Figure \ref{fig:filter_system}).

We followed the same steps for the comparison as we did for Method Ib. After the STRIPE82 calibration we compared the external calibration ZPs to those of Method Ia to find the offsets that need to be corrected to bring the magnitudes to the Method Ia reference. These offsets are also shown in Table \ref{tab:offsets}. Except for the redder filters, the offsets are systematically higher than those observed for Method Ib.

\begin{figure}
\begin{center}
\includegraphics[width=.48\textwidth]{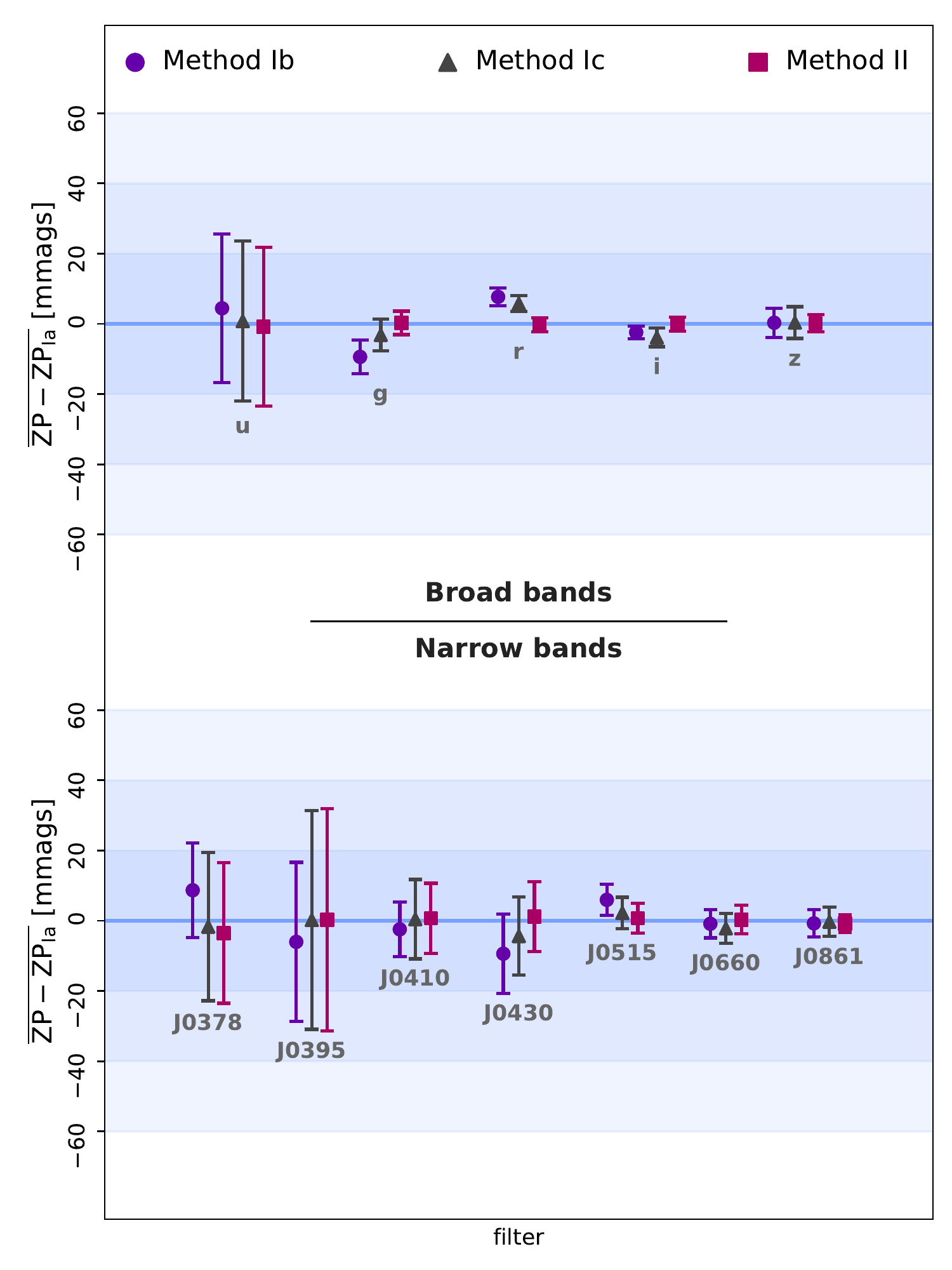}
\caption{\label{fig:DZP_refs} Comparison between the use of different references relative to the SDSS calibration of the STRIPE82 data. The compared ZPs correspond to the Final Zero-Points obtained for each calibration. The points represent the average differences between the ZP obtained from a given reference and that obtained from SDSS. The error bars correspond to the standard deviation of the differences. The three references considered were ATLAS$_\mathrm{RefCat}$/GALEX (purple circles), ATLAS$_\mathrm{RefCat}$/stellar locus technique (black triangles), and Skymapper (red squares).}
\label{fig:ref_comp_final}
\end{center}
\end{figure}

The comparison between the final ZPs of Method Ic and Method Ia is represented in Figure \ref{fig:ref_comp_final} as black triangles and also presented in Table \ref{tab:offsets}. In terms of the final offsets, Method Ic is slightly closer than Method Ib to the results of Method Ia. But since these offsets are smaller than 0.01 mag for all filters in both cases, and the scatter for filters J0378, J0395 and J0410 are significantly lower for Method Ib, in DR2 we chose to favour this method over Method Ic for the calibration. Nevertheless, we kept method Ic in our analysis as it shows that, once the offsets are corrected, the zero-points of most filters are barely affected by the choice of the reference catalogue, and the errors are still smaller than 0.04 mag in the worst cases. 

\begin{figure}
\begin{center}
\includegraphics[width=.48\textwidth]{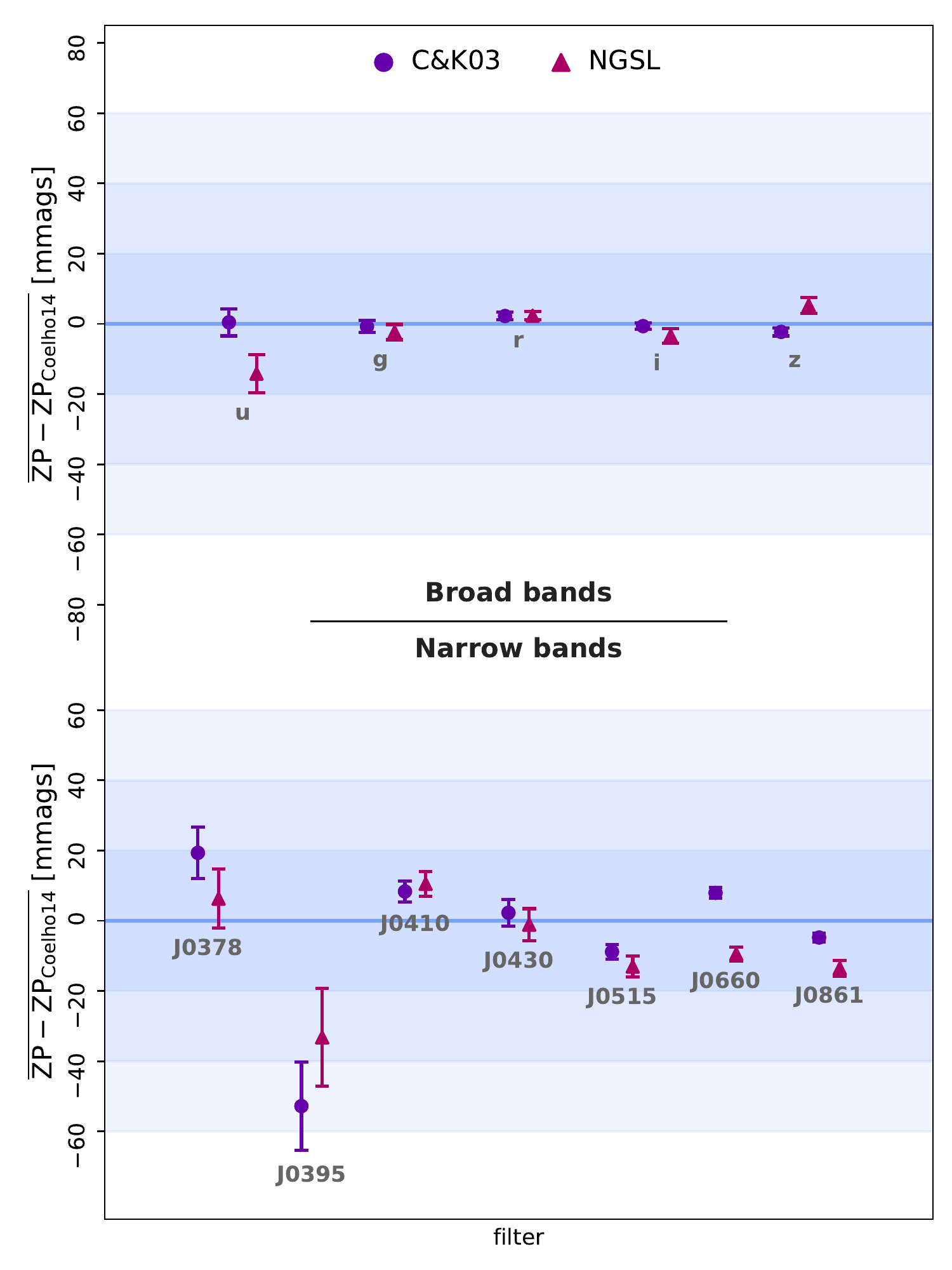}
\caption{\label{fig:DZP_models}. Analogous to Figure \ref{fig:DZP_refs} for the comparison between the use of different spectral libraries relative to the Coelho14 set. The two other libraries are the \citep{Castelli+Kurucz2003} synthetic models (C\&K03) and the NGSL empirical library}.
\label{fig:mod_comp_partial}
\end{center}
\end{figure}

\subsubsection{Method II - ATLAS Refcat2 and Stellar Locus}

Although the GALEX DR6/7 catalogue is almost all-sky, there are still some gaps in the footprint that cause some of the S-PLUS fields to have only a few dozens to none stars with measured NUV magnitudes. In these cases, we find that the ATLAS Refcat2 alone is not able to reliably constrain the models for the 5 bluer filters. This led us to include the intermediary stellar locus calibration between the external and internal calibration steps, as described in Section \ref{sec::Calibration_description}. We denominate this calibration strategy as Method II since it is the one that differs the most from the previous methods. It is important to emphasize that this is not a very common scenario, occurring only for 24 out of the 514 fields in DR2.

We compared Method II zero-points, obtained for the 170 STRIPE82 fields, to those of Method Ia to correct the offsets and characterise the uncertainties of the final zero-points. In this particular case, we find that the offsets correlate with the mean ISM extinction of the field:

\begin{equation}
    \delta_{ZP} = C + \alpha \, \mathrm{E}_{\mathrm{B-V}} \,.
\end{equation}
where the mean extinction is estimated from the average $E_{\mathrm{B-V}}$ of the best fit models for the stars used in the calibration.

This can be understood by the fact that we did not take into account the ISM extinction when comparing the stellar locus of the field to that of the reference Method Ia calibration, which are both obviously differently affected by the reddening. We still chose not to include the extinction corrections to keep the pipeline independent from any extinction maps. We argue that the correction does not seem to be necessary since, after the offset corrections, Method II already provides results that are similar to those obtained from Methods Ib and Ic. The offsets between Methods II and Ia, as a function of $E_\mathrm{B-V}$, are shown in Table \ref{tab:offsets}.

When comparing the final results (Table \ref{tab:ref_comp_final} and Figure \ref{fig:ref_comp_final}, magenta squares), Method II does a better job than Methods Ia and Ib in terms of the final offset against Method Ia but displays a similarly larger rms for filters J0378, J0395 and J0410 as Method Ic. Another advantage of Method II over Method Ic is that the ATLAS RefCat2, used by Method II, is the same one used in Method Ib, saving the job of obtaining and crossmatching an additional reference catalogue to the S-PLUS photometric catalogues.

Considering all the results in Figure \ref{fig:ref_comp_final} and Table \ref{tab:ref_comp_final}, when the application of more than one method is possible for a given DR2 field, we set the order of priorities to choose between the methods as: Method Ia $>$ Method Ib $>$ Method II $>$ Method Ic.

\subsection{Influence of chosen spectral library}
\label{sec:spectral_library_compaison}

We also investigate how the choice of the reference spectral library interferes with the results. In addition to the \citetalias{Coelho14} synthetic spectral library, we have calibrated the STRIPE82 region using the ATLAS9 synthetic models of \citetalias{Castelli+Kurucz2003} (not to be confused with the ATLAS reference catalogue) and the empirical Next Generation Spectral Library (NGSL, \citealp{Gregg+2006, Heap2007}).

The C\&K03 library consists of 3808 synthetic spectra that cover a similar region of the $T_\mathrm{eff}$ versus $\log g$ space as the 3727 Coelho14 models. One of the main differences is that the Coelho14 models extend only to $T_\mathrm{eff} < 30000$ K, while C\&K03 extends to $T_\mathrm{eff} < 50000$ K. Although, due to the rarity of these stars in contrast with the field dwarfs, the lack of a reference model in the spectral library to properly fit them does not interfere with the calibration. Another important difference is the coverage in composition space. The Coelho14 models cover 8 values for [Fe/H] between -1.3 and 0.2, while the C\&K03 models also cover 8 values of [Fe/H], but between -2.5 and 0.5. 

On the other hand, the NGSL library consists of 379 empirical spectra covering the UV and the optical, observed using the Hubble Space Telescope Imaging Spectrograph. The library is designed to be equally divided among four metallicity groups: very low ($\mathrm{[Fe/H]} < -1.5$), low ($-1.5 < \mathrm{[Fe/H]} < -0.5$), near-solar ($-0.5 < \mathrm{[Fe/H]} < 0.1$), and super-solar ($\mathrm{[Fe/H]} > 0.1$), sampling the entire HR diagram in each bin.

As we did for the Coelho14 models, we first introduced 20 different values of $E_\mathrm{B-V}$ extinction, between 0.025 and 1.00 mag, to the C\&K03 and NGSL libraries, following a \citet{Cardelli+1989} dust law. These models were then pre-convolved to the S-PLUS and SDSS filter systems before proceeding with the calibration. To evaluate the results, we once more rely on the comparison between different calibrations of the STRIPE82 region. All the calibrations were done using Method Ia, and the only changes between them are the spectral libraries themselves. 

We present the offsets and rms between the C\&K03 and NGSL and the Coelho14 ZPs in the right columns of Table \ref{tab:ref_comp_final} as well as in Figure \ref{fig:mod_comp_partial}, which is analogous to Figure \ref{fig:ref_comp_final}. The difference between this and the reference catalogue comparison is that in this case, we do not correct for any observed offsets, as our goal here is to compare the use of different models instead of correcting one calibration to the same scale as the other. Regarding the comparison between the Coelho14 and the C\&K03 models, we see that the broad-band ZPs are practically indistinguishable. We observe a small systematic offset for most narrow bands but almost no scatter, which can be explained by slight differences in the models at the key spectral features measured by these filters. For most cases, these differences are smaller than 0.01 mag, which is within the flux measurement errors. However, in the case of filters J0378 and J0395, the differences of 0.019 and -0.053 mag, respectively, are much more significant.

It is no surprise that the filter with the highest discrepancies happens to be the one that is the most sensitive to metallicity, given that the most significant difference between the two synthetic libraries is the way they cover the abundance space. At the time of DR2, it is still not perfectly clear if the finer grid in metallicity of Coelho14 or the larger interval in metallicity of C\&K03 is to be preferable to derive the ZPs for the filter J0395. We argue in favour of Coelho14 because stars as metal-poor as [Fe/H] = $-2.5$ are not abundant \citep{Beers+2005}, specially in the magnitude interval selected for the calibration. Also, since the $\chi^2$ minimisation employed in the model fitting does not take any priors into account, the number of stars that are assigned a metal-poor template is certainly overestimated in the C\&K03 fitting and are likely to bias the predicted values in the J0395 filter. Nevertheless, we call attention to the systematic 0.05 mag offset observed for this filter between the use of the two synthetic libraries and plan to revisit this issue in future data releases. 

In the case of the NGSL, the offsets are likely related to the small number of spectra in this empirical library: only 379, which is an order of magnitude lower than the theoretical models and consequently provides only limited coverage of the parameter's space. Given the limitations of these models to provide the correct template for a significant number of reference stars, it is quite surprising that the results are still very similar to those of the theoretical libraries, which highlights the robustness of the technique employed to characterise the zero-points.

\subsection{Internal consistency of the photometry}
\label{sec:internal_crossmatch}

We analysed the internal consistency of the pipeline by comparing the calibrated magnitudes of the same sources in different observations. This approach requires significant overlap between adjacent fields, otherwise only the sources observed in the borders of the CCD would be taken into account. It is also important to consider the sources observed across the whole surface of the CCD to properly evaluate the application of the ZP inhomogeneities correction maps~\ref{sec:CCD_correction}. To achieve this, we limited this analysis to a small region of the STRIPE82 where, by the design of the S-PLUS survey, the overlaps between adjacent fields are more significant ($\sim$50 per cent of the area). This region covers 47 fields and is highlighted in panel c of Figure~\ref{fig:footprint_reference}.

We used the \texttt{STILTS} \citep{Taylor2006} tool to do an internal crossmatch between these fields (with a maximum error of 1 arcsecond). We selected only the pairs observed with S/N between 100 and 1000 in the 3-arcsec diameter aperture and only included the sources with \texttt{SExtractor}'s PhotoFlag equal zero in the detection image photometry. This selection results in $\sim500-10000$ pairs to analyse, depending on the filter. 

\begin{figure}\textcolor{white}{[h!]}
\begin{center}
\includegraphics[width=.48\textwidth]{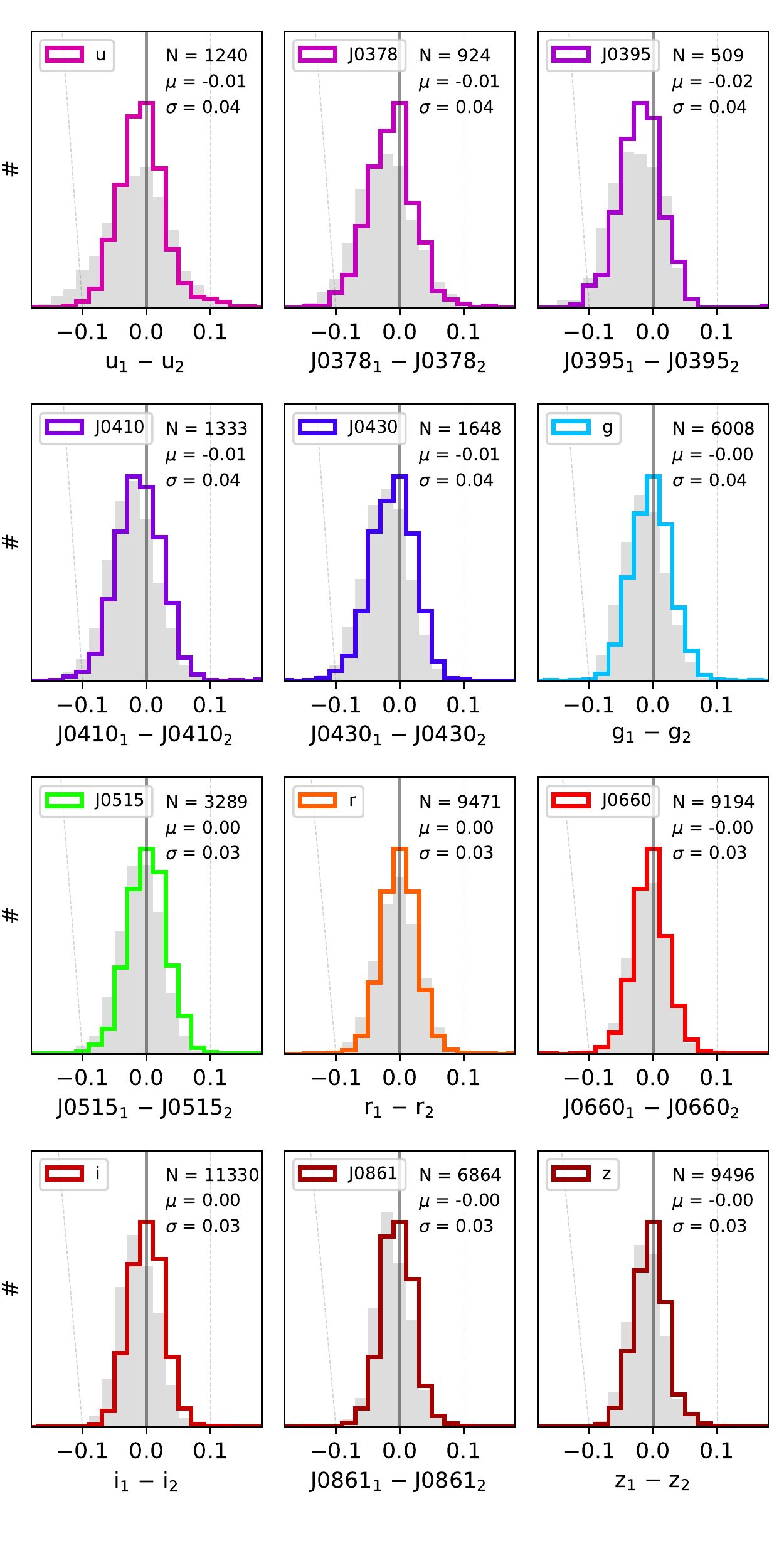}
\caption{\label{fig:hist_overlapping_fields} Internal consistency of the S-PLUS filters characterised by the differences between the final calibrated magnitudes of the same sources observed twice in adjacent fields. The coloured histograms correspond to the final magnitudes obtained by following all the steps described for the pipeline, while the grey histograms are obtained without taking into account the correction of the ZP variation across the field. We only consider for this analysis the sources with S/N > 200 and that are confined within the shaded region in Figure \ref{fig:footprint_reference}, panel c. The number of pairs, the average and the standard deviation for the coloured histograms are shown in the upper right corner of each plot.}
\end{center}
\end{figure}

In Figure \ref{fig:hist_overlapping_fields} we show the distribution of differences between the calibrated magnitudes of the selected sources observed in adjacent fields. The number of pairs in each sample, as well as the mean and standard deviation, are also shown in each panel. The grey solid histograms correspond to the calibration that was done without taking into account the correction maps. In contrast, the coloured histograms correspond to the calibration in which we applied the correction maps right after the photometry and before beginning the external calibration. Comparing both histograms in each panel, it is clear that the correction maps improve the internal consistency of the S-PLUS photometry for all filters. This is true even for the narrow bands, which were corrected using the correction map of the closest broad band (in terms of effective wavelength). 

The most significant differences are found for filter J0395 and indicate that one side of the CCD is systematically overestimating the ZPs by 0.02 mag in relation to the other. This is also the filter with the least number of stars available to estimate these differences. In the future, the increase of the number of S-PLUS fields with significant adjacent overlaps will provide enough reference stars for us to derive the correction maps directly from the S-PLUS data specifically for each filter, allowing us to improve this result in the next data-releases. Although it is important to note that, even in the case of J0395, the corrected magnitudes show a smaller offset and scatter in relation to the non-corrected case.

The internal comparison is consistent, in terms of scatter and offsets, with the external comparison with SDSS shown in Figure~\ref{fig:SPLUS_SDSS_comparison}. In this case, our analysis also includes the narrow bands. Overall, the average offsets are smaller than 0.01 mag for all the filters redder than J0430 and smaller than 0.02 mag for the remaining (except J0395). This analysis shows that the ZPs are correctly estimated for different observations and that it is homogeneous across the field after the correction maps have been applied.

\subsection{Expected uncertainty of photometry and zero-points}

The results discussed in this section allows us to characterise the expected uncertainties of the photometry (flux measurement + zero-points), as well as to isolate the zero-point uncertainties. The external comparison between S-PLUS and SDSS, as well as the internal comparison between adjacent S-PLUS fields, indicates that the upper limit (which ignores uncertainties in the reference catalogue) of the photometric errors for the bright stars (magnitude lower than 17 in the respective filters) is of 40 mmags for filters $u$, J0378, J0395, J0410, J0430 and $g$, and of 30 mmags for filters F515, $r$, J0660, $i$, J0861 and $z$. When considering only the ZPs, the comparison of the results obtained from the use of different reference catalogues and spectral libraries indicates that the errors are smaller than $\sim$10 mmags for filters J0410, J0430, $g$, J0515, $r$, J0660, $i$, J0861 and $z$, smaller than $\sim$15 mmags for filter J0378 ($\sim$20 mmag for Method II) and finally, smaller than $\sim$25 mmags for filters $u$ and J0395 ($\sim$30 mmags when using Method II).  The uncertainties are very similar to the results reported by \citetalias{LopesSanjuan+2019} for the J-PLUS calibration using a different technique.

\section{DR2 photometry characterisation}
\label{sec::DR2}

In this Section we present and characterise the photometry of the S-PLUS DR2. From now on, the results refer to all 514 DR2 fields,  calibrated using the technique described in Sections \ref{sec::Calibration_description} and \ref{sec::Calibration_validation}. This analysis aims to provide the necessary information to guide the user of the S-PLUS data for different scientific applications. We characterise the ZP distribution for each filter; estimate the photometric depths for different S/N thresholds; provide the average S/N for each filter as a function of the magnitude; and characterise the expected completeness of the observations in each filter, compared to the $r$ band, as a function of magnitude.

\subsection{Calibration Strategy}

\begin{figure}
\centering
\includegraphics[width=.45\textwidth]{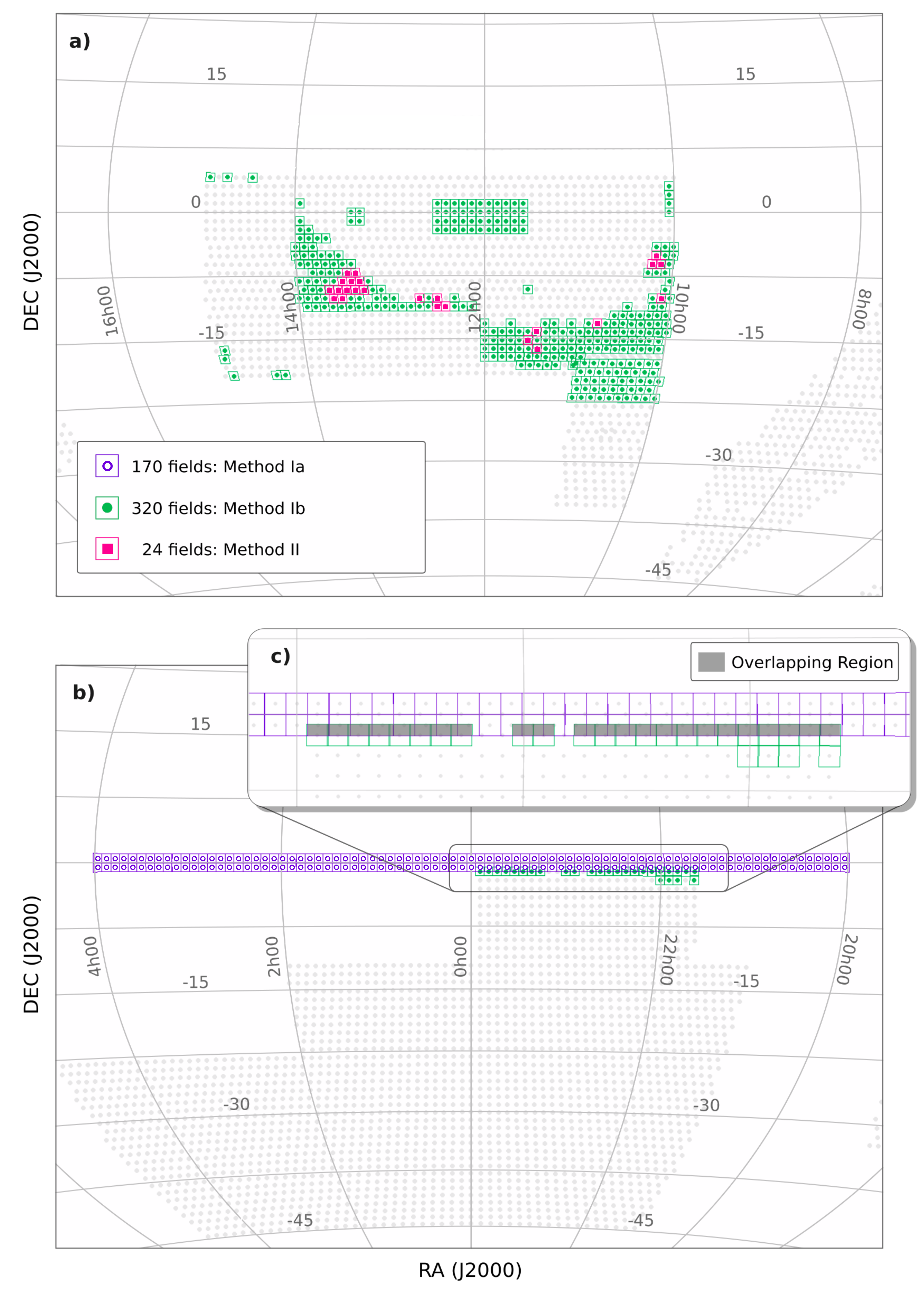}
\caption{The S-PLUS DR2 footprint coloured by the reference strategy used for the calibration of each field. The fields calibrated using SDSS as a reference  (Method Ia) are represented in purple, while those that use ATLAS$_RefCat$ + GALEX (Method Ib) are shown in green. The pink fields were calibrated using the ATLAS$_RefCat$ + stellar locus technique (Method II). Panel (a) is centred in the S-PLUS DR2 fields that are in the Northern Galactic Hemisphere, while Panel (b) is centred at those in the Southern Galactic Hemisphere. Panel (c) highlights the region with large overlaps between fields that was selected for the analysis of the internal consistency of the calibration.}
\label{fig:footprint_reference}
\end{figure}

As discussed in Section \ref{sec::Calibration_validation}, four different strategies were considered for the DR2 calibration: Methods Ia, Ib, Ic and Method II (see the details in Table \ref{tab:methods}). The most suitable method for each field was employed to ensure that the available resources are optimally used to provide the best possible accuracy and precision for that field. For instance, 170 fields (consisting of the STRIPE82 region) were calibrated using Method Ia, while Method Ib was used for 320 fields and Method II was used in the remaining 24 fields. In Figure \ref{fig:footprint_reference}, we present the DR2 footprint coloured according to the method used in the calibration. In particular, the fields calibrated using Method II are the ones whose number of reference stars with measured NUV magnitudes in the GALEX catalogue is smaller than 400.

\begin{figure}
\begin{center}
\includegraphics[width=.47\textwidth]{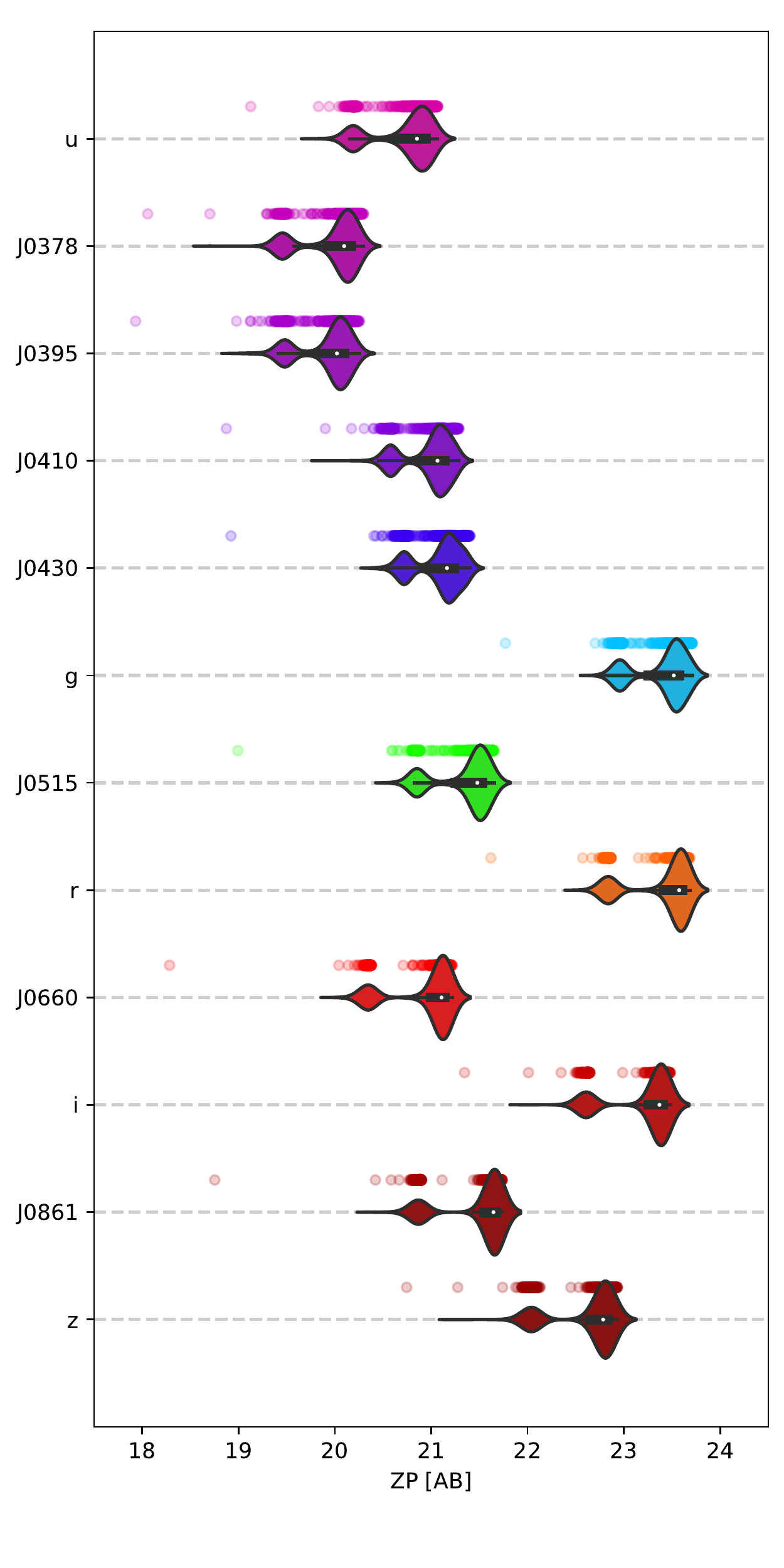}
\caption{\label{fig:ZP_dist} Distribution of the calibration ZPs of all DR2 observations for each filter. The bi-modal structure of the distributions is caused by a change in the observational mode of the camera at the end of 2016. The only significant outlier is field STRIPE82-0094, which stands out for being observed at worse sky transparency.}
\end{center}
\end{figure}

\subsection{Zero-point distribution}

The S-PLUS main survey observations are performed only during photometric nights, but even in this case, the range of the ZPs for each filter may exceed 1 magnitude. The main reason for this large spread was a change in the camera's observational mode near the end of 2016, which affects the gain of the CCD. This change explains the bi-modal structure observed in the ZP distribution of each filter, which we show in Figure \ref{fig:ZP_dist}. This effect is simply the result of the pipeline naturally finding the correct ZP for each specific gain.

It is important to mention that more than half of the overlapping pairs used in our analysis of the internal consistency of the pipeline (Section \ref{sec:internal_crossmatch}) consists of fields observed with different modes. The fact that the internal consistency of the photometry is equally robust under these circumstances allows us to conclude that the spread in the ZPs distribution is real and that the ZPs are correctly assigned. 
 
The clear outlier that can be observed in the distribution of all filters is the field STRIPE82-0094. This field was observed when the sky transparency was not as good as the others, and we plan to re-observe it in the future. Nevertheless, this field was kept in DR2 because the pipeline was able to correctly account for this difference in transparency and provide the proper ZPs for its calibration (although it is expected to have a much shallower photometric depth in comparison to the other DR2 fields).

\begin{figure}
\begin{center}
\includegraphics[width=.47\textwidth]{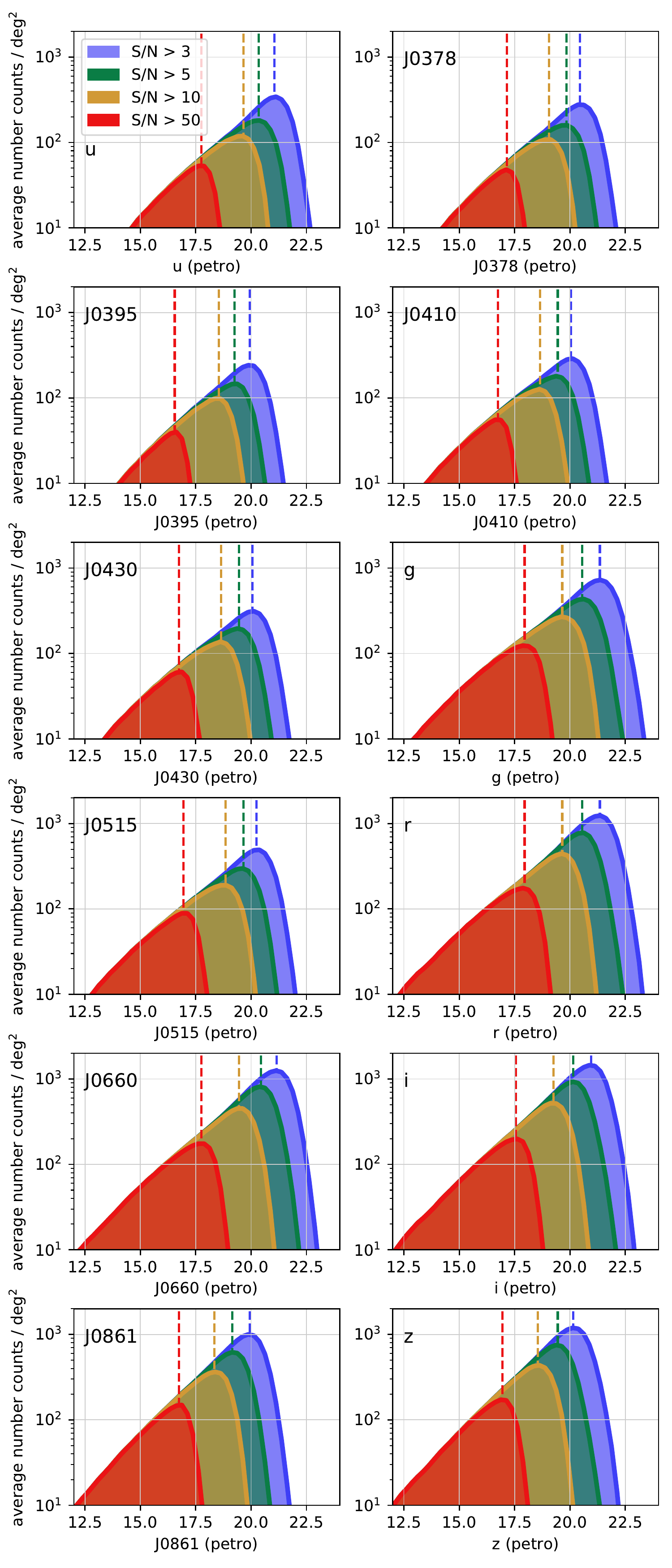}
\caption{\label{fig:depth} Average distribution of Petrosian magnitudes per square degrees for different S/N thresholds (S/N > 3, blue; S/N > 5, green; S/N > 10, yellow; S/N > 50, red) for each filter. The whole DR2 data is taken into account. The photometric depths at each S/N threshold are characterised as the peak of each distribution and are represented by the respective coloured dashed lines.}
\end{center}
\end{figure}

\subsection{Depth}

The characterisation of the photometric depth of our images is important to understand the selection effects present in our catalogues and to help ascertain if a particular science is possible to be done using the limitations of the data. We have used the whole dataset of the S-PLUS DR2 to do a filter-by-filter analysis of the photometric depths. By design, different filters will have different photometric depths. Filter J0660, for example, reaches much deeper magnitudes in comparison to the other narrow bands due to a larger exposure time (see Table 5 of \citealp{MendesDeOliveira+2019}), which was chosen to allow S-PLUS to produce H$\alpha$ emission line maps.  Nevertheless, except for the variations in the gain, the S-PLUS Main Survey observations are all done under similar photometric conditions. We do not expect significant variations of photometric depths for different pointings of the same filter.

In Figure \ref{fig:depth} we show, for each filter, the average number counts per square degree in magnitude bins of 0.25 mag for four different S/N thresholds (3, 5, 10, 50). The Petrosian aperture was chosen for this analysis because it is a measurement of the total magnitude for both point and extended faint sources. The smoothness of the curves is a consequence of averaging the 514 fields and highlights the uniformity of the DR2 photometry.

\begin{table}
\caption{\label{tab:depth} Average photometric depths of each filter for the whole DR2 data for different S/N thresholds (S/N > 3, S/N > 5, S/N > 10 and S/N > 50). The depths are characterised as the magnitude of the peak of the magnitude distribution (i.e. the turnover point or the derivative equals zero) for each S/N threshold.}
\centering
\begin{tabular}{lcccc}
\hline \hline
filter & S/N > 50 & S/N > 10 & S/N > 5 & S/N > 3 \\ \hline
$u$ & 17.7 & 19.6 & 20.3 & 21.0 \\
J0378 & 17.1 & 19.0 & 19.8 & 20.4 \\
J0395 & 16.5 & 18.5 & 19.2 & 19.9 \\
J0410 & 16.7 & 18.6 & 19.4 & 20.0 \\
J0430 & 16.7 & 18.6 & 19.4 & 20.0 \\
$g$ & 17.9 & 19.6 & 20.5 & 21.3 \\
J0515 & 16.9 & 18.8 & 19.6 & 20.2 \\
$r$ & 17.9 & 19.6 & 20.5 & 21.3 \\
J0660 & 17.7 & 19.4 & 20.4 & 21.1 \\
$i$ & 17.5 & 19.2 & 20.1 & 20.9 \\
J0861 & 16.7 & 18.3 & 19.1 & 19.9 \\
$z$ & 16.9 & 18.5 & 19.4 & 20.1 \\
\hline \hline
\end{tabular}
\end{table}

\begin{figure}\textcolor{white}{[h!]}
\begin{center}
\includegraphics[width=.47\textwidth]{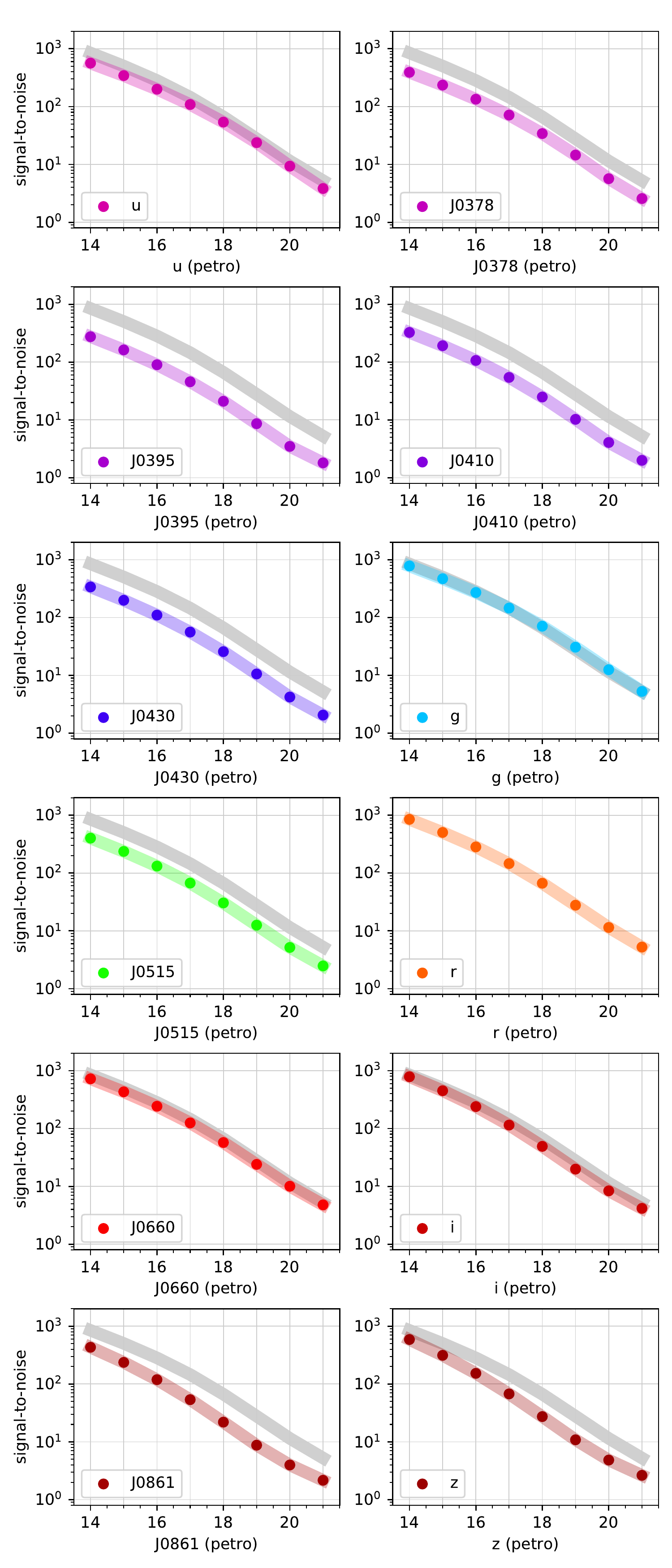}
\caption{\label{fig:s2n} Average S/N at different bins of Petrosian magnitude for each filter, where the whole DR2 data is taken into account. The $r$ filter, which has the overall highest S/N at all magnitudes (together with $g$ and J0660), is represented in all panels as a grey line for comparison.}
\end{center}
\end{figure}

\begin{figure}\textcolor{white}{[h!]}
\begin{center}
\includegraphics[width=.47\textwidth]{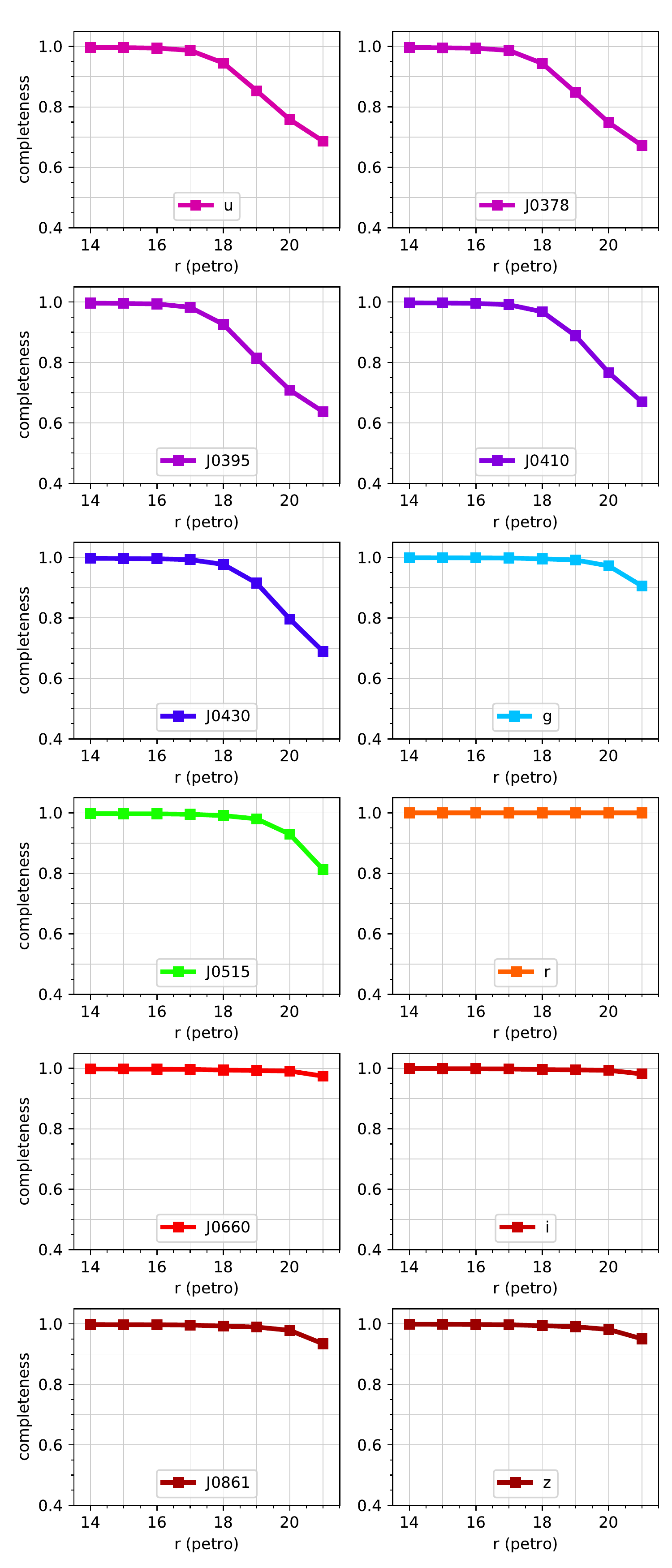}
\caption{\label{fig:completeness} Fraction of the observations in each filter, in relation to the number of observations in the $r$ band, for different bins of $r$ Petrosian magnitudes. The $r$ band was chosen as the reference because it is the filter with the highest number of detections in DR2. A source is considered observed in a given filter if \texttt{SExtractor} attributed a value other than 99 for this filter (i.e. it has a S/N higher than 1.1 in the respective filter).}
\end{center}
\end{figure}

We characterise the photometric depths at each S/N threshold as the magnitude corresponding to the peak of the distribution (i.e., the derivative is zero), at which detections start to rapidly decline. This is done by producing a histogram with a bin size of 0.1 magnitudes and finding the bin with the highest number of counts. The calculated depths for each filter and S/N threshold are represented by the dashed lines in Figure \ref{fig:depth} and are summarised in Table \ref{tab:depth}. In general, filters $g$, $r$, and J0660 have deeper photometric depths, while the narrow bands J0395, J0410, J0430 and J0861 are shallower.

\subsection{Signal-to-noise}

Another important relation that we analyse is how the average signal-to-noise correlates to the magnitude in each filter. This relation allows the users of the S-PLUS data to evaluate if a given source is expected to have a desirable signal-to-noise given its magnitude in each filter.

As we did for the photometric depths, the whole DR2 data was taken into account. In this case, the data was divided into bins of 1~magnitude. In Figure \ref{fig:s2n}, we show the average signal-to-noise as a function of magnitude for each filter. The grey line corresponds to the result obtained for the $r$ band and is shown in each panel for comparison. 

The filters with higher S/N per magnitude bin are again filters $g$, $r$ and J0660, while the lower S/N is found for filters J0395, J0410, J0430 and J0861, which is also a reflection of the designed exposure times. Among the broad bands, the $z$ band is the one that presents the lowest S/N per bin of magnitude.

\subsection{Completeness in relation to the r-band}

Finally, we evaluate the completeness of each filter in relation to the $r$ band for different bins of magnitudes. In other words, we calculate the fraction of sources that are observed in a particular bin of $r$ magnitude that are also observed in each of the other filters. The $r$ band was chosen as the reference because it is the filter with the highest number of detections. This is another important relation to characterise the selection effects present in DR2. In particular, this relation must be taken into account when working with simulated S-PLUS data, for example. 

In Figure \ref{fig:completeness}, we show the average completeness for the 12 bands in relation to the $r$ filter in bins of 1 magnitude. The whole DR2 data was taken into account. An important conclusion that can be drawn from this analysis is that for $r > 18$~mag, an increasing fraction of the sources is not detected in the blue filters ($u$, J0378, J0395, J0410 and J0430). This is an important selection effect that must be considered by all studies interested in objects in this magnitude range. This effect is negligible for the remaining red bands, except for $g$, which does not detect $\sim 10$ per cent of the sources for $r \sim 21$~mag, and for J0515 which does not detect around 20 per cent of the sources in this bin. It is important to remember that the detection image used in the photometry consists of a weighted sum of the $g$, $r$, $i$ and $z$ images and is likely biased towards the detection of red sources and partially responsible for the aforementioned effect.

\section{Catalogues and data access}
\label{sec::Catalogs}

\subsection{Column descriptions}

In this section we provide a short description of each column present in the DR2 catalogues. For simplicity, we divide the columns, shown in Table \ref{tab:catalog_columns}, in 4 groups: designation and astrometry, morphology, flags and classification, and photometry. In some cases, the represented column name contains the placeholders \{filter\} to indicate that the column is present for any filter: \texttt{u}, \texttt{g}, \texttt{r}, \texttt{i}, \texttt{z}, \texttt{J0378}, \texttt{J0395}, \texttt{J0410}, \texttt{J0430}, \texttt{J0515}, \texttt{J0660} and \texttt{J0861}; and \{aperture\}, which represents one of the possible six apertures: \texttt{auto}, \texttt{petro}, \texttt{iso}, \texttt{aper\_3}, \texttt{aper\_6} and \texttt{PStotal}.

\subsubsection{Designation and Astrometry}

The first two columns of the catalogues refer to the identification of the sources: \texttt{Field} is the name of the S-PLUS field corresponding to the observation, while \texttt{ID} corresponds to an observation identifier attributed for this target in DR2. This \texttt{ID} has no relation to the DR1 \texttt{ID}. The same source may be observed in two adjacent fields and may be associated with multiple entries in the catalogue with different IDs (and necessarily different Field column value as well).

The \texttt{RA} and \texttt{DEC} positions are in the J2000 epoch and are given in degrees. We also provide the \texttt{X}, \texttt{Y} position of the target in the reduced images. These positions do not translate to the exact physical pixels in the CCD because the images are re-aligned during the co-addition process in the reduction pipeline.

\subsubsection{Morphology}

The parameters describing the morphology of the sources are taken from the \texttt{SExtractor}'s detection image catalogue and include the columns \texttt{ISOarea}, \texttt{MU\_MAX}, \texttt{A}, \texttt{B}, \texttt{THETA}, \texttt{ELLONGATION}, \texttt{ELLIPTICITY}, \texttt{FLUX\_RADIUS}, and \texttt{KRON\_radius}, as well as the measured FWHM assuming a Gaussian core, both for the detection image (\texttt{FWHM}) and for each of the 12 filters (\texttt{FWHM}\_\{filter\}). Finally, \texttt{FWHM\_n} is obtained by dividing the column \texttt{FWHM} by the average FWHM of a selection of only stars with \texttt{CLASS\_STAR} $>$ 0.9 and S/N between 100 and 1000 in the auto aperture of the detection image.

\subsubsection{Flags and Classification}

This category includes a flag indicating the method used for the calibration, the photometric quality flags of \texttt{SExtractor}, as well as \texttt{SExtractor}'s star classification parameter. The \texttt{calibration\_flag} in DR2 contains 6 flag bits that indicate the reference catalogues employed for the calibration of the field and informs if the stellar locus step was necessary for the fields. The values are:
\begin{align} 
1&: \mathrm{use\,\,of\,\,Stellar\,\,Locus} \nonumber \\ 
2&: \mathrm{SDSS\,\,(DR12)} \nonumber \\ 
4&: \mathrm{SDSS\,\,(Ivezic07)} \nonumber \\ 
8&: \mathrm{ATLAS\,\,RefCat2} \nonumber \\ 
16&: \mathrm{GALEX\,\,(DR6/7)}  \nonumber \\ 
32&: \mathrm{Skymapper\,\,(DR2)}. \nonumber
\end{align}

The actual value of the flag is the sum of powers of 2 coded in decimal, just like the bitmasks used in SDSS. In terms of the nomenclature used for the different methods in this paper, Method Ia can be indicated as flag 2 or 4 (depending on the origin of the SDSS data), Method Ib has flag 24, Method Ic has flag 32, and Method II has a calibration flag of 9.

The columns \texttt{PhotoFlag\_\{filter\}} for each filter correspond to \texttt{SExtractor} \texttt{FLAGS} column, which indicates, among other things, the possible contamination of neighbouring sources, the existence of saturated pixels and truncated objects. We direct the user to the \texttt{SExtractor} documentation for more information. In particular, we suggest the selection of \texttt{PhotoFlag\_\{filter\}} = 0 or 2 (isolated or deblended target with no reported problems) for the selection of targets with good photometry in a particular filter. The \texttt{FLAGS} column of the detection image is also included as \texttt{PhotoFlagDet}.

Finally, the \texttt{CLASS\_STAR}\_\{filter\} column is a value between 0 and 1 representing how likely a given source is to be a point source based on its PSF morphology in that particular filter. The column \texttt{CLASS\_STAR} represents the value corresponding to this classification in the detection image.

\subsubsection{Photometry}

Our catalogues include the magnitudes measured in six different apertures for each filter. We refer to the \texttt{SExtractor} documentation for details about the two variable elliptical apertures (\texttt{auto} and \texttt{petro}) and the isophotal aperture (\texttt{iso}). Apertures \texttt{aper\_3} and \texttt{aper\_6} correspond to the fixed circular apertures with diameters of 3 and 6 arcsec. Finally, we also include the aperture denominated \texttt{PStotal}, which is the circular aperture of 3 arcsec that is corrected for the fraction of the flux that falls outside of this diameter and therefore corresponds to the best representation of the total magnitude of point sources.

The instrumental magnitudes in all different apertures are calibrated from the ZPs obtained using the \texttt{PStotal} magnitudes and are in the AB system. Table \ref{tab:catalog_columns} describes the photometric columns in the catalogues. The column \{filter\}\_\{aperture\} corresponds to the AB magnitude measured for each filter in the given aperture. The errors (\texttt{e}\_\{filter\}\_\{aperture\}) are estimated by \texttt{SExtractor} and the S/N (\texttt{s2n}\_\{filter\}\_\{aperture\}) is defined as the flux divided by its error. For non-detection in a given filter, the magnitude is replaced by 99, the error by the $2\sigma$ upper limit in the field and the S/N is set to -1. The column \texttt{nDet}\_\{aperture\} indicates the number of filters in which the source was detected for each particular aperture. 

\begin{table}\textcolor{white}{[h!]}
\caption{Column names, description and units of the DR2 catalogues. The placeholder \{filter\} indicates that the column is present for any filter: \texttt{u}, \texttt{g}, \texttt{r}, \texttt{i}, \texttt{z}, \texttt{J0378}, \texttt{J0395}, \texttt{J0410}, \texttt{J0430}, \texttt{J0515}, \texttt{J0660} and \texttt{J0861}; while \{aperture\} represents one of the possible six apertures: \texttt{auto}, \texttt{petro}, \texttt{iso}, \texttt{aper\_3}, \texttt{aper\_6} and \texttt{PStotal}.}
\centering
\label{tab:catalog_columns}
\begin{tabular}{rll}
\hline \hline

\multicolumn{3}{c}{Identification and Position} \\ \hline

\texttt{\textbf{Field}}    & Name of the S-PLUS field & \\
                           & of the observation       & \\
\texttt{\textbf{ID}}       & Observation ID in DR2    &\\
\texttt{\textbf{RA}}       & Right Ascension (J2000)  & [deg] \\
\texttt{\textbf{DEC}}      & Declination (J2000)      & [deg] \\
\texttt{\textbf{X}}        & CCD X-axis position      & \\
                           & (reduced image)          & [pixel]\\
\texttt{\textbf{Y}}        & CCD Y-axis position      & \\
                           & (reduced image)          & [pixel]\\
\hline \hline

\multicolumn{3}{c}{Morphology} \\ \hline

\texttt{\textbf{ISOarea}}          & Isophotal area above    & \\ 
                                   & 1.1 sigma threshold     & \\
\texttt{\textbf{MU\_MAX}}          & Peak surface brightness & [mag/arcsec$^2$] \\
                                   & above background        & \\
\texttt{\textbf{A}}                & Isophotal image         & [pixel] \\
                                   & major axis              & \\
\texttt{\textbf{B}}                & Isophotal image         & [pixel] \\
                                   & minor axis              & \\
\texttt{\textbf{THETA}}            & Isophotal image         & [deg] \\
                                   & position angle          & \\
\texttt{\textbf{ELONGATION}}       & A/B                     & \\
\texttt{\textbf{ELLIPTICITY}}      & 1 - B/A                 & \\
\texttt{\textbf{FLUX\_RADIUS}}     & Radius containing       & [pixel] \\
                                   & (0.2,0.5,0.7,0.9)       & \\
                                   & fraction of the light   & \\
\texttt{\textbf{KRON\_RADIUS}}     & Kron apertures          & \\
                                   & in units of A or B      & \\
\texttt{\textbf{FWHM}}             & FWHM assuming a         & [pixel] \\
                                   & Gaussian core           & \\
\texttt{\textbf{FWHM\_n}}          & Normalized FWHM         & \\
\textbf{\texttt{FWHM}\_\{filter\}} & FWHM at each filter     & [pixel] \\
\hline \hline

\multicolumn{3}{c}{Flags and classification Columns} \\ \hline

\textbf{\texttt{calibration\_flag}}       & Indication of the       & \\
                                          & reference catalogue       & \\
                                          & for calibration         & \\
\textbf{\texttt{PhotoFlag}\_\{filter\}}   & SExtractor 'FLAGS'      & \\
\textbf{\texttt{PhotoFlagDet}}            & SExtractor 'FLAGS'      & \\
                                          & in the  Detection image & \\
\textbf{\texttt{CLASS\_STAR}\_\{filter\}} & Star classification     & \\
\textbf{\texttt{CLASS\_STAR}}             & Star classification     & \\
                                          & in the Detection image  & \\
\hline \hline

\multicolumn{3}{c}{Photometry Columns} \\ \hline

\textbf{\{filter\}\_\{aperture\}}               & AB magnitude            & \\
\textbf{\texttt{e}\_\{filter\}\_\{aperture\}}   & Magnitude error         & \\
\textbf{\texttt{s2n}\_\{filter\}\_\{aperture\}} & Source's S/N            & \\
\textbf{\texttt{nDet}\_\{aperture\}}            & Number of detections    & \\
\textbf{\texttt{s2n\_Det}\_\{aperture\}}        & Source's S/N            & \\ 
                                                & in the detection image  & \\

\end{tabular}
\end{table}

\subsection{Bright Star Masks}
The presence of bright stars affects the S-PLUS images in a number of ways that may be relevant, e.g. bright stars are saturated themselves and thus not useful; they increase the background level over a large area, and noise peaks in this background can be misclassified as real objects; they also generate diffraction spikes that can be detected as sources. Bright stars block out a relatively large area on the CCD, thereby significantly affecting the effective survey area within the observation tiles.

In order to quantify the impact of these bright stars on the surrounding object number counts, we performed a radial number count analysis around these objects. Because we are working with bright stars that are often saturated in the S-PLUS data, we used the Guide Star Catalog version $1.2$ \citep[GSC 1.2, ][]{Morrison+2001} as a reference for the locations of bright stars and their magnitudes. This allowed us to recover the maximum radius of effect of these stars and, therefore, the area that should be masked around them. Further details about the masking process will be presented in a forthcoming paper.

\subsection{Star/Galaxy/Quasar classification}
The S-PLUS DR2 is provided with class labels (0 = quasar, 1 = star, 2 = galaxy) for all sources, along with probabilities. The classification is done using a Random Forest algorithm trained with the 12 S-PLUS bands, \texttt{FWHM\_n}, \texttt{A}, \texttt{B}, and \texttt{KRON\_RADIUS}. For sources with the WISE counterpart, the algorithm was trained with the same features along with W1 and W2 from WISE, improving purity and completeness for all classes. Details about the classification study and their performance metrics can be found in Nakazono et al. (submitted).

\subsection{Photo-zs}

\begin{figure*}
\begin{center}
\includegraphics[width=\textwidth]{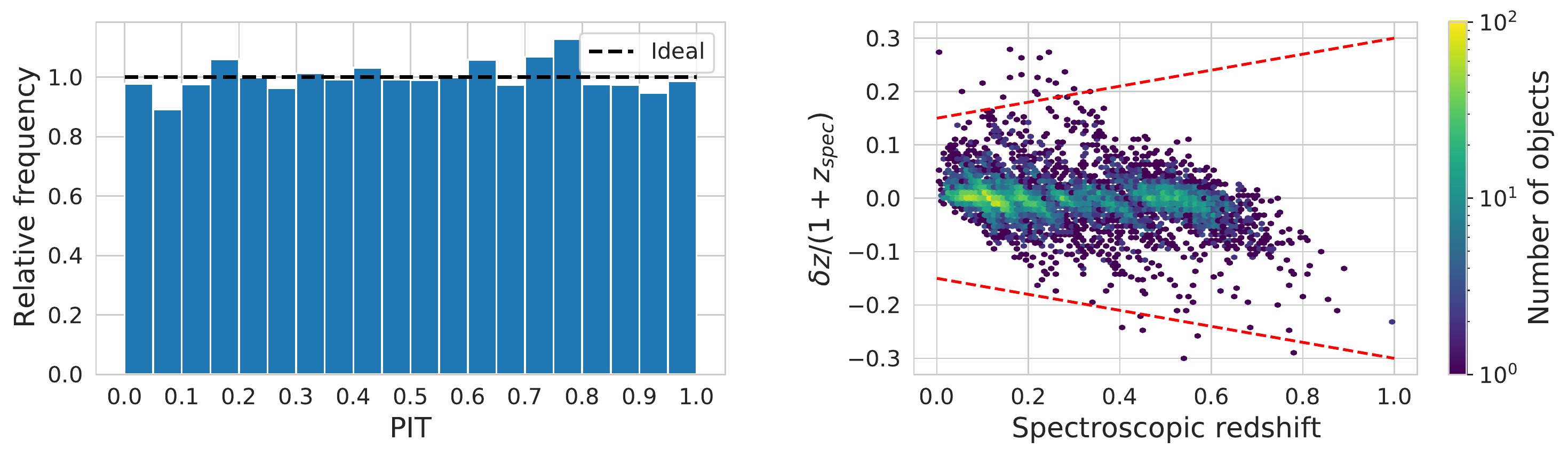}
\caption{\label{fig:photo-zs} Photometric redshift results for the testing sample. (Left) Probability Integral Transform with a nearly uniform distribution, indicating well-calibrated PDFs. (Right) Point-like density scatter plot for the photometric redshift obtained, where $\delta z = z_\text{spec} - z_\text{spec}$ and the red dashed lines represents the catastrophic errors with  $\delta z > 0.15(1+z_\text{spec})$. The colour scale represents the number of objects for each point.}
\end{center}
\end{figure*}

DR2 provides photometric redshifts and their respective probability distribution functions (PDFs)  obtained via deep-learning, using a Mixture Density Network \citep[Vinicius-Lima et al., ][]{bishop1994mixture}.

The network is trained using the 12-band photometry from S-PLUS, cross-matched with the unWISE catalogue \citep[][for the W1 and W2 magnitudes]{schlafly2019unwise}, the 2MASS catalogue \citep[][J, H, and K magnitudes]{cutri20032mass}, and the GALEX catalogue \citep[][FUV and NUV magnitudes]{bianchi2017galex}. The targets are SDSS DR16 spectroscopic redshifts \citep[see, for example,][]{du2020completed}; a total of 105830 spec-z were used as a calibration set. This sample is split randomly into training/validation and testing samples, containing 70\% and 30\% of the total number of objects, respectively. During the training process, the  training/validation sample is divided into 75\% for training and 25\% for validation using 4-fold cross-validation.

The combination of broad- and narrow-band photometry provided accurate photometric redshifts and well-calibrated PDFs, with negligible bias and outlier fractions. Indeed, in the magnitude interval 14 $\leqslant$ \texttt{r\_aper\_6} $\leqslant$ 21, the median normalized bias is $-0.0015$, the scatter ($\sigma_\text{NMAD}$) is 0.02, and the fraction of outliers is 0.91\% (see figure \ref{fig:photo-zs}). Further details about the method and obtained results can be found in Vinicius-Lima et al. (submitted).

\subsection{Data access}
The S-PLUS web server allows different ways to access the data, given that it is based on API endpoints. The main interface that interacts with the server is the website \href{https://splus.cloud}{https://splus.cloud}. In addition, we provide a python package called \texttt{splusdata}\footnote{\href{https://github.com/Schwarzam/splusdata}{https://github.com/Schwarzam/splusdata}}, referred to in the same website, which performs the same tasks as the web interface. The description on how to use these tools is available at the documentation section in the \href{https://splus.cloud}{splus.cloud} website.

\section{Summary}
\label{sec::Summary}

We developed a new pipeline for the photometric calibration of S-PLUS, which is also suitable for any other wide-field multi-filter survey. We performed several different calibrations of the STRIPE82 region in order to validate the different calibration strategies adopted in S-PLUS DR2. Even though we had to correct for systematic offsets found between the different methods, the final scatter shows an excellent agreement between the different strategies, as long as the offsets are taken into account. By adopting the SDSS strategy as the reference calibration of S-PLUS, we find that the expected ZP errors are $\lesssim$~10~mmags for the filters: J0410, J0430, $g$, J0515, $r$, J0660, $i$, J0861 and $z$; $\lesssim$~15~mmags for filter J0378 and $\lesssim$~25~mmags for filters $u$ and J0395.

We also find good agreement between the calibrated magnitudes of the same sources observed in adjacent overlapping fields, especially when we take into account the correction of ZP inhomogeneities across the CCD through previously obtained correction maps. This internal comparison, as well as the external comparison against SDSS, provide an upper limit for the photometric errors of 0.04 mag for the blue bands and 0.03 mag for the red bands. In most cases, except for J0395, the offsets are smaller than 0.01 mag. Filter J0395 presents a particularly challenging calibration, with an average internal systematic offset of 0.02 mag. Additional S-PLUS data are needed, particularly in fields with large overlaps to allow for the estimation of correction maps directly from S-PLUS.

We tested the dependency of the pipeline on the choice of the spectral library that provides the stellar templates by also re-calibrating the STRIPE82 region using different sources for our models. In particular, we tested the libraries of \citet{Castelli+Kurucz2003} and NGSL \citep{Heap2007}, in addition to the adopted \citet{Coelho14} models. The only significant differences were found for the J0395 filter (displaying an offset of -0.053 mag), and to a lesser extent, for J0378 (whose offset is 0.019 mag). We believe that the differences arise from the dependency of these filters on metallicity, which is differently covered in the grid of both spectral libraries. We argue in favour of Coelho14 models since we believe the lower metallicities present in \citet{Castelli+Kurucz2003} are causing our pipeline to systematically underestimate metallicities. Nevertheless, future research is needed to ensure that the best models are employed in the next data releases.

The new pipeline was applied to 514 S-PLUS fields. The measured calibrated magnitudes in six different apertures, together with astrometry and other photometric parameters, are also published here and constitute the Second Data Release of S-PLUS.  We show that three of the four different strategies discussed in this paper had to be employed in the calibration of DR2 depending on the region of the sky. DR2 includes the 170 STRIPE82 fields of DR1 and 344 new fields, which increases the sky coverage of the published S-PLUS data by a factor of three.

\section*{Acknowledgments}

The S-PLUS project, including the T80-South robotic telescope and the S-PLUS scientific survey, was founded as a partnership between the Funda\c{c}\~{a}o de Amparo \`{a} Pesquisa do Estado de S\~{a}o Paulo (FAPESP), the Observat\'{o}rio Nacional (ON), the Federal University of Sergipe (UFS), and the Federal University of Santa Catarina (UFSC), with important financial and practical contributions from other collaborating institutes in Brazil, Chile (Universidad de La Serena), and Spain (Centro de Estudios de F\'{\i}sica del Cosmos de Arag\'{o}n, CEFCA). We further acknowledge financial support from the São Paulo Research Foundation (FAPESP), the Brazilian National Research Council (CNPq), the Coordination for the Improvement of Higher Education Personnel (CAPES), the Carlos Chagas Filho Rio de Janeiro State Research Foundation (FAPERJ), and the Brazilian Innovation Agency (FINEP).

The authors who are members of the S-PLUS collaboration are grateful for the contributions from CTIO staff in helping in the construction, commissioning and maintenance of the T80-South telescope and camera. We are also indebted to Rene Laporte and INPE, as well as Keith Taylor, for their important contributions to the project. From CEFCA, we particularly would like to thank Antonio Mar\'{i}n-Franch for his invaluable contributions in the early phases of the project, David Crist{\'o}bal-Hornillos and his team for their help with the installation of the data reduction package \textsc{jype} version 0.9.9, C\'{e}sar \'{I}\~{n}iguez for providing 2D measurements of the filter transmissions, and all other staff members for their support with various aspects of the project.

FA-F, FRH, CEB, MLB, LMN, HDP, LAG-S, TS-S acknowledge funding for this work from FAPESP grants 2009/54202-8, 2011/51680-6, 2014/10566-4, 2016/12331-0, 2018/06822-6, 2018/09165-6, 2018/20977-2, 2018/21250-9, 2018/21661-9, 2019/01312-2, 2019/10923-5, 2019/11910-4, 2019/23388-0, 2019/26412-0, 2019/26492-3. EVRL, CMdO and LSJ acknowledge funding for this work from CNPq grants 169181/2017-0, 309209/2019-6 and 115795/2020-0. EVRL acknowledges funding for this work from CAPES grant 88887.470064/2019-00. GOS acknowledges a PIBITI grant. The work of V.M.P. is supported by NOIRLab, which is managed by the Association of Universities for Research in Astronomy (AURA) under a cooperative agreement with the National Science Foundation. J.A.-G. acknowledges support from Fondecyt Regular 1201490 and from ANID – Millennium Science Initiative Program – ICN12\_009 awarded to the Millennium Institute of Astrophysics MAS.

The authors  would like to thank Ulisses Manzo Castello, Marco Antonio dos Santos and Luis Ricardo Manrique for all the support in infrastructure matters. The authors would like to thank Stavros Akras, Gustavo Luis Baume, Alvaro Alvarez-Candal, Amanda Reis Lopes, Paulo Lopes, Carolina Queiroz and Kanak Saha for useful suggestions and comments.

The authors made use and acknowledge TOPCAT \footnote{\url{http://www.starlink.ac.uk/topcat/ (TOPCAT)}} tool to analyse the data.



\bibliographystyle{mnras}
\bibliography{bibliography} 




\appendix

\section{Characterization of zero-points}
\label{ap:zp_characterization}

We used simulated data, following the same usual distribution of the instrumental magnitudes in each filter, to find the best way to characterise the ZPs from the comparison between the predicted and the instrumental magnitudes. We were particularly interested in finding the best estimator that could still provide the correct ZP even when a significant number of outliers is introduced. These outliers represent stars that may have a misattributed stellar template during the model fitting, which can show systematic trends if some specific stellar property is being substantially favoured. This is the case, for example, for the high number of stars that are fitted by a giant stellar template when the reference catalogue only has broad bands, which systematically underestimates the magnitudes of filters J0378 and J0515. 

\begin{figure}
\begin{center}
\includegraphics[width=.4\textwidth]{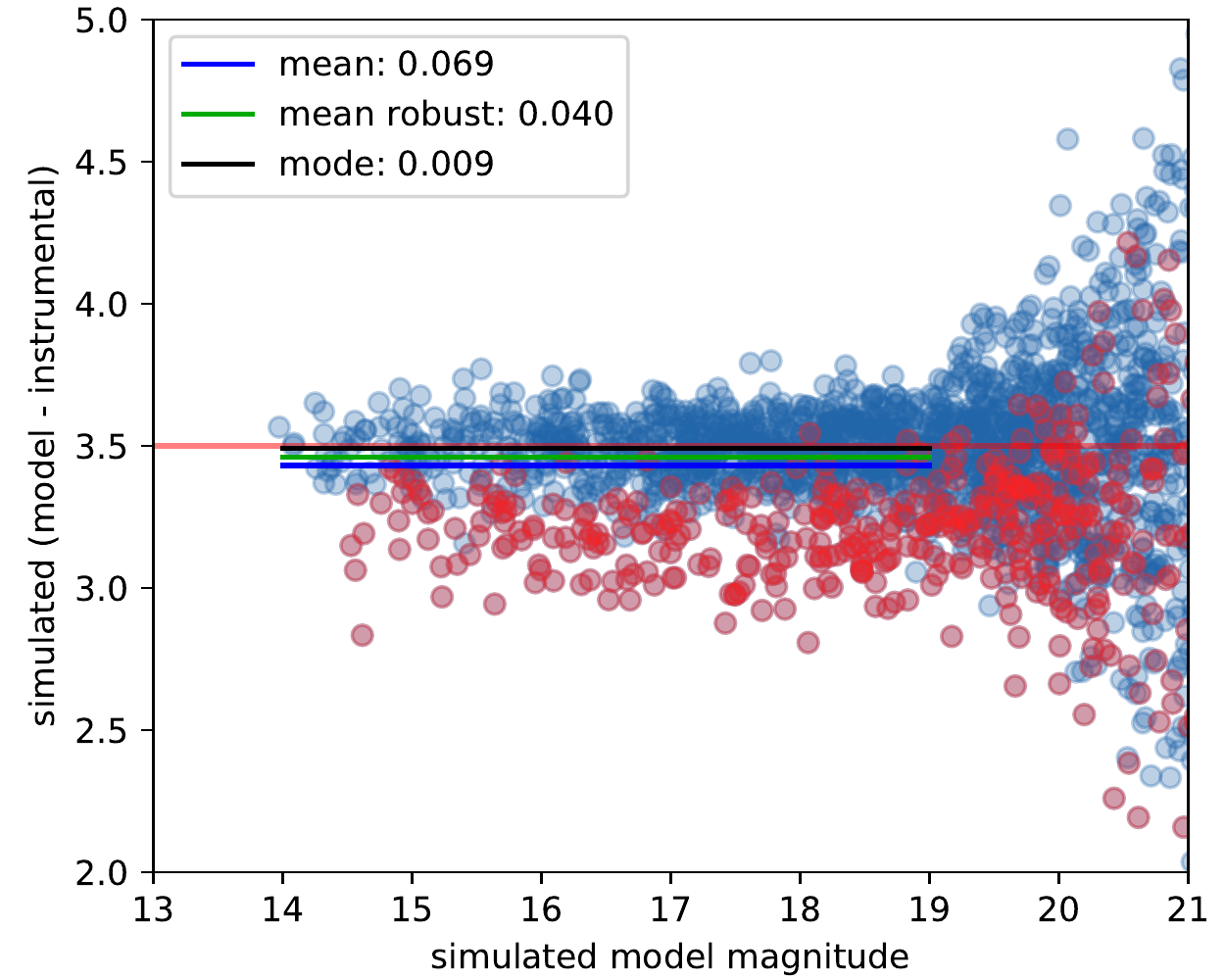}
\caption{\label{fig:sim_zp_fitting_ex} Simulated data that represents the characterisation of ZPs from the difference between model-predicted and instrumental magnitudes. The true data is shown in blue, while the red points correspond to a simulated population of outliers. The ZPs characterised from different estimators are shown as horizontal lines. The numbers in the legend correspond to the difference between the estimated and the true ZP in each case.}
\end{center}
\end{figure}

The simulation consists of introducing an expected ZP for each filter and a fraction of outliers from 10 to 40 per cent. The distribution of the instrumental magnitudes in each filter is taken from a typical S-PLUS observation. We also change the mean offsets and the scatter of the outliers from 0.1 to 0.4 mag. One of these simulations, for the characterisation of the ZP for filter $g$, is shown in Figure \ref{fig:sim_zp_fitting_ex}. In this case, we simulate an external calibration ZP of 3.5, with a population of outliers (red points) corresponding to 20 per cent of the total number of stars, with an offset of 0.3 mag and a scatter of 0.01 mag. We considered three different estimators for the ZPs: the mean of the differences between model and instrumental magnitudes; a robust mean, which includes a 1$\sigma$ clipping; and the characterisation through the mode of a kernel density estimated distribution. In this case, the mode (black line) is the estimator that provides the ZP that is closest to the simulated value.

We extensively repeated this analysis for different filters and properties of outliers. In Table \ref{tab:mean_vs_mode} we show the results of a set of simulations of different outlier populations. In each case, the numbers represent the difference between the estimated and the true ZP for each of the three estimators. We find that the mode estimator provides the best ZPs in all the considered scenarios. We also used these simulations to find the best interval to select the magnitudes for the ZP estimation (between 14 and 19 mag) and to find the best bandwidth for the Gaussian kernels (0.05 mag).

\begin{table}
\caption{\texttt{SExtractor} configuration used in DR2}
\centering
\label{tab:mean_vs_mode}
\includegraphics[width=.48\textwidth]{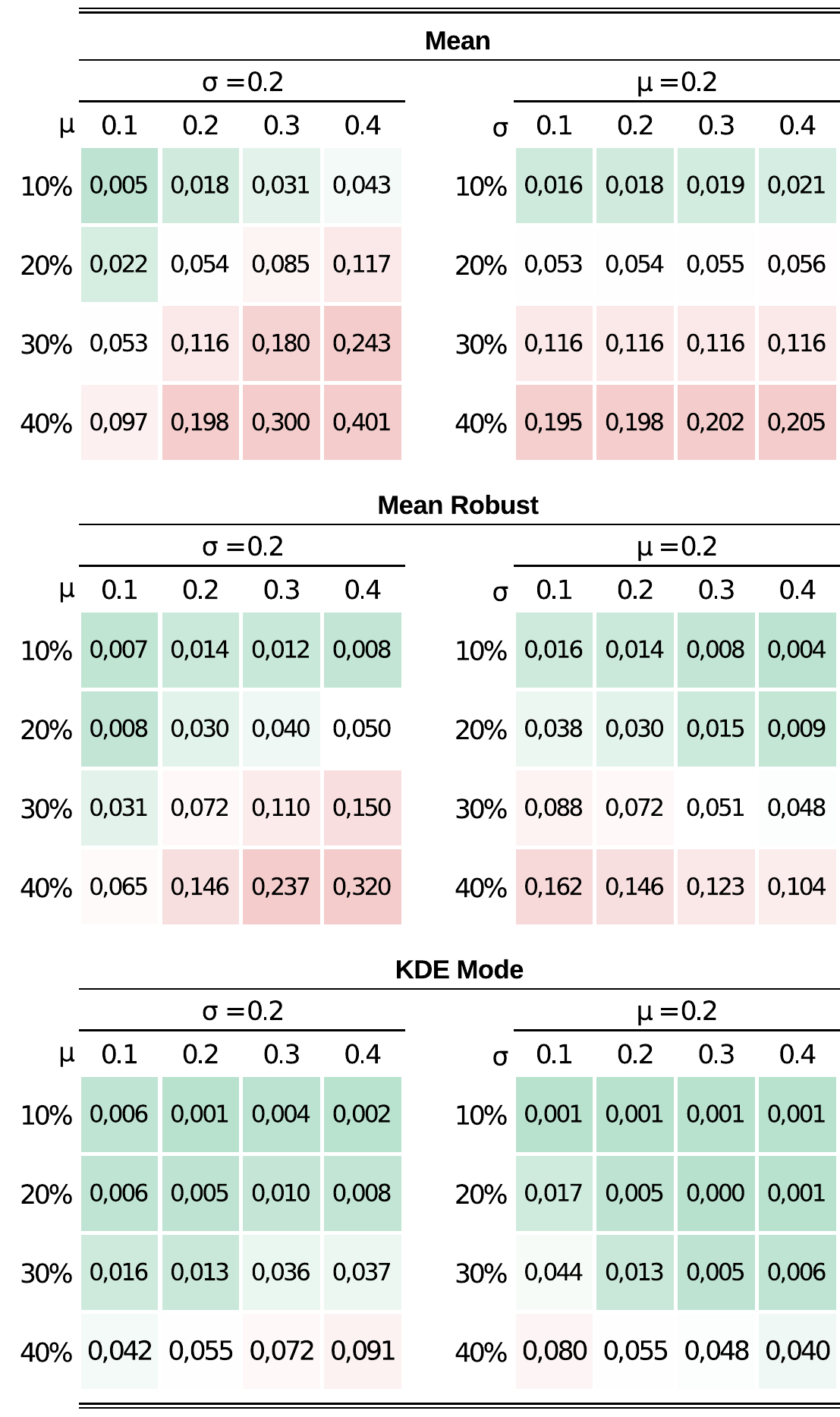}
\end{table}

\section{SExtractor configuration}
\label{ap:SExtractor}

In Table \ref{tab:SEx_config} we present the \texttt{SExtractor} configuration parameters used for the photometric measurements in DR2. Configurations that depend on the observation are marked as ``*VARIABLE*''. The magnitudes are measured in the 32 fixed apertures defined in PHOT\_APERTURES (radius in units of pixels) because they are necessary to construct the growth curves and estimate the aperture corrections. The only fixed apertures included in the final catalogues are the 3- and 6-arcsec diameter apertures.

\begin{table}
\caption{\texttt{SExtractor} configuration used in DR2}
\centering
\label{tab:SEx_config}
\begin{tabular}{rl}
\hline \hline

\multicolumn{2}{c}{DR2 SExtractor configuration} \\ \hline

\textbf{CATALOG\_TYPE}             & FITS\_1.0 (ASCII\_HEAD)\\
\textbf{DETECT\_TYPE}              & CCD \\
\textbf{DETECT\_MINAREA}           & 4 \\
\textbf{DETECT\_THRESH}            & 1.1 \\
\textbf{ANALYSIS\_THRESH}          & 3. \\
\textbf{FILTER}                    & Y \\
\textbf{FILTER\_NAME}              & tophat\_3.0\_3x3.conv \\
\textbf{DEBLEND\_NTHRESH}          & 64 \\
\textbf{DEBLEND\_MINCONT}          & 0.0002 \\
\textbf{CLEAN}                     & Y \\
\textbf{CLEAN\_PARAM}              & 1. \\
\textbf{MASK\_TYPE}                & CORRECT \\
\textbf{PHOT\_APERTURES}           & 1.81818, 3.63636, 5.45455 \\
                                   & 7.27273, 10.90909, 14.54545, \\
                                   & 18.18182, 21.09091, 24.00000, \\
                                   & 26.90909, 29.81818, 32.72727,\\
                                   & 35.63636, 38.54545, 41.45455,\\
                                   & 44.36364, 47.27273, 50.18182, \\
                                   & 53.09091, 56.00000, 58.90909, \\
                                   & 61.81818, 64.72727, 67.63636,\\
                                   & 70.54545, 73.45455, 76.36364,\\
                                   & 79.27273, 82.18182, 85.09091,\\
                                   & 88.00000, 90.90909\\
\textbf{PHOT\_AUTOPARAMS}          & 3.0, 1.82\\
\textbf{PHOT\_PETROPARAMS}         & 2.0, 2.72\\
\textbf{PHOT\_FLUXFRAC}            & 0.2, 0.5, 0.7, 0.9\\
\textbf{SATUR\_LEVEL}              & *VARIABLE*\\
\textbf{MAG\_ZEROPOINT}            & 20\\
\textbf{MAG\_GAMMA}                & 4.0\\
\textbf{GAIN}                      & *VARIABLE*\\
\textbf{PIXEL\_SCALE}              & 0.55\\
\textbf{SEEING\_FWHM}              & *VARIABLE*\\
\textbf{STARNNW\_NAME}             & default.nnw\\
\textbf{BACK\_SIZE}                & 256\\
\textbf{BACK\_FILTERSIZE}          & 7\\
\textbf{BACKPHOTO\_TYPE}           & LOCAL\\
\textbf{BACKPHOTO\_THICK}          & 48\\
\textbf{MEMORY\_OBJSTACK}          & 15000\\
\textbf{MEMORY\_PIXSTACK}          & 2600000\\
\textbf{MEMORY\_BUFSIZE}           & 4600\\

\hline \hline
\end{tabular}
\end{table}

\section{Affiliations}

$^{6}$Centro Brasileiro de Pesquisas F\'isicas, Rua Dr. Xavier Sigaud 150, CEP 22290-180, Rio de Janeiro, RJ, Brazil\\
$^{7}$Centro Federal de Educa\c{c}\~ao Tecnol\'ogica Celso Suckow da Fonseca, Rodovia M\'ario Covas, lote J2, quadra J, Distrito Industrial de Itagua\'i, \\ Itagua\'i, RJ, CEP 23810-000, Brazil\\
$^{8}$Community Science and Data Center/NSF’s NOIRLab, 950 N. Cherry Ave., Tucson, AZ 85719, USA \\
$^{9}$Centro de Astronom\'{i}a (CITEVA), Universidad de Antofagasta, Av. Angamos 601, Antofagasta, Chile\\
$^{10}$Millennium Institute of Astrophysics, Nuncio Monse\~nor Sotero Sanz 100, Of. 104, Providencia, Santiago, Chile\\
$^{11}$Departamento de F\'isica, Universidade Federal de Santa Catarina, Florian\'{o}polis, SC, 88040-900, Brazil\\
$^{12}$NOAO, P.O. Box 26732, Tucson, AZ 85726\\
$^{13}$GMTO Corporation 465 N. Halstead Street, Suite 250 Pasadena, CA 91107


\bsp	
\label{lastpage}
\end{document}